\documentclass[usenatbib,fleqn,twocolumn,trackchanges]{aastex62}

\submitjournal{ApJ}
\accepted{05/18/2020}

\usepackage{mathtools}

\usepackage{threeparttable}
\usepackage{lipsum}

\usepackage{amsmath,amssymb}
\usepackage{cases}
\usepackage{mathrsfs}
\usepackage{graphicx}
\usepackage{epstopdf}
\graphicspath{{./figures/}}
\usepackage{url}
\usepackage{aas_macros}

\newcommand\lsim{\mathrel{\rlap{\lower4pt\hbox{\hskip1pt$\sim$}}
    \raise1pt\hbox{$<$}}}
\newcommand\gsim{\mathrel{\rlap{\lower4pt\hbox{\hskip1pt$\sim$}}
    \raise1pt\hbox{$>$}}}

\newcommand{\Msun}{{\rm M_\odot}}

\shorttitle{The Hills mechanism and the S-stars}
\shortauthors{Generozov \& Madigan}
\begin{document}

\title{The Hills Mechanism and the Galactic Center S-stars}

\correspondingauthor{Aleksey Generozov}
\email{alge9397@colorado.edu}

\author[0000-0001-9261-0989]{Aleksey Generozov}
\affiliation{JILA and Department of Astrophysical and Planetary Sciences at CU Boulder, Boulder, CO 80309, USA}

\author[0000-0002-1119-5769]{Ann-Marie Madigan}
\affiliation{JILA and Department of Astrophysical and Planetary Sciences at CU Boulder, Boulder, CO 80309, USA}

\begin{abstract} 
  Our Galactic center contains young stars, including the few million year old clockwise disk between 0.05 and 0.5 pc from the Galactic center, and the S-star cluster of B-type stars at a galactocentric distance of $\sim$0.01 pc. Recent observations suggest the S-stars are remnants of tidally disrupted binaries from the clockwise disk. In particular, \citet{koposov+2019} discovered a hypervelocity star that was ejected from the Galactic center 5 Myr ago, with a velocity vector consistent with the disk. We perform a detailed study of this binary disruption scenario. First, we quantify the plausible range of binary semimajor axes in the disk.  Dynamical evaporation of such binaries is dominated by other disk stars rather than by the isotropic stellar population. For the expected range of semimajor axes in the disk, binary tidal disruptions can reproduce the observed S-star semimajor axis distribution. Reproducing the observed thermal eccentricity distribution of the S-stars requires an additional relaxation process. The flight time of the Koposov star suggests that this process must be effective within 10 Myr. We consider three possibilities: (i) scalar resonant relaxation from the observed isotropic star cluster, (ii) torques from the clockwise disk, and (iii) an intermediate-mass black hole. We conclude that the first and third mechanisms are fast enough to reproduce the observed S-star eccentricity distribution. Finally, we show that the primary star from an unequal-mass binary would be deposited at larger semimajor axes than the secondary, possibly explaining the dearth of O stars among the S-stars. 
 \end{abstract}


\section{Introduction}
The Galactic center contains a population of young stars. This population
includes a disk of O, B, and Wolf--Rayet stars between 0.05 and 0.5 pc
from the center (the clockwise disk; \citealt{levin&beloborodov03,paumard+2006}) as well as an isotropic cluster of B
stars at $\sim 0.01$ pc (the S-stars; \citealt{gillessen+2017}).
The presence of young stars on these scales is a challenge for current
theories of star formation, as strong tidal forces would shred molecular
clouds at the present location of the clockwise disk \citep{ghez+2003}. One proposed solution is
that the disk stars formed from a Toomre unstable gas accretion disk
\citep{levin2007}. However, it is unlikely that this instability could extend
to the present-day location of the S-stars (in particular, to the present-day orbit of S2; \citealt{nayakshin&cuadra2005}).

Instead, the S-stars may have migrated from larger scales, either via
interactions with a gas disk \citep{levin2007}, or via the Hills mechanism
\citep{hills1988, ginsburg&loeb2006, perets+2007, lockmann+2009, madigan+2009, dremova+2019}. In the latter case, binaries are scattered close to the
central supermassive black hole (SMBH) and tidally disrupted. One of the stars
in the binary escapes at high velocity, while the other remains bound to the
central SMBH. The bound stars 
can steepen the stellar density profile in the Galactic center
and enhance the rate of stellar tidal disruption events and extreme mass ratio inspirals  \citep{fragione&sari2018,sari&fragione2019}. Binaries may originate from large ($\gsim$ pc) scales \citep{perets+2007}, disruption of a young star cluster \citep{fragione+2017}, or from the clockwise disk itself \citep{madigan+2009}. The latter scenario is particularly attractive, as recent observations indicate that
the age of the S-stars is consistent with the age of the clockwise disk
\citep{habibi+2017}. Furthermore, a recently discovered hypervelocity star
also points to a disk origin for the S-stars \citep{koposov+2019}. 
This star is noteworthy because, of all of the known hypervelocity stars, it has the strongest
case for a Galactic center origin (see Figure 5 of \citealt{koposov+2019}).
This star has a velocity vector consistent with the clockwise disk.
Also, the flight time of this star from the Galactic center (4.8 Myr) is comparable to
the disk's age. While the evidence for a disk origin for this star is not quite definitive, the most parsimonious explanation for these observations is that some binaries from the disk
underwent disruptions a few million years ago, which would have left a
population similar to the observed S-stars.

Other models for the origins of
the S-stars include compression and fragmentation of an outflowing spherical
shell \citep{nayakshin&zubovas2018} and scattering of binaries by stellar mass
black holes \citep{trani+2019}. However, neither of these models would account
for the observed hypervelocity star.

Within the Hills scenario, it is difficult to reproduce the observed thermal
eccentricity distribution of the S-stars \citep{gillessen+2017}. The Hills mechanism would place
stars on highly eccentric orbits (e.g. \citealt{antonini&merritt2013}). Thus,
a subsequent relaxation process must be invoked to reproduce the
observations.

In this paper, we revisit the viability of the Hills mechanism as a source of
the S-stars. Motivated by the similar ages of the disk and S-stars, we focus
on the case where the source binaries originate in the clockwise disk as in
\citealt{madigan+2009}. First, we quantify the plausible range of binary
parameters within this disk, including the effects of dynamical evaporation.
We find that the disk stars dominate this process, even though they are less
numerous than the old, isotropic population in the Galactic center. We then
simulate a large ensemble of close encounters between disk binaries and the
central SMBH. Remnants from these encounters can account for the semimajor
axis distribution of the S-stars, but their eccentricities are too high.

We consider three possible mechanisms that could reproduce the observed
S-star eccentricities: (i) scalar resonant relaxation (SRR) due to the
surrounding isotropic cluster, (ii) torques from the clockwise disk and
(iii) an intermediate-mass black hole (IMBH) near the S-stars. SRR occurs in highly symmetric potential
where long-term correlations between stellar orbits lead to a build--up of
coherent torques that change the stars' eccentricity
\citep{rauch&tremaine1996}. Previous works (e.g.
\citealt{antonini&merritt2013}) find that the SRR timescale near the S-stars
can be as short as $\sim 10^7$ years if the effects of stellar mass black
holes are included. Here we revisit this estimate in light of new theoretical
developments in the theory of resonant relaxation. Specifically, we use the
formalism of \citet{bar-or&fouvry2018}, who derived diffusion coefficients for
resonant relaxation from the orbit-averaged Hamiltonian of a star in a
spherically symmetric star cluster. In agreement with \citet{antonini&merritt2013}, we find that a realistic population of
stellar mass black holes can reproduce the eccentricity distribution 
of the S-stars within $10^{7}$ years. This timescale is comparable to the ages of the S-stars
and flight time of the \citet{koposov+2019} star.
We find that torques from the clockwise disk are quenched by
general relativistic precession and would not be able to reproduce the
S-stars' eccentricity distribution. Previously, IMBHs have been considered as
a mechanism for reproducing the orbital properties of the S-stars
\citep{merritt+2009}, though this work did not consider stars on highly
eccentric orbits as would be expected from the Hills mechanism. Here we show that an
$\sim 10^3 M_{\odot}$ IMBH at $\sim$0.01 pc thermalizes the S-star orbits
within a few Myr, starting from an initial eccentricity that is
$\gsim$0.97.

The remainder of this paper is organized as follows. We
review the disk instability that pushes binaries to tidal disruption in
$\S$~\ref{sec:instability}. In $\S$~\ref{sec:gc} we
discuss the plausible range of binary properties within the clockwise disk.  We then simulate a large number of close encounters
between disk binaries and the Galactic center SMBH, as summarized in
$\S$~\ref{sec:ensemble} and $\S$~\ref{sec:results}. We explore various processes that could reproduce the
observed S-star eccentricity distribution in $\S$~\ref{sec:relax}. Finally,
we consider other consequences of tidal encounters between binaries and SMBHs
in $\S$~\ref{sec:disc}.

\section{Eccentric disk instability}
\label{sec:instability}
If the clockwise disk started as a high-eccentricity ($e\gsim 0.6$), 
lopsided structure,
some of its stars and binaries could have been excited to nearly radial
orbits and tidally disrupted as described in \citet{madigan+2009}. Briefly,
stars in the disk precess retrograde to their orbital angular momenta due to
the influence of the surrounding star cluster \citep{madigan+2011,merritt2013}. Stars with
higher eccentricities precess more slowly and end up behind the bulk of the
disk, which then torques them to even higher eccentricities. Eventually, such
stars (or binaries) may be tidally disrupted.
\citet{Bonnell2008} and \citet{mapelli+2012} found that such a lopsided, eccentric 
disk could arise from the disruption of a molecular cloud in the Galactic center (see, e.g.
Figure 2 in the latter).

To illustrate the development of this instability, we perform N-body
simulations of an eccentric, apsidally aligned disk, closely following
\citet{madigan+2009} for our initial conditions. Specifically, we initialize a
disk with 100 point masses of $100\, M_{\odot}$ each (for a total disk mass of
$10^4 M_{\odot}$), orbiting a $4\times 10^{6}\, M_{\odot}$ SMBH. The disk stars
have semimajor axes between 0.05 pc and 0.5 pc with an $r^{-2}$ surface
density profile like the clockwise disk. All stars start with an initial
eccentricity of 0.7 and nearly aligned eccentricity and angular momentum
vectors. The simulations also include the potential of an $r^{-1.5}$ stellar
density profile, with $4\times 10^6 M_{\odot}$ of stars within 4 pc.

We integrate the disk forward in time with the IAS15 integrator of the \texttt{REBOUND}
N-body code
\citep{rein.liu2012,rein.spiegel2015}.\footnote{\url{https://github.com/hannorein/rebound}}
We also include an approximate treatment of general relativity (GR) effects via \texttt{REBOUNDX}
\citep{tamayo+2019}.\footnote{Available at
\url{https://github.com/dtamayo/reboundx}; We use the ``GR'' effect.} Any
particles that pass within $3\times 10^{-4}$ pc of the central SMBH are
recorded as binary disruptions. In general, these stars are not removed from
the simulation to keep the disk potential as constant as possible, but we only
record one disruption per particle.\footnote{We found it necessary to remove particles 
passing within 10 gravitational radii of the central black hole to avoid numerical 
instability.}

\begin{figure}
\includegraphics[width=8.5cm]{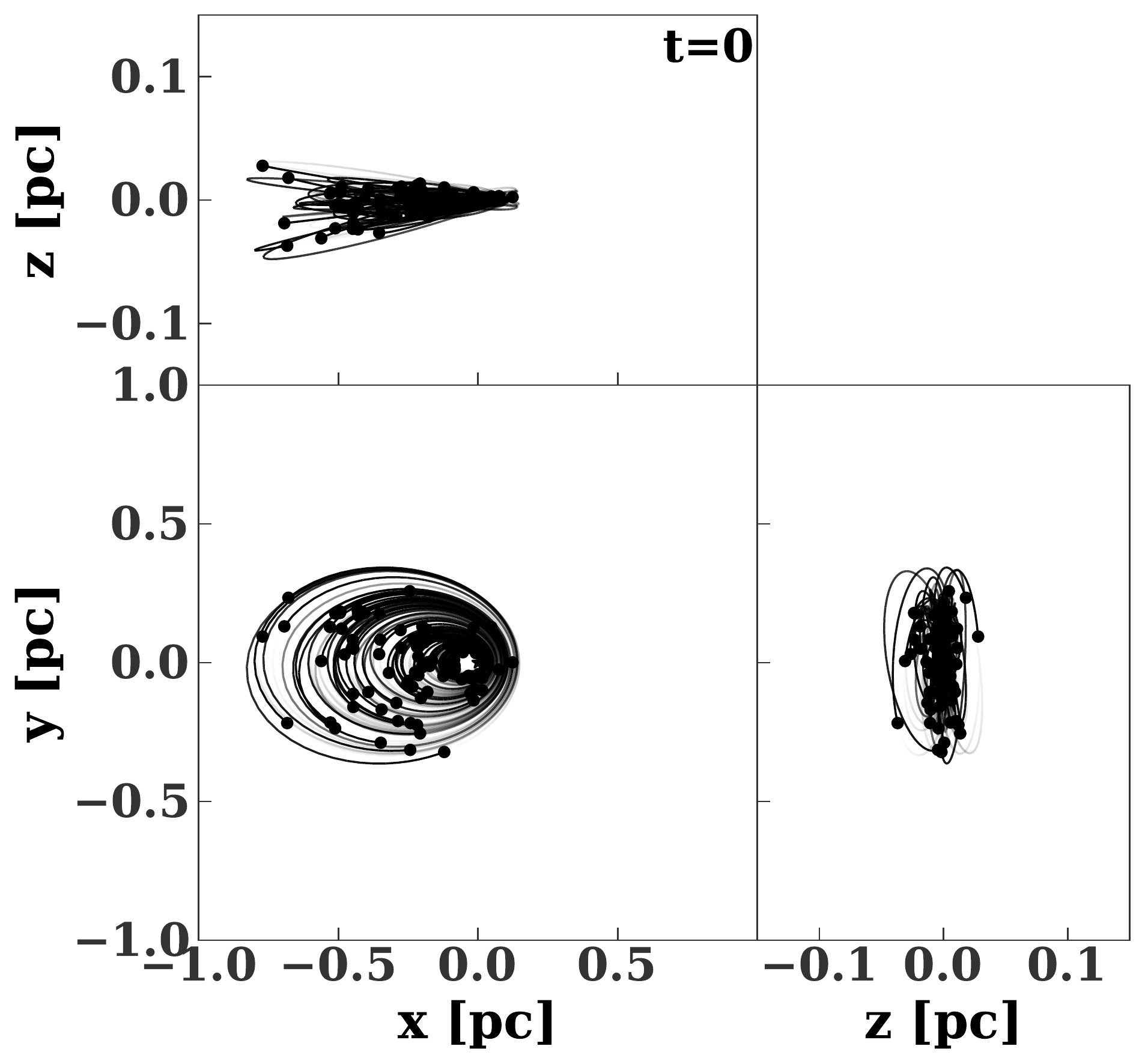}
\includegraphics[width=8.5cm]{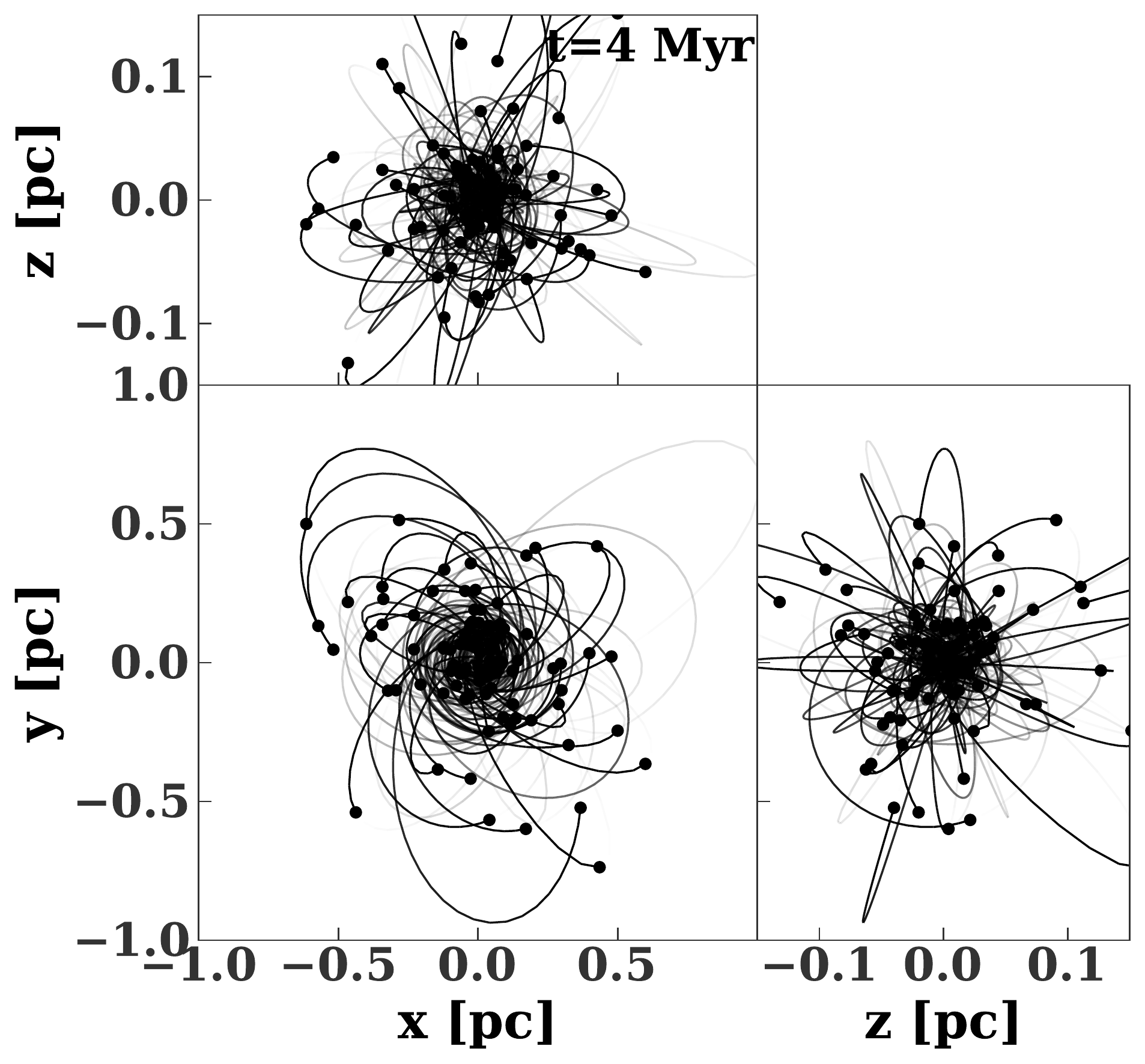}
\caption{\label{fig:nbodySim} Top panel: initial orbits for an N-body
simulation of an eccentric, apsidally aligned disk in the Galactic center.
Bottom panel: The disk after 4 Myr of evolution, after the eccentric
disk instability has developed.}
\end{figure}

\begin{figure}
\includegraphics[width=8.5cm]{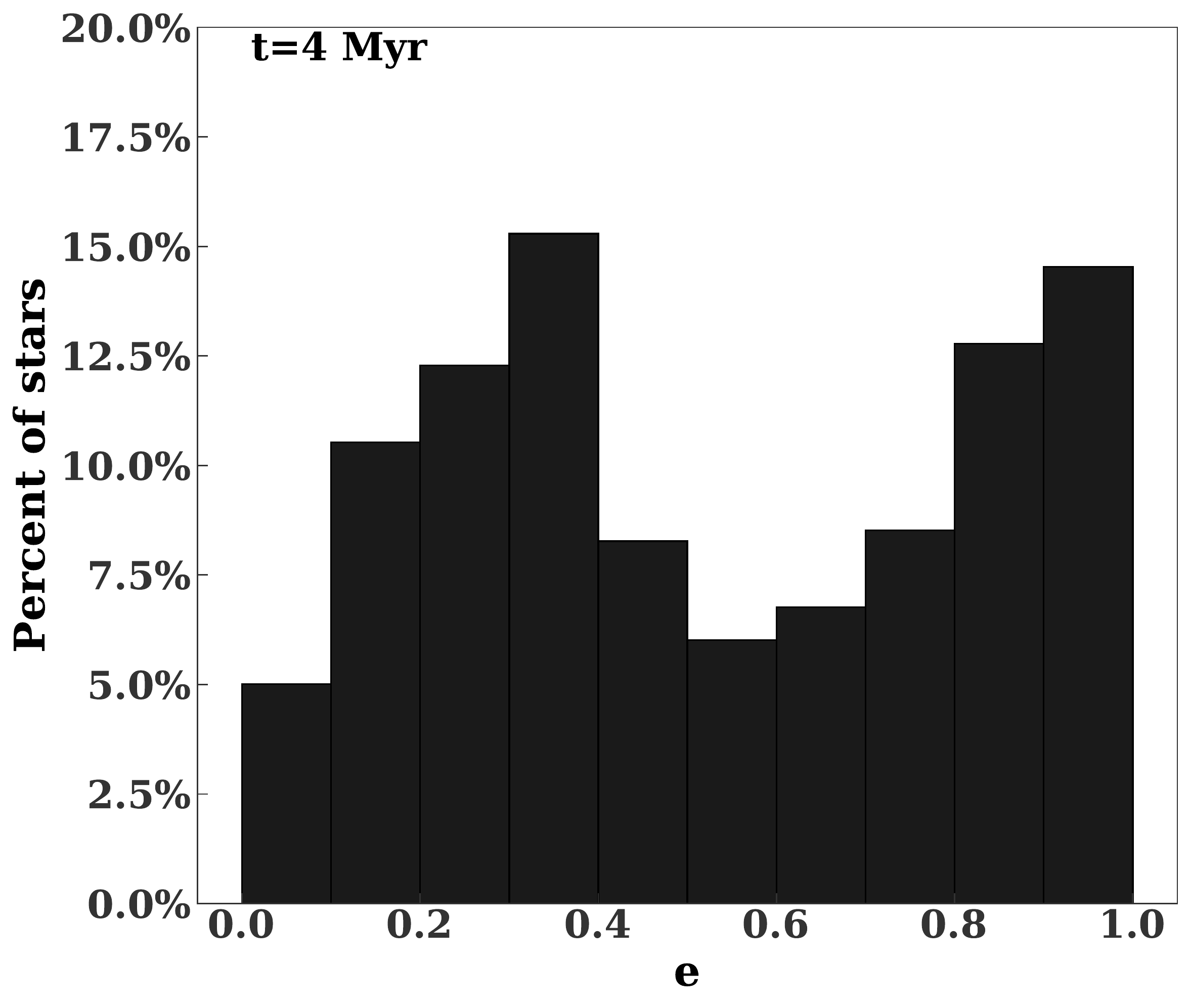}
\caption{\label{fig:eccHist} Bimodal eccentricity distribution of an 
evolved, eccentric disk in the Galactic center. 
This distribution is constructed by stacking the results 
of four different simulations (like the one shown in Fig.~\ref{fig:nbodySim}). The initial
eccentricity distribution is a delta function at $e=0.7$.}
\end{figure}

The top panel of Fig.~\ref{fig:nbodySim} shows the initial orbits of disk
stars for a particular simulation, while the bottom panel shows the disk
orbits after $\sim 4$ Myr. By this time, the orbits are spread out and have a
bimodal eccentricity distribution, as shown in Fig.~\ref{fig:eccHist} (see also
\citealt{madigan+2009,gualandris+2012}). \citet{bartko+2009} claimed detection of a bimodal
eccentricity distribution within the clockwise disk. However, subsequent work
has argued that these measurements are contaminated by nondisk stars and favor a
unimodal eccentricity distribution with a mean eccentricity of $0.27\pm 0.07$
\citep{yelda+2014}. The latter observation would disfavor the eccentric disk
scenario. However, caution is warranted in interpreting these observations, as
disk membership and the observed disk eccentricities may be contaminated by binary
stars \citep{naoz+2018}.

Only a few percent of the particles in our simulations undergo disruptions
compared to $\sim30$ percent for simulations with similar initial conditions
in \citet{madigan+2009}. However, the total number of disruptions is sensitive
to the slope of the stellar density profile, and the total mass of the disk.
In particular, the disk may have been more massive earlier in its history,
when gas was still present. Doubling the mass of the disk to $\sim 2\times
10^4 M_{\odot}$, and taking a flatter $r^{-1.1}$ stellar density profile
(motivated by \citealt{schodel+2017}), increases the disrupted fraction up to
17 $\pm$ 2 percent.

\begin{figure}
\includegraphics[width=8.5cm]{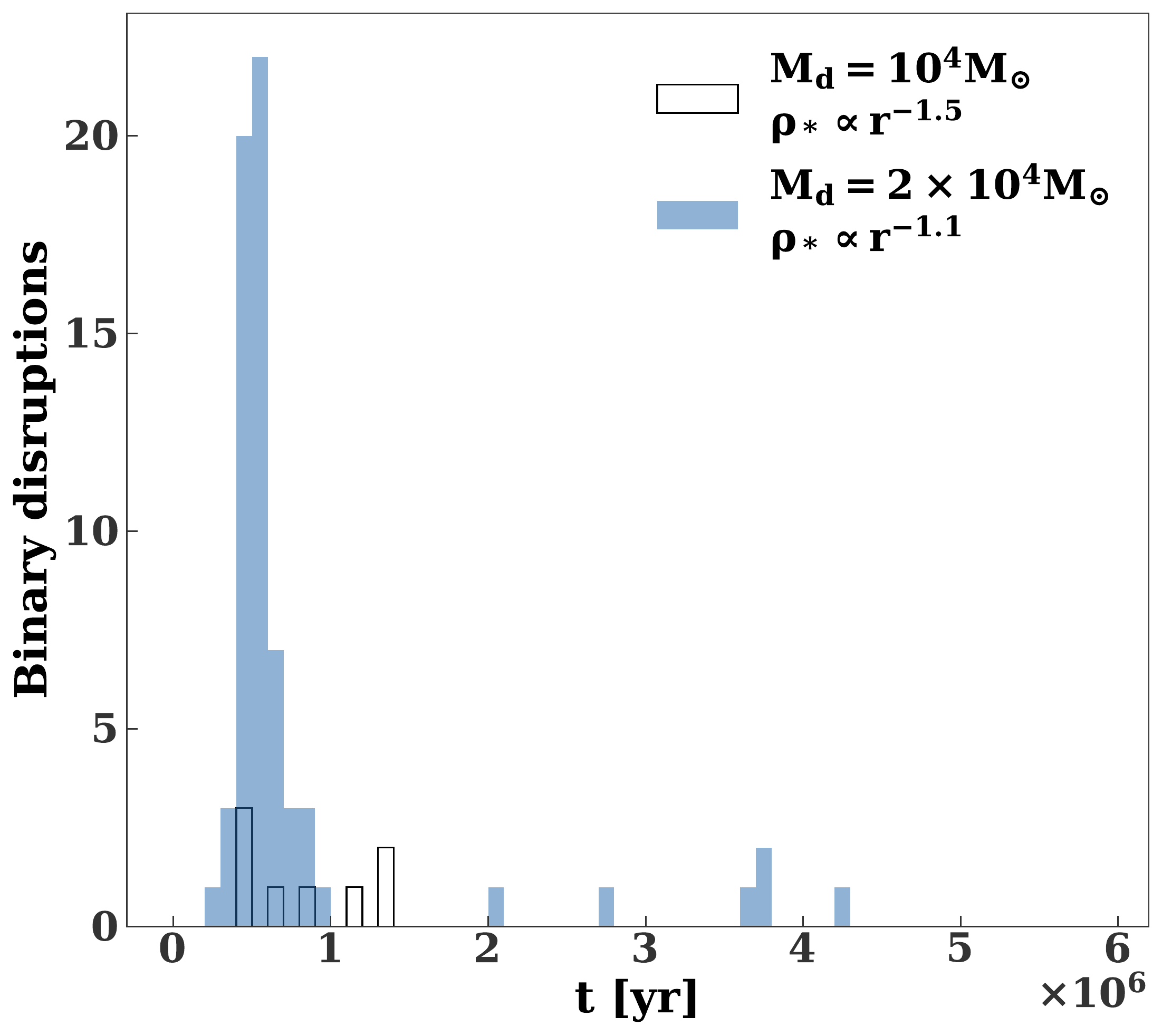}
\caption{\label{fig:disTime}  Times when binaries are disrupted in
our disk simulations ($t=0$ is the start of the simulation). The blue filled
histogram shows the distribution of disruption times for simulations with a
$2\times 10^4 M_{\odot}$ disk and an $r^{-1.1}$ background stellar density
profile, while the black open histogram shows the distribution of disruption
times for simulations with a $10^4 M_{\odot}$ disk and a $r^{-1.5}$ stellar
background. Binary disruptions begin after a secular time ($1-2\times 10^5$
years). Each distribution is constructed by stacking the results of four
different simulations.}
\end{figure}

Fig.~\ref{fig:disTime} shows binary disruptions occur promptly after the
disk forms. In particular, disruptions begin at a few times the disk's secular time, viz.

\begin{align}
t_{\rm sec}=\frac{M}{M_d} P ,
\end{align}
where $M$ and $M_d$ are the SMBH and disk mass respectively, and $P$ is the orbital period.
Here $t_{\rm sec}\approx 1-2\times 10^5$ years at 0.05 pc (the inner edge of the clockwise disk).

We note that although our simulations have approximately an order of magnitude 
fewer stars than the real disk, \citet{madigan+2018} found that the number of disruptions in eccentric 
disk simulations depends weakly on the number of particles. This is expected, because the 
secular torques from the disk would not depend on the number of stars it contains. We have also run some simulations with 300 particles for 1.4 Myr and find a similar disrupted fraction. For example, we find a disrupted fraction of 14 $\pm$ 1\% (17 $\pm$ 2\%) for 300 (100) particle simulations with an $r^{-1.1}$ stellar background and a $2\times 10^4 M_{\odot}$ disk.

Overall, these simulations show that an eccentric, apsidally aligned disk can produce a large number of binary disruptions via a secular instability. In particular, these simulations show the clockwise can reproduce the observed number of S-stars (see $\S$~\ref{sec:ns}) soon after its formation.

\subsection{Full or empty loss cone?}
\label{sec:lc}
In order to be tidally disrupted, a binary has to enter into a loss cone
of low angular momentum orbits that pierce the tidal radius. Binaries can be in
either the empty or full loss cone regime. In the former case, binaries
would diffuse into the loss cone over many orbital periods. In the latter
case, binaries can jump into and out of the loss cone in a single orbital
period. Here we show that disk binaries are in the empty loss cone regime.

The typical torque per unit mass from the disk is 
\begin{equation}
\tau_d \sim \frac{G M_{d}}{a_d},
\end{equation}
where $M_{\rm d}\approx 10^4 M_{\odot}$ is the disk mass and $a_d$ is a
characteristic semimajor axis for a disk orbit. The change in angular momentum per
orbit due to this torque is 

\begin{align}
&\Delta j =\tau_d P(a_d) \equiv  Q j_{\rm lc} \implies\nonumber\\
&Q= \frac{2 \pi}{\sqrt{2}}\frac{M_{\rm disk}}{M} \sqrt{\frac{a_d}{r_t}}\nonumber\\
&Q\approx 0.27
 \left(\frac{a_{\rm bin}}{1 \,{\rm au}}\right)^{-1/2}
\left(\frac{a_d}{0.1 \,{\rm pc}}\right)^{1/2} \left(\frac{m}{20\, M_{\odot}}\right)^{1/6}
\end{align}
where $P(a_d)$ is the orbital period and $j_{\rm lc}$ is the loss cone angular
momentum (see also \citealt{wernke&madigan2019} equation 12). Thus, disk
binaries are typically in the empty loss cone regime ($Q<1$), and approach the
tidal radius gradually over several orbits.

\section{Galactic center Binary Population}
\label{sec:gc}
An SMBH can tidally disrupt a binary into two stars. The postdisruption orbits
of these stars depend on the initial properties (e.g. the semimajor axis and
eccentricity) of the binary. In this section, we quantify the plausible range of
binary properties of the young stellar population in the clockwise disk in the
Galactic center.

\subsection{Binary dynamics}
\label{sec:bd}
Binaries with a semimajor axis

\begin{align}
&a_{\rm bin}> a_{\rm hs}=\frac{G m_1 m_2}{\sigma^2 <m>},\nonumber\\
\label{eq:ahs}
\end{align}
are soft and will evaporate over time. Here $m_1$ and $m_2$ are the
masses of the primary and secondary star in the binary, respectively; $<m>$
and $\sigma$ are the mean mass and the velocity dispersion of the surrounding
stellar population. The timescale for evaporation is 

\begin{align}
t_{\rm evap}=0.07 \frac{v_{12}^2 \sigma}{G^2 n <m^2> \ln \Lambda_{12}},
\label{eq:tevap}
\end{align}
where $v_{12}$ is the (internal) orbital velocity of the binary, $<m^2>$ is
the second moment of the stellar mass function, $n$ is the stellar number
density, and $\Lambda_{12} \approx2 \sigma^2/v_{12}^2$ (see equation 3 of
\citealt{alexander&pfuhl2014}). From Equation~\eqref{eq:tevap}, soft binaries at a galactocentric radius $r$ with

\begin{align}
&a_{\rm bin}>a_{\rm evap}=e^{W(t_a/t)} \frac{G m}{2 \sigma^2}\\
&t_a=\frac{0.14 \sigma^3}{G^2 n <m^2>},
\label{eq:aevap}
\end{align}
would have evaporated after time $t$. Here $m$ is the total mass of the binary
and $W$ is the Lambert $W$ function. 

Gravitational wave emission and finite stellar radii set a lower limit on the
binary semi-major axis. Within time $t$, binaries with

\begin{align}
&a_{\rm bin}<a_{\rm GW}=\left[\frac{5 c^5}{256 G^3 m^3 t} \frac{(1+q)^2}{q} f(e_{\rm bin})\right]^{-1/4}\\
&f(e)=\frac{(1-e^2)^{7/2}}{1+\frac{73}{24}e^2+\frac{37}{96}e^4}
\label{eq:agw}
\end{align}
would have coalesced. Here $e$ and $q$ are the binary eccentricity and mass ratio (see 
\citealt{peters1964} equation 5.6). Binaries with 

\begin{align}
&a_{\rm bin} (1-e_{\rm bin})<a_{\rm Roche}\nonumber\\
&={\rm max}[R_1/h(m_1/m_2), R_2/h(m_2/m_1)]\nonumber\\
&h(q)=\frac{0.49 q^{2/3}}{0.6 q^{2/3}+\log(1+q^{1/3})},
\label{eq:aroche}
\end{align}
will experience Roche lobe overflow \citep{eggleton1983}. Here $m_1$ and $m_2$ 
are the masses of the individual stars, and $R_1$ and $R_2$ are their radii. We consider $a_{\rm
roche}$ to be a lower limit to the binary separation. For the young stars in
the clockwise disk, $a_{\rm GW}<a_{\rm Roche}$, and gravitational wave inspiral
is unimportant.

Kozai--Lidov oscillations can also be important for binaries in the
Galactic center \citep{lidov1962, kozai1962, antonini&perets2012,prodan+2015,stephan+2016,fragione&antonini2019}. The quadrupole Kozai--Lidov timescale at the inner edge of the clockwise disk is 

\begin{equation}
t_{\rm KL}= 2.1 \times 10^5 {\rm yr} \left(\frac{m}{20 M_{\odot}}\right)^{1/2} \left(\frac{a_{\rm bin}}{1 {\rm au}}\right)^{-3/2}.
\label{eq:tkl}
\end{equation}
The octupole Kozai--Lidov timescale is (see the review by \citealt{naoz2016})

\begin{align}
    t_{\rm KL, oct} & \approx   \frac{t_{\rm KL}}{\epsilon} \nonumber\\
    & =  \frac{a_{d}}{a_{\rm bin}} \frac{1-e_{d}^2}{e_{d}} t_{\rm KL} \nonumber\\
     \approx & 10^4 \left(  \frac{a_{\rm bin}} {1 {\rm au}}  \right)^{-1} \frac{1-e_d^2}{e_d} t_{\rm KL},
\end{align}
where the subscript ``d'' denotes properties of the disk binaries' outer orbits, and in the third line we take $a_{d}=0.05$ pc. Thus, $t_{\rm KL, oct}$ is typically $\gsim 10^9$ years, and octupole level effects can be neglected for the few million years old disk stars. Therefore, Kozai--Lidov oscillations would only occur for highly inclined, wide binaries ($a_{\rm bin}\gsim 1$ AU) for which the precession timescale due to GR is longer than the Kozai--Lidov timescale. We expect the disk binaries to have low inclinations (i.e. their internal angular momentum is aligned with the angular momentum of their outer orbit around the SMBH). Vector resonant relaxation may realign the outer orbits of binaries in the disk on $\sim$Myr timescales \citep{kocsis&tremaine2011}. Nonetheless, the eccentric disk instability is also quite fast and can produce binary disruptions on a similar timescale (see Fig.~\ref{fig:disTime}). Moreover, the binaries that are disrupted would be precisely those with small inclinations, as secular torques from the disk would be aligned with their angular momentum vectors \citep{madigan+2018, foote+2020}.

\subsection{The Galactic center}
There are two populations that can perturb binaries in the $\sim 4$ Myr old
\citep{lu+2013} clockwise disk: (1) the surrounding old, isotropic star
cluster and (2) the disk stars themselves. 

To see which is more important, we compare the evaporation timescales of the two populations (Equation~\ref{eq:tevap}). The disk stars will have a shorter evaporation timescale and will dominate evaporation if

\begin{align}
\frac{n_{\rm iso} <m^2>_{\rm iso}}{\sigma_{\rm iso}}\leq \frac{n_d <m^2>_d}{\sigma_d},
\label{eq:diso}
\end{align}
where subscripts ``$d$'' and ``${\rm iso}$'' indicate the disk and isotropic populations
respectively. Note that $\sigma_d/\sigma_{\rm iso}\approx (H/r)$, where $H$ is the 
scale height of the disk.

There are $\sim$1000 stars\footnote{This number is sensitive to
the assumed lower bound of the disk mass function. Here we assume $1
M_{\odot}$.} in the disk with an $r^{-2}$ surface density profile between $\sim$0.05
and $\sim$0.5 pc \citep{paumard+2006,do+2013}. Thus,

\begin{align}
n_{\rm d}\sim 6.3\times 10^4 \left(\frac{H}{r}\right)^{-1} \left(\frac{r}{0.1 {\rm pc}}\right)^{-3} {\rm pc}^{-3}. 
\label{eq:nd}
\end{align}
For the observed $M^{-1.7}$ \citep{lu+2013} mass function in the disk 

\begin{align}
<m^2>_d^{1/2}\approx 11 M_{\odot}.
\label{eq:m2d}
\end{align}
We assume that this mass function extends from 1 $M_{\odot}$ to 60 $M_{\odot}$
(the main sequence turn-off mass for a 4 Myr old stellar population).

The stellar density in the isotropic component can be constrained by existing
observations. For example, \citet{schodel+2017} found that the stellar density
is $\sim 1-2 \times 10^5 M_{\odot}$ pc$^{-3}$ one parsec from the Galactic
Center with an $r^{-1.13\pm 0.03_{\rm model}\pm 0.05_{\rm sys}}$ profile.
However, the mean square mass will be a strong function of the number of
stellar mass black holes in the isotropic component. We adopt the
Fiducial$\times 10$ model from \citet{generozov+2018} for the black hole
density profile. The black holes in this model are formed near the present day
clockwise disk. The present epoch is assumed to be typical, so $\sim$300
massive stars form every few million years, and become black holes and neutron
stars. This model approximately reproduces the observed stellar density
profile in the Galactic center, as well as the recently discovered population
of black hole X-ray binaries in the central parsec
\citep{hailey+2018,mori+2019} via tidal capture of main-sequence stars. This model
implicitly assumes that the initial mass function in the disk is truncated near $10 M_{\odot}$,
which is problematic, as the hypervelocity star observed by
\citet{koposov+2019} is an A-type star with a few solar masses. Nonetheless,
the black hole density profile in the Fiducial$\times 10$ model is within a
factor of a few of previous estimates on the scales of interest, which assume
that black holes and low-mass stars form as a single population with a
standard mass function (see the review by \citealt{alexander+2017} and the
references therein).

In the Fiducial$\times 10$ model
\begin{align}
n_{\rm iso}\approx 5\times 10^6 \left(\frac{r}{0.1 \,\,{\rm pc}}\right)^{-1.5} {\rm pc}^{-3}
\label{eq:niso}
\end{align}
and 
\begin{align}
<m^2>_{\rm iso}^{1/2}\approx 7 \left(\frac{r}{0.1 {\rm pc}}\right)^{-0.3} M_{\odot}
\label{eq:m2iso}
\end{align}
between $0.01$ pc and $0.1$ pc. The number density is dominated by
main sequence stars, but stellar mass black holes
dominate the two--body relaxation on these scales. From
equations~\eqref{eq:diso} the disk stars dominate the evaporation of binaries
for disk aspect ratios
\begin{align}
\frac{H}{r} \lesssim 0.18 \left(\frac{r}{0.1 \,{\rm pc}}\right)^{-0.45}.
\end{align}
This inequality should be satisfied in the clockwise disk, which has an observed
$H/r \approx 0.1$ \citep{paumard+2006}.

\section{Binary disruption ensemble}
\label{sec:ensemble}
We simulate an ensemble of encounters between stellar binaries and a $4\times
10^6 \Msun$ SMBH, where the binaries pass near their tidal radius.
The binary properties in this ensemble are summarized in Table~\ref{tab:bin}.

\begin{deluxetable}{lccc}
\tablecaption{\label{tab:bin} Summary of Parameters for our Monte Carlo Simulations
of Binary--SMBH Encounters.}
\tablewidth{0pt}
\tabletypesize{\small}
\tablehead{\colhead{Parameter}  & \colhead{Distribution}  & \colhead{Lower Bound}  & \colhead{Upper Bound}} 
\startdata
$a_{\rm bin}$  & $a_{\rm bin}^{-1}$ & Equation~\ref{eq:amin} & Equation~\ref{eq:amax} \\
\hline
$m_1$     & $m_1^{-1.7}$ & $8 M_{\odot}$  & $15 M_{\odot}$   \\ 
\hline
$e$       & 0, thermal   & 0              & 1   \\ 
\enddata
\tablecomments{The three rows are the binary semimajor
axis, the mass of each star, and the eccentricity. The mass ratio is unity and the pericenter is the effective tidal radius (Equation~\ref{eq:rteff}).}
\end{deluxetable}

We draw the binary semimajor axis from a log-uniform distribution. The maximum semimajor axis is set by dynamical evaporation, while the minimum semimajor axis is set by finite stellar radii and gravitational wave emission. In particular, the semimajor axis distribution extends from 

\begin{align}
a_{\rm min}={\rm max}\left[a_{\rm gw}, a_{\rm roche}\right]
\label{eq:amin}
\end{align}
to 
\begin{align}
a_{\rm max}={\rm min}\left[{\rm max}\left[a_{\rm hs}, a_{\rm evap}\right], 100 {\,\rm au} \right].
\label{eq:amax}
\end{align}
The semimajor axes on the right-hand side of the above equations are
defined in $\S$~\ref{sec:bd} (see equations~\ref{eq:ahs},~\ref{eq:aevap},~\ref{eq:agw}, and~\ref{eq:aroche}).
These quantities are functions of the binary mass, mass ratio, and eccentricity. They also depend on the
galactocentric radius and the age of the binaries.
We evaluate $a_{\rm min}$ and $a_{\rm max}$ assuming 4 Myr old stars at a
galactocentric radius of 0.1 pc. Note we use the disk properties (number density, velocity dispersion, and mean square mass)
to evaluate $a_{\rm hs}$ and $a_{\rm evap}$. In particular, we use equations~\eqref{eq:nd} and~\eqref{eq:m2d} for the number 
density and mean squared mass respectively. The velocity dispersion is $\sqrt{G M/r} (H/r)$, $H/r$ is the disk aspect ratio (0.1 here).

\begin{figure}
\includegraphics[width=8.5cm]{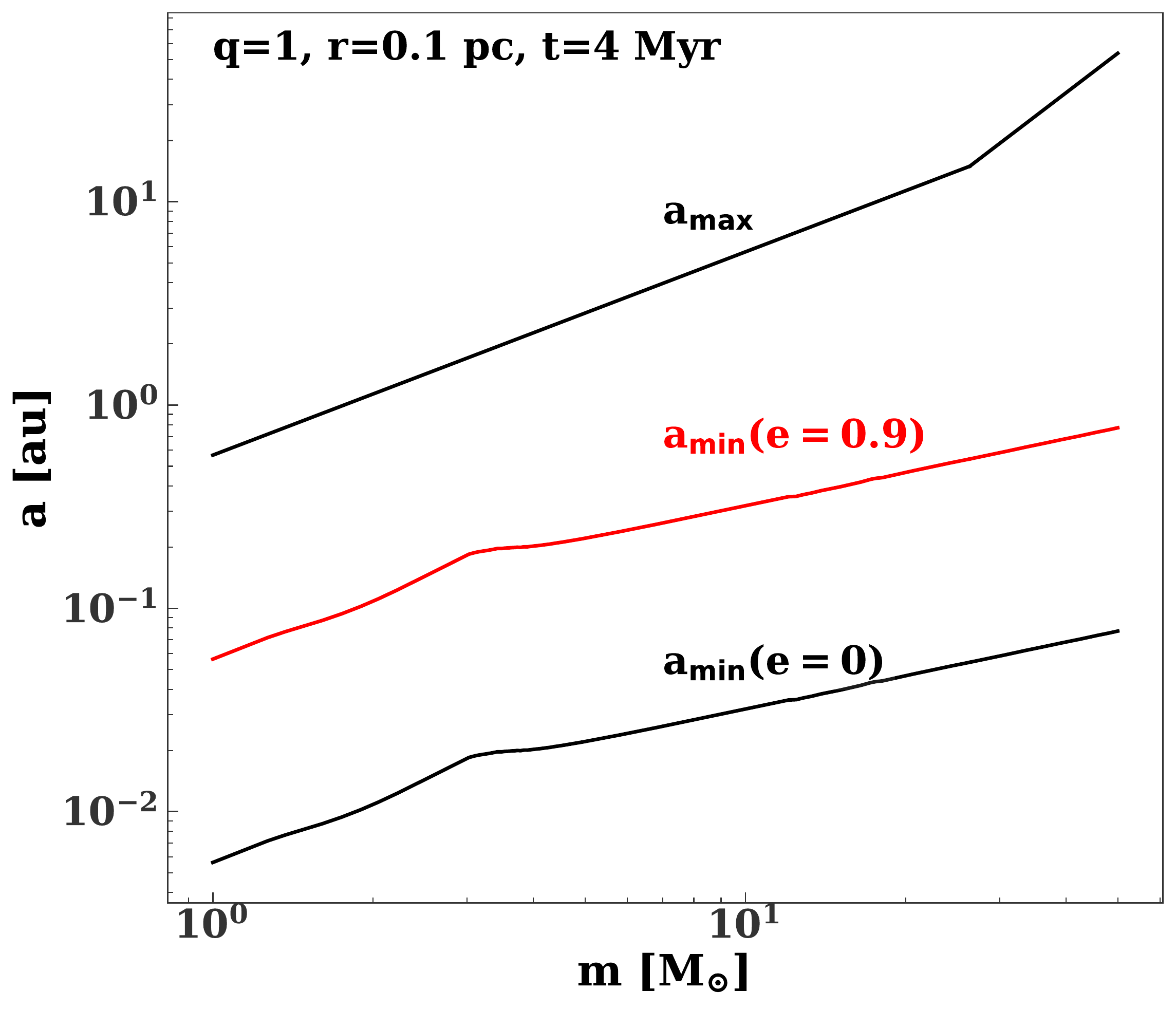}
\caption{\label{fig:sma} Minimum and maximum possible semi-major axis 
as a function of mass for binaries within the clockwise disk at a
galactocentric distance of 0.1 pc. The stars in the binary have equal masses 
($q=1$).}
\end{figure}

Fig.~\ref{fig:sma} shows the minimum
and maximum semimajor axes as function of binary mass for two different binary eccentricities.

The stellar masses are drawn from the observed mass function of the disk
\citep{lu+2013}. As the observed S-stars are all B--stars we consider only
stars between 8 and 15 $M_{\odot}$ (these are approximately minimum and
maximum masses of the eight S-stars studied by \citealt{habibi+2017}).\footnote{Note that the \citet{habibi+2017} stars
are among the brightest S-stars; the dimmest observed S-stars are $\sim 3 M_{\odot}$ \citep{cai+2018}.} The
mass ratio of the binary is taken to be one for simplicity. We use two
different distributions for the internal binary eccentricity (circular and
thermal). 

As noted in \S~\ref{sec:lc}, binary disruptions will likely be in the empty loss cone
regime. In this regime, the pericenter distribution of tidally 
disrupting single stars is strongly peaked at the tidal radius. The binary
case is more complicated, as the disruption probability gradually increases as
the pericenter decreases.

\begin{figure}
\includegraphics[width=8.5cm]{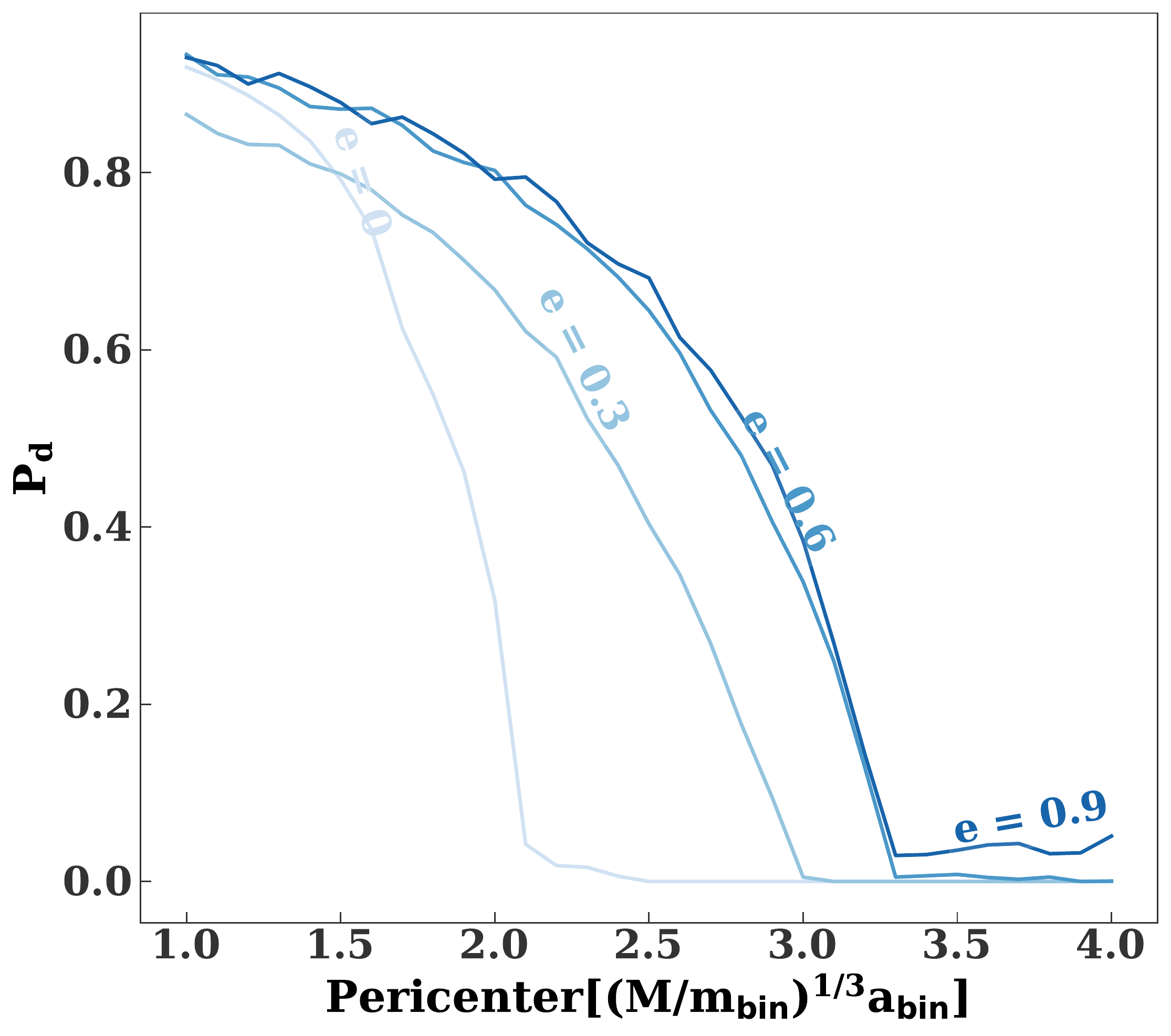}
\includegraphics[width=8.5cm]{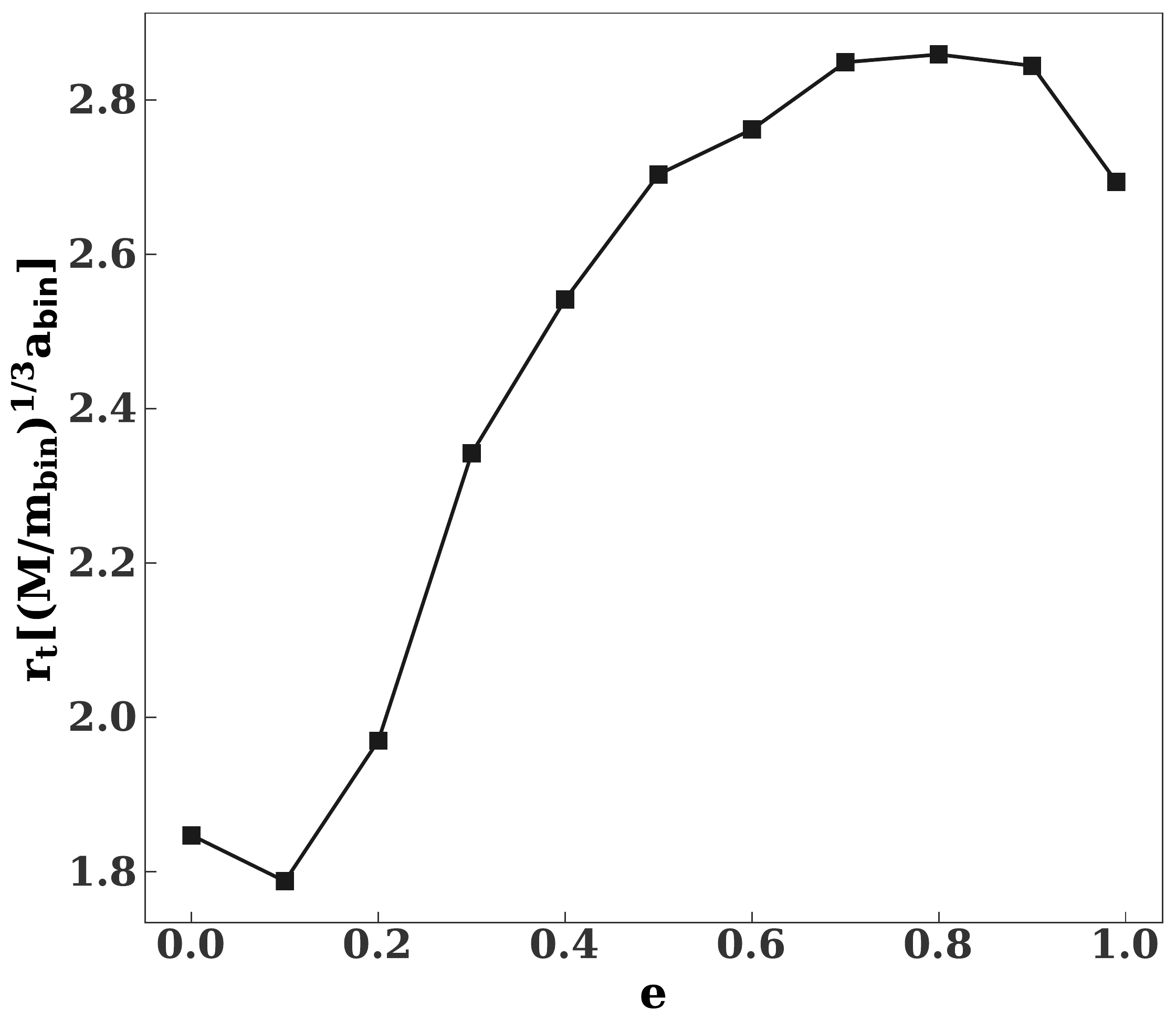}
\caption{\label{fig:dis_prob} Top panel: probability 
of binary disruption as a function of the binary pericenter and internal
eccentricity. Binary disruption is probabilistic as it depends on the phase of
the binary. This calculation assumes the binary's center-of-mass orbit is parabolic.
Bottom panel: effective tidal radius as a function of binary eccentricity (see
Equation~\ref{eq:rteff} and the surrounding discussion).}
\end{figure}

The top panel of Fig.~\ref{fig:dis_prob} shows the probability for a binary to disrupt as a
function of its pericenter and its internal eccentricity. For any combination
of these parameters, the outcome of a close encounter between a binary and an
SMBH is determined by the binary's phase. The disruption probability first
becomes nonzero between $\sim2$ and $3 r_{t,o}$, where

\begin{align}
r_{t,o}&\equiv\left(\frac{M}{m_{\rm bin}}\right)^{1/3} a_{\rm bin}\nonumber\\
&=2.8\times 10^{-4} \left(\frac{m_{\rm bin}}{20 M_{\odot}}\right)^{-1/3} \left
(\frac{a_{\rm bin}}{1 \,\, \rm au}\right) {\rm\,\, pc}.
\label{eq:rt}
\end{align}
The probabilistic nature of binary disruptions implies that there would be a
spread in the pericenters of disrupting binaries even in the empty loss cone
regime. Here we do not attempt to model this distribution but assume that all
binaries are disrupted at an effective tidal radius

\begin{equation}
r_t\equiv \chi r_{t,o}= \chi \left(\frac{M}{m_{\rm bin}}\right)^{1/3} a_{\rm bin}.
\label{eq:rteff}
\end{equation}
The prefactor $\chi$ is a function of the internal binary eccentricity. We define $\chi$ so that a binaries have a 50\% disruption probability at the effective
tidal radius. The bottom panel of Fig.~\ref{fig:dis_prob} shows $\chi$ as a
function of the binary eccentricity.

\subsection{Method} 
We solve for the positions of the stars using both the parabolic Hills
approximation \citep{sari+2010} as well as direct three-body integration with
\texttt{AR--Chain} \citep{mikkola&merritt2008}. In both cases post-Newtonian
effects are neglected (\citealt{antonini+2010} conclude that the latter would
be negligible except for the closest binaries). The binary's center of mass
orbit is approximated as parabolic unless otherwise noted.

The presented results in the main text are from \texttt{AR--Chain} integrations. 
However, we find that the Hills approximation is generally 
accurate, as discussed in Appendix~\ref{app:hills}.

\section{Results}
\label{sec:results}

\subsection{Distributions of orbital elements}
\label{sec:dist}
Taylor expanding the potential energy about the tidal radius gives the spread in orbital energy across the binary when it is disrupted, viz.
\begin{align}
\Delta E = k' \frac{G M m_{\rm bin} a_{\rm bin} q}{r_t^2 (1+q)^2},
\end{align}
where $q$, $m_{\rm bin}$, and $a_{\rm bin}$ are the mass ratio of the secondary to the primary, total mass,
and semimajor axis of the binary and $k'$ is a factor of order unity \citep{hills1988,yu&tremaine2003}. After the
disruption, one of the stars is left on a bound orbit with energy

\begin{align}
E_s=\frac{G M m_s}{2 a_s},
\end{align}
where $m_s$ is the mass of the star, and $a_s$ is its postdisruption semimajor axis. For a parabolic disruption, $E_s=\Delta E$. Therefore, 

\begin{align}
a_s&= \underbrace{\frac{\chi^2}{k'}}_{1/k} \frac{m_s}{m_{\rm bin}} \frac{(1+q)^2}{2 q} \left(M/m_{\rm bin}\right)^{2/3} a_{\rm bin}\nonumber\\ 
&=\frac{1}{k} \left(M/m_{\rm bin}\right)^{2/3} a_{\rm bin}\nonumber\\
&\approx0.016 \left(\frac{m_{\rm bin}}{20 M_\odot} \right)^{-2/3}\left(\frac{a_{\rm bin}}{1 \,{\rm au}}\right) {\rm pc},
\label{eq:as}
\end{align}
where $k\approx 0.3-1$ depends on the phase and eccentricity
of the binary. In the second line of Equation~\eqref{eq:as}, we assume the binary stars
are equal in mass.

\begin{figure}
\includegraphics[width=8.5cm]{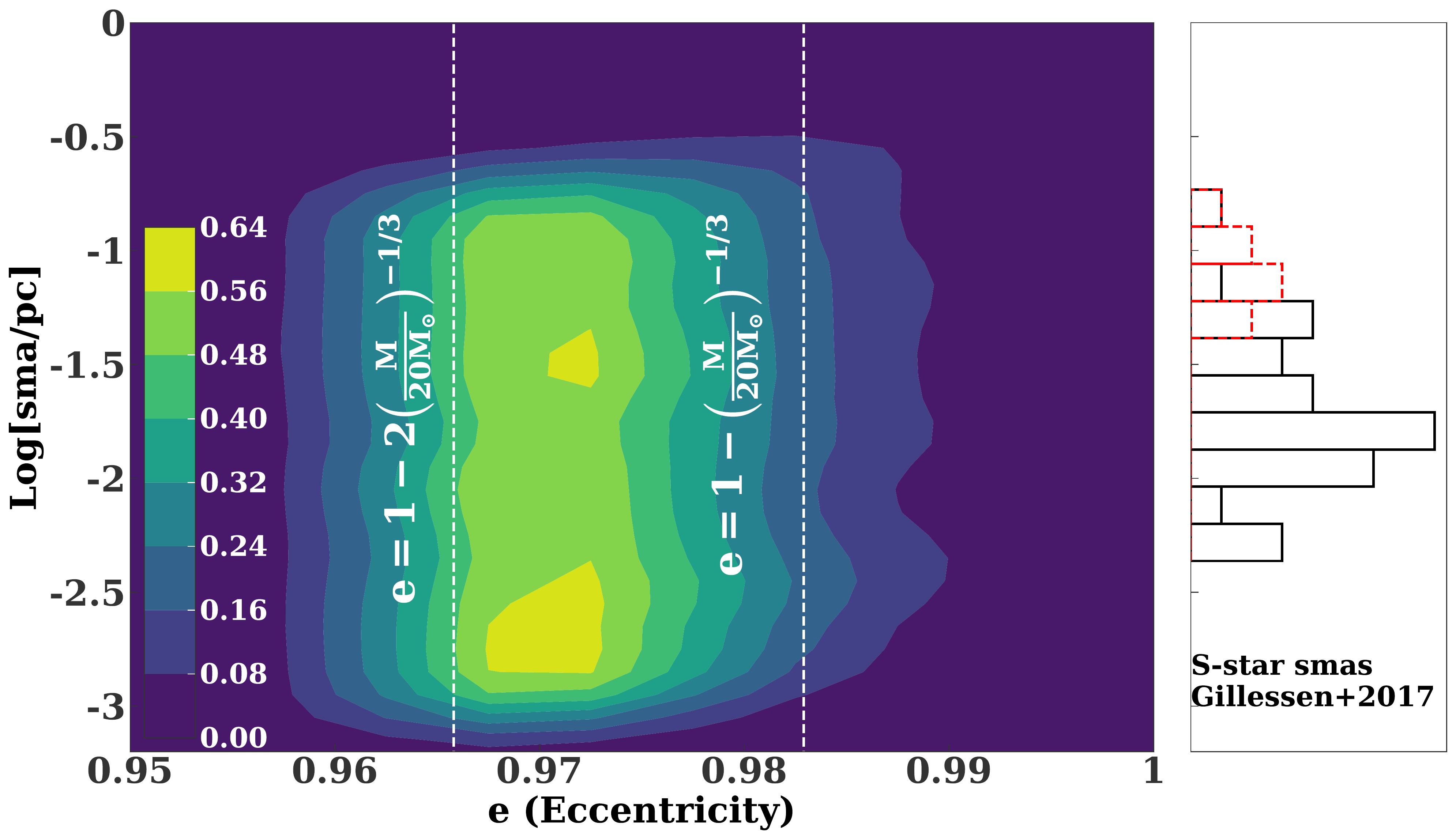}
\includegraphics[width=8.5cm]{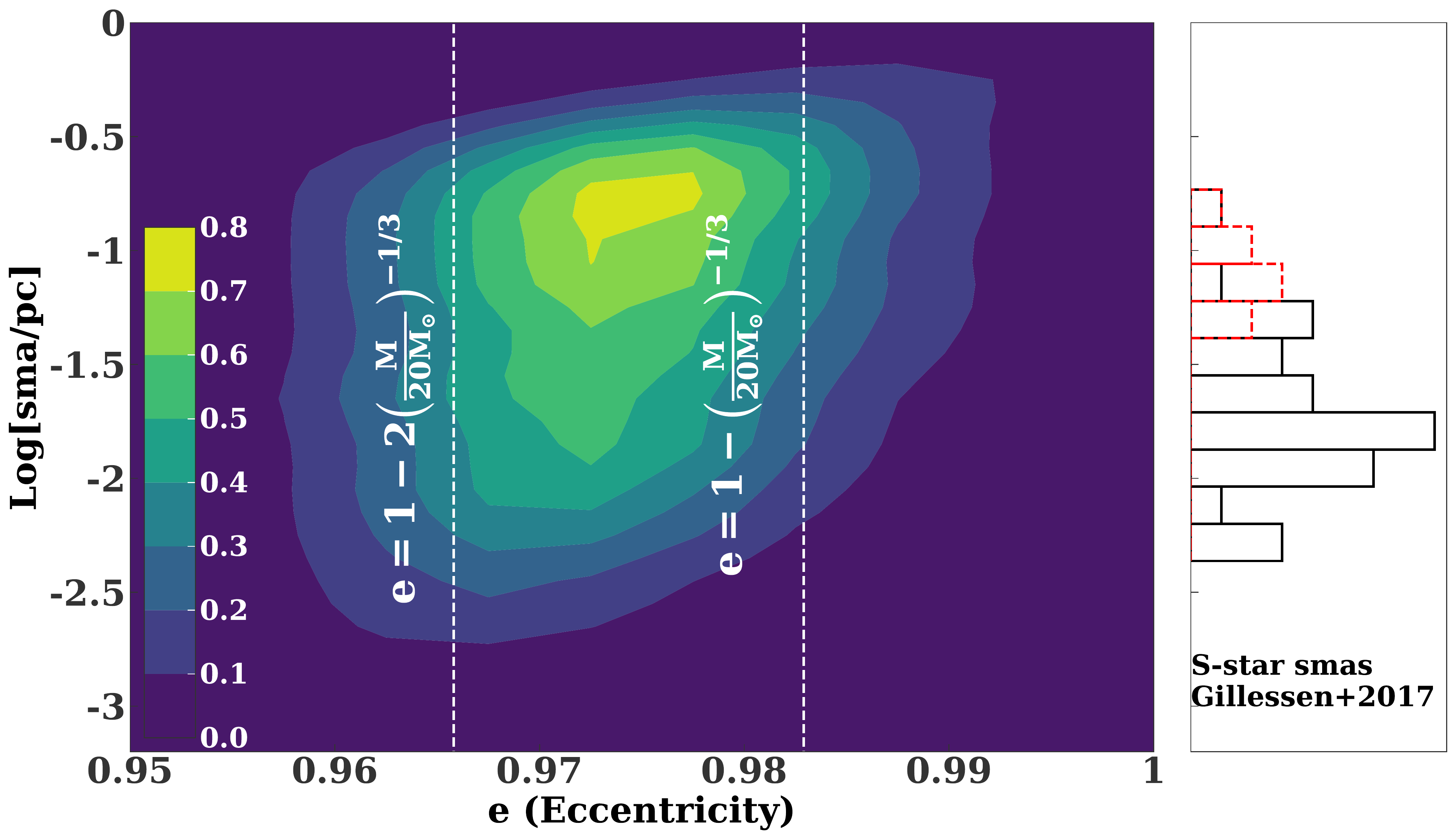}
\caption{\label{fig:dist} Postdisruption semimajor axis and eccentricity
distribution for circular binaries (top) and thermal eccentricity
binaries (bottom) from our Monte Carlo ensemble of binary--SMBH
encounters. In this case the binary's center-of-mass orbit is approximated to
be parabolic. The dashed, vertical lines show the range of eccentricities 
predicted analytically (see Equation~\ref{eq:ebound}).
The histogram on the right shows the observed distribution of
S-star semimajor axes from \citet{gillessen+2017}; the eight S-stars 
that are members of the clockwise disk are shown in red. } 
\end{figure}

The bound star will have approximately the same pericenter as the original binary orbit. Therefore, its eccentricity is 

\begin{equation}
e_s \approx 1-\chi k\left(\frac{M}{m_{\rm bin}}\right)^{-1/3},
\label{eq:ebound}
\end{equation}
where $\chi k \approx 1-2$. Fig.~\ref{fig:dist} shows the postdisruption
semimajor axis and eccentricity distributions for the bound stars in our
ensemble of simulated binary--SMBH encounters. As expected, the bound stars are on highly
eccentric orbits ($e\gsim 0.96$) with semimajor axes between $\sim 10^{-3}$
and $1$ pc. For thermal eccentricity binaries, the semimajor axis distribution of remnant stars inside the inner edge of the
disk ($\sim$ 0.05 pc) is statistically consistent with the semimajor axis
distribution of S-stars in this region.\footnote{The KS (Anderson-Darling)
  test probability is 0.21 ($>$0.25).}The two distributions are not
statistically consistent in the case of circular binaries. However, this can
be fixed by taking a flatter binary semimajor axis distribution (e.g.
$\frac{dN}{da_{\rm bin}} \propto a_{\rm bin}^{-0.5}$).

\begin{figure}
\includegraphics[width=8.5cm]{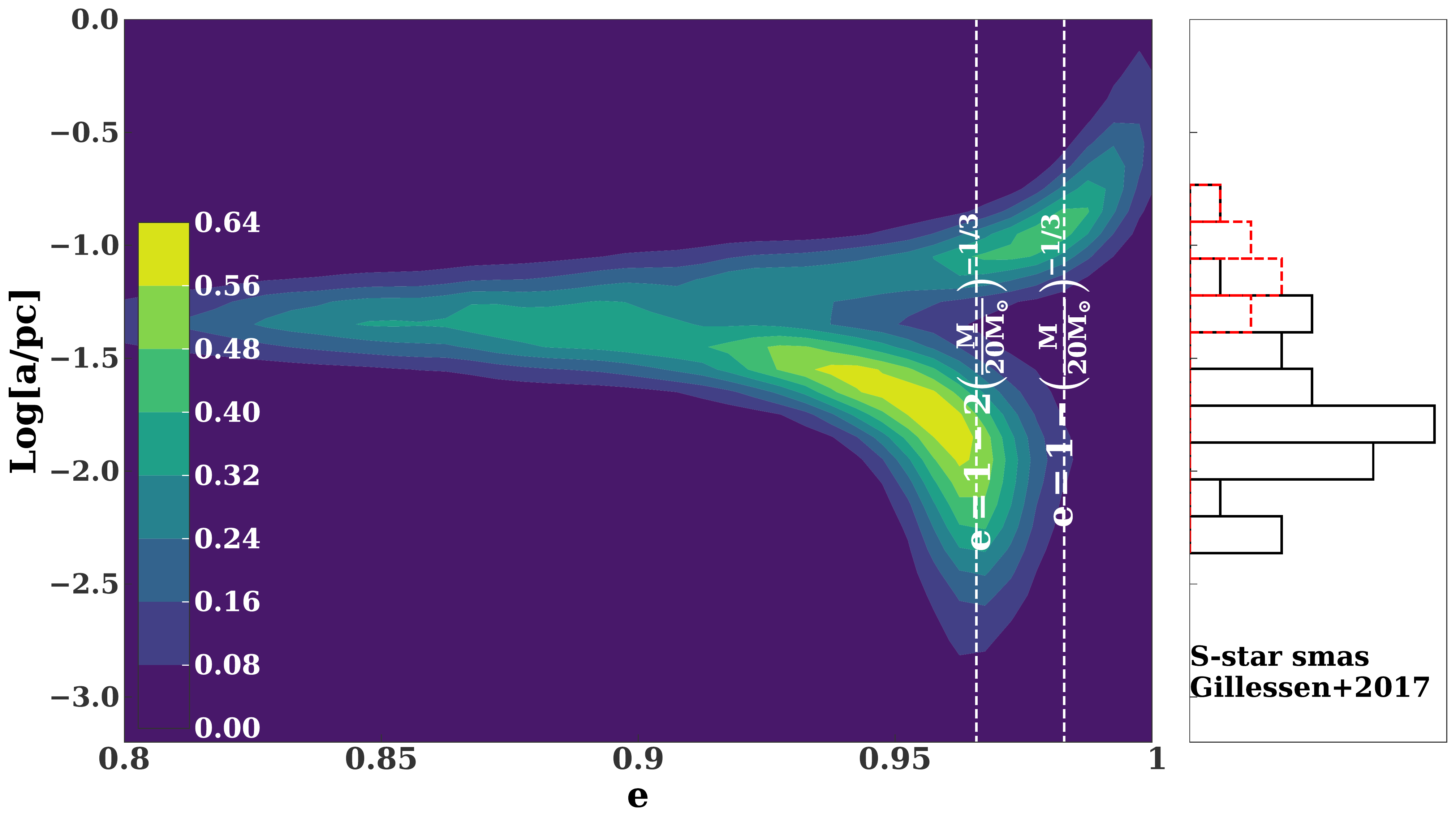}
\caption{\label{fig:paraDist} Same as the bottom panel of Fig.~\ref{fig:dist}, 
except the binary's center-of-mass is no longer approximated as parabolic. Instead,
each binary's center-of-mass orbit has an eccentricity $e=1-\frac{r_t}{a_d}$, where $r_t$ is the effective tidal radius (Equation~\ref{eq:rteff}) and $a_d=0.05$ pc is the inner edge of the clockwise disk. Note that in this case both stars can remain bound to the SMBH after a binary disruption.}
\end{figure}

The distributions in Fig.~\ref{fig:dist} extend significantly outside of the inner edge of 
the clockwise disk. This is an artifact. The distributions in Fig.~\ref{fig:dist}
are constructed by simulating close encounters between binaries on parabolic orbits and an SMBH. Parabolic orbits 
are an approximation, which breaks down when the energy of the center-of-mass orbit is greater than the 
spread in energy across the binary. In Fig.~\ref{fig:dist}, the energy of the bound star 
is equal to the spread in energy across the binary. Evidently, the parabolic approximation is not self-consistent when the bound star has a semimajor axis greater than 0.05 pc (the semimajor axis of the binary center of mass in our simulations). When a realistic eccentricity for the binary center of mass is included, the semimajor axis distribution of the bound stars becomes less extended (as shown in Fig.~\ref{fig:paraDist}). There is still a population of stars at larger semimajor axes. These are stars that get a positive energy kick during the encounters with the SMBH, but remain bound. Their companion gets a negative energy kick, ending up at smaller semimajor axes. This results in a bifurcation in the semimajor axis distribution.

As noted in \citet{antonini&merritt2013}, the Hills mechanism deposits stars on
highly eccentric orbits and cannot reproduce the observed thermal
eccentricity distribution of the S-stars. \citet{subr&haas2016} claimed that binary
disruption can result in a thermal eccentricity distribution if the binary
approaches the tidal radius gradually. This would only be the
case if the binary's center-of-mass orbit is not nearly parabolic.

Fig.~\ref{fig:paraDist} shows the postdisruption semimajor axis and
eccentricity distribution of bound stars for binaries that approach the SMBH with realistic eccentricities. 
Interestingly, there is a tail of lower-eccentricity stars at larger semimajor axes. Nonetheless,
inside of a few$\times 10^{-2}$ pc, the eccentricity distribution is similar to
what we found with the parabolic approximation.

 We conclude that an additional relaxation mechanism is required to
reproduce the observed eccentricity distribution of the S-stars. In
\S~\ref{sec:relax} we investigate three different possibilities.

\subsection{Reproducing the number of S-stars}
\label{sec:ns}
We now check whether the eccentric disk instability could reproduce 
the observed number of S-stars. Following the logic of the Drake equation \citep{drake1965},
we expect

\begin{equation}
N_s=   N_d f_d (1-f_c) 
\end{equation}
S-stars inside the inner edge of the clockwise disk. Here $N_d$ is the number
of binaries in the disk, and $f_d$ is the fraction of disk stars that
have their pericenter pushed below the binary tidal radius. Based 
on our eccentric disk simulations $f_d\lesssim 0.2$.
$f_c$ is the fraction of binaries that would collide instead of disrupting. 
Based on our Monte Carlo simulations of binary--SMBH 
encounters, $f_c \approx 0.2$.

We must also account for completeness effects. Dim, low-mass S-stars 
around $\sim 1 M_{\odot}$ are not currently observable. Based on the observed spectra and K-band luminosities \citep{habibi+2017, cai+2018}, the S-stars have masses between $\sim$3 and 15 $M_{\odot}$. We expect $f_{\rm mf} N_s$ S-stars within this mass range. For the $m^{-1.7}$ disk mask function \citep{lu+2013}, $f_{\rm mf}\approx 0.33$.\footnote{Assuming that this mass function extends between $1$ and $60 M_{\odot}$.}

The total mass of the clockwise disk is inferred to be between $\sim 1.4\times
10^4 M_{\odot}$ and $3.7\times 10^4 M_{\odot}$ \citep{lu+2013}. This would
imply between 2300 and 6200 stars in the disk (or between $1100$ and $3100$
binaries for a binary fraction of unity). For $N_d=1000$,
the number of S-stars between 3 and 15 $M_{\odot}$  would be
$\lsim 50$. Thus, the eccentric disk instability can reproduce the observed number of 
S-stars.

\section{Reproducing the observed eccentricity distribution of the S-stars}
We have argued that if the S-stars are sourced by the Hills mechanism, a
relaxation process is needed to reproduce their present-day eccentricity
distribution. The timescale over which this relaxation occurs is tightly
constrained by (i) the observed ages of the S-stars ($\sim 10$ Myr; \citealt{habibi+2017}), (ii) the age of the clockwise disk ($2.5-5.8$ Myr; 
\citealt{lu+2013}), and (iii) the flight time\footnote{Spectroscopy indicates that the age of the hypervelocity star in
	\citet{koposov+2019} is $10^{7.72_{-0.33}^{+0.25}}$ years. Thus, although this
	star was ejected from the Galactic center $\sim$5 Myr ago (approximately
	when the clockwise disk was forming), it originated in an earlier epoch.
	This may indicate that this star and its companion were entrained into the
	clockwise disk after they formed.} of the \citet{koposov+2019}
hypervelocity star from the Galactic center (4.8 Myr). 

In this section, we consider three possible relaxation mechanisms: (a)
scalar resonant relaxation (b) secular torques from the clockwise disk and (c) an IMBH. 
We find that the first and third mechanisms can reproduce the observed eccentricity
distribution of the S-stars within $10^7$ years, as required by observations.

\label{sec:relax}
\subsection{Resonant relaxation}
Resonant relaxation occurs in potentials with a high degree of symmetry, like a
Keplerian potential. In a nearly Keplerian potential, orbits will be fixed
over many periods. This leads to a coherent buildup of torques on each orbit
\citep{rauch&tremaine1996}. These torques can change either the direction or
the magnitude of an orbit's angular momentum. The former is vector resonant
relaxation (VRR), and the latter is scalar resonant relaxation (SRR). We focus
on SRR, as VRR cannot change an orbit's eccentricity. However, VRR likely
plays an important role in the dynamics of the S-star cluster, as it can
change the orientations of the S-stars' orbits within a few million years
\citep{kocsis&tremaine2011}.

The SRR time depends on the precession rate of the stellar orbits. After
orbits start to precess, the torques no longer add coherently and the angular
momentum evolution becomes a random walk. Orbital precession can be caused by
general relativity or by the distributed mass in the star cluster. When orbital precession is dominated by the distributed
mass, the SRR time is
\begin{align}
&t_{\rm rr}\propto \frac{M}{\widetilde{m}} P\nonumber\\
&\widetilde{m}\equiv \frac{\overline{m^2}}{\overline{m}}
\end{align}
where $M$ is the mass of the SMBH, $P$ is the orbital period, and $m$ is the
stellar mass. Horizontal bars denote averages
\citep{rauch&tremaine1996,alexander2017}. Adding massive perturbers, like
black holes, increases $\widetilde{m}$, and decreases the resonant relaxation time.
Physically, a clumpier mass distribution increases the rms
torque on stellar orbits.

There have been a number of estimates of the SRR time in the
Galactic center. Notably, \citet{antonini&merritt2013} derived the resonant
relaxation time from Monte Carlo simulations, concluding a population of
stellar mass black holes can reduce the resonant relaxation time near the
present-day location of the S-stars to $\sim 10^7$ years, roughly consistent
with their observed ages and the time since the hypervelocity star from
\citet{koposov+2019} was ejected from the Galactic center.

Here we revisit these estimates in light of new developments in the theory of
resonant relaxation. In particular, \citet{bar-or&fouvry2018} derived
diffusion coefficients for SRR in a spherical, isotropic
cluster from first principles. In their work, the stochastic perturbations of
the stellar background on a star are encapsulated as a sum of spherical
harmonic noise terms in its orbit-averaged Hamiltonian. The angular momentum
diffusion coefficients are related to the autocorrelation function of these
terms. \citet{bar-or&fouvry2018} found that SRR is strongly
quenched for rapidly precessing eccentric orbits by the adiabatic invariance
of the angular momentum, i.e. by the divergence of the relativistic precession frequency as orbits get more and more eccentric.

\begin{figure}
    \centering
    \includegraphics[width=8.5cm]{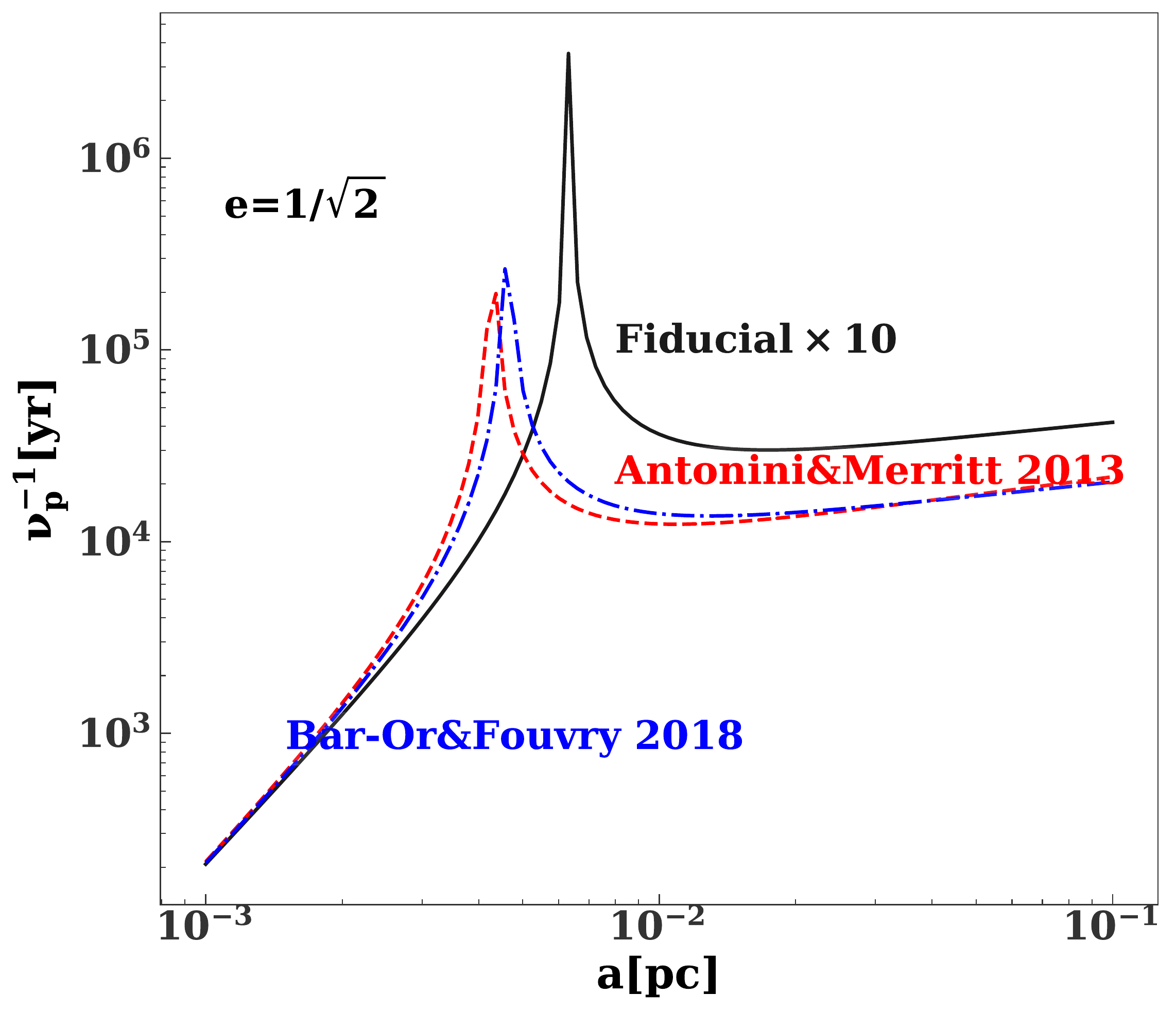}
    \caption{\label{fig:precRate} Orbital precession timescale as a function of semimajor axis for different models of the Galactic center. The orbital eccentricity is $1/\sqrt{2}$. The peak in the precession timescale is due to cancellation between prograde GR precession (which dominates on small scales) and retrograde precession due to the distributed mass in the Galactic center (which dominates on large scales).}
\end{figure}

\begin{figure*}
\gridline{
          \fig{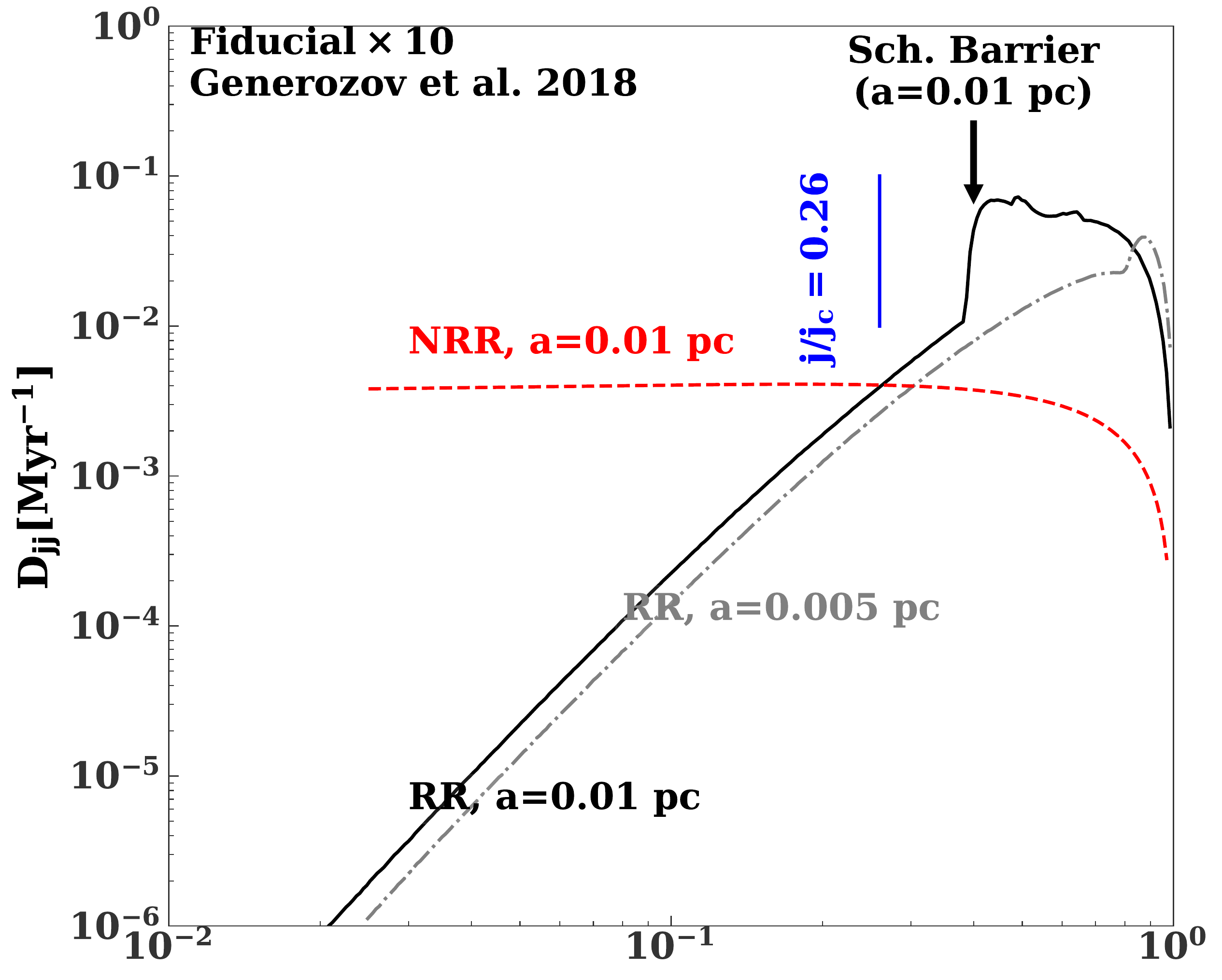}{0.4\textwidth}{}
          \fig{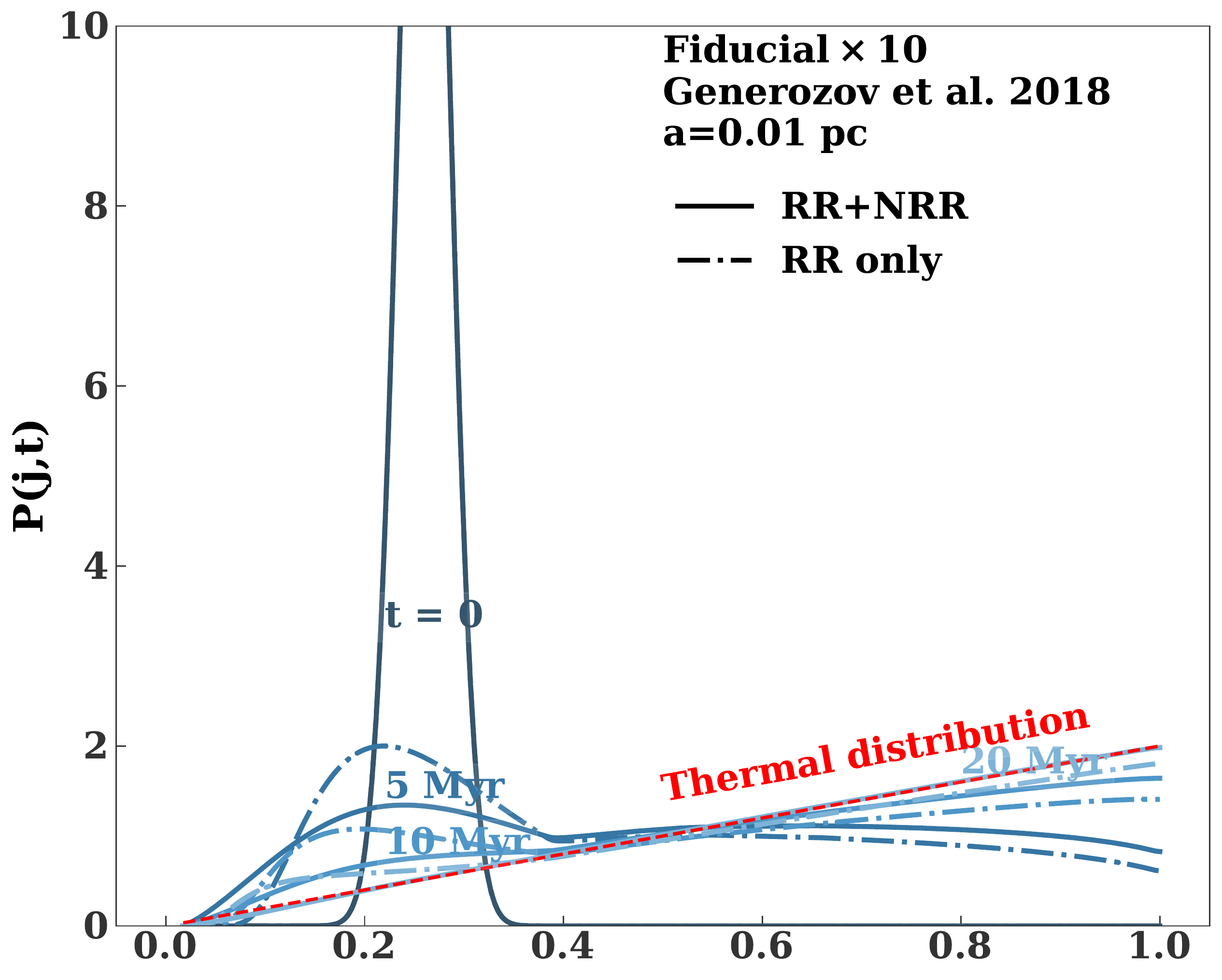}{0.4\textwidth}{}
         }
\gridline{
          \fig{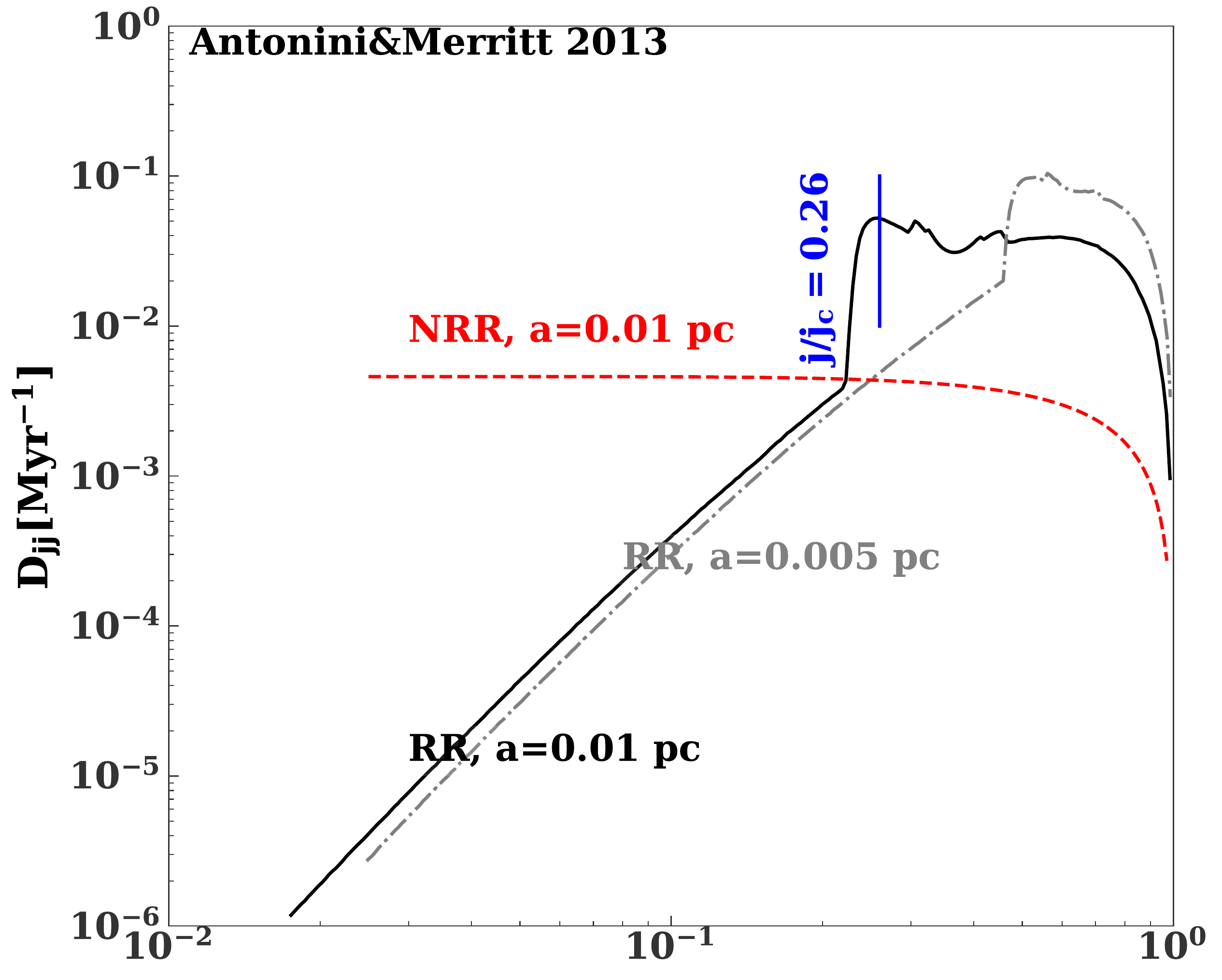}{0.4\textwidth}{}
          \fig{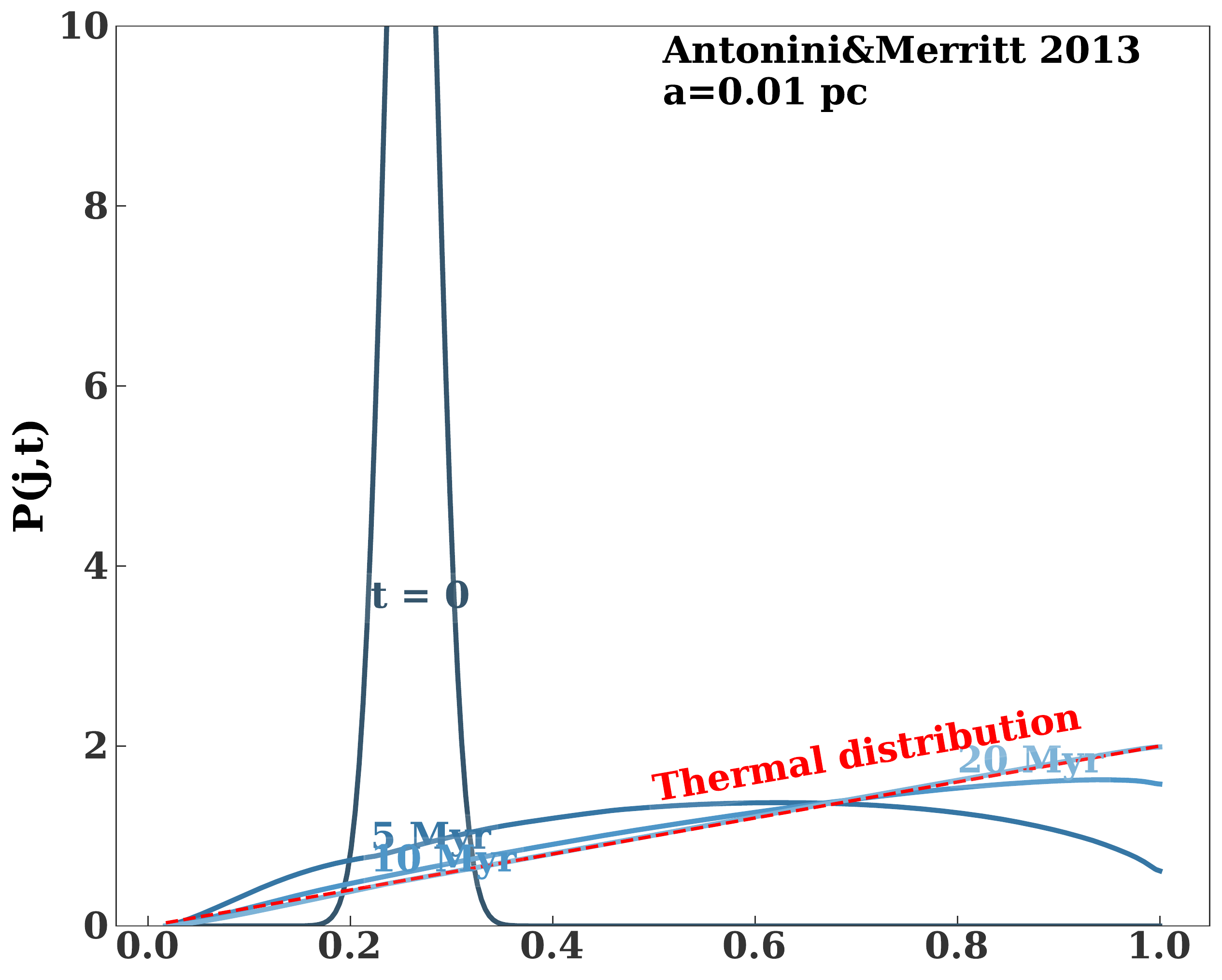}{0.4\textwidth}{}
         }
\gridline{
          \fig{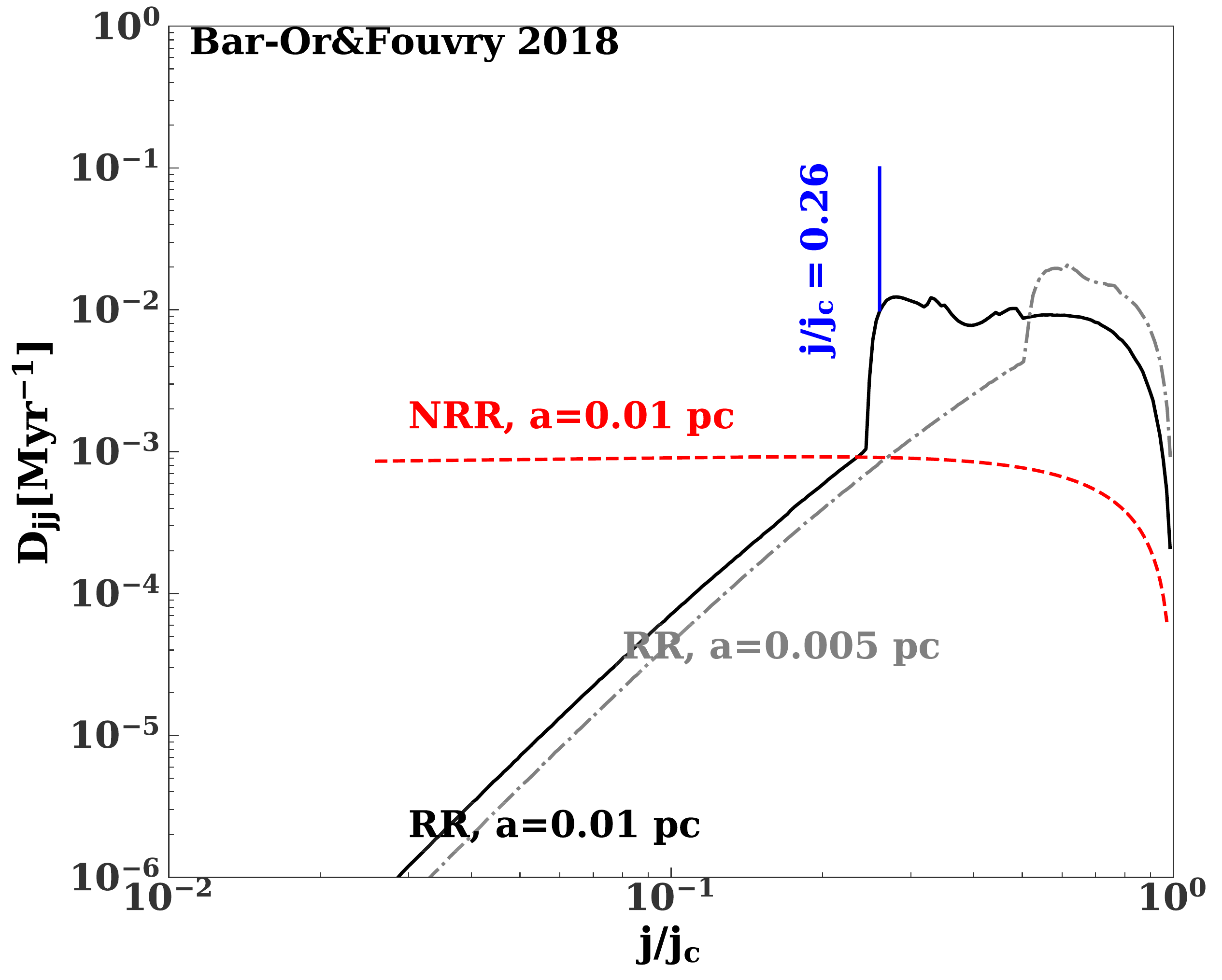}{0.4\textwidth}{}
          \fig{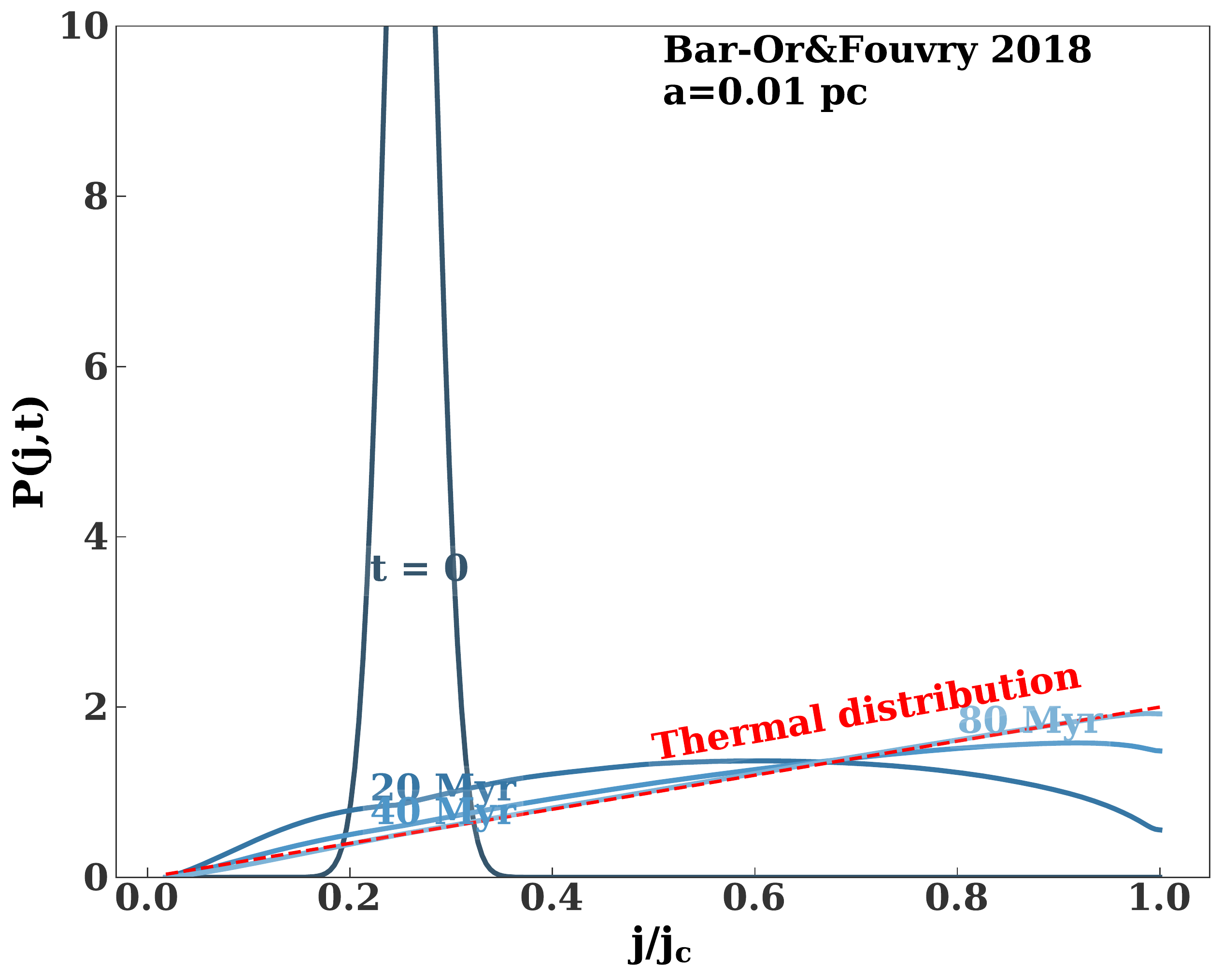}{0.4\textwidth}{}
         }
\caption{\label{fig:rr_diff} Left panels: resonant relaxation diffusion coefficient at $a=0.01$ pc (solid black line) and 0.005 pc (dash-dotted gray line) for different models of the Galactic center. The dashed red lines are the second-order nonresonant angular momentum diffusion coefficient. For small angular momentum (inside the Schwarzschild barrier), the angular momentum diffusion coefficient decreases sharply due to general relativistic precession. The vertical, blue lines show the mean angular momentum of bound stars with semimajor axes less than or equal to 0.01 pc in the binary disruption simulations in Figure~\ref{fig:paraDist}. 
Right panels: evolution of an example angular momentum distribution function due to resonant and nonresonant relaxation for different Galactic center density profiles. In each case we evolve the angular momentum distribution using Equation~\eqref{eq:trra} with the appropriate diffusion coefficients
  from the left panels. The dash--dotted lines in the top, right panel show the effects of excluding nonresonant relaxation. The initial distribution function is a Gaussian centered on $j/j_c$=0.26 (corresponding to an eccentricity of
  0.97, the mean eccentricity expected from binary disruptions). The angular momentum distribution evolves toward a thermal distribution (red dashed line). Note the different snapshot times in the bottom right panel. 
}
\end{figure*}

The isotropically averaged resonant relaxation time is

\begin{equation}
t_{\rm rr}=\frac{1}{\int_{j_{\rm lc}}^1 2 j D_{jj,  RR} (j) dj} , 
\label{eq:trra}
\end{equation}
where $j$ is the angular momentum, $D_{jj, RR}$ is the (second-order) SRR
diffusion coefficient, and the subscript ``{\rm lc}'' denotes the loss cone \citep{bar-or&fouvry2018}. The
angular momentum is normalized to that of a circular orbit with the same
energy. The angular momentum diffusion coefficient $D_{jj, RR} (j)$ can be
calculated with the publicly available software package
\texttt{SCRRPY}.\footnote{We have generalized this code to include the effects
of mass spectrum. The original (modified) code is available at 
\url{https://github.com/benbaror/scrrpy} (\url{https://github.com/alekseygenerozov/scrr_multimass}).}

In practice, the timescale for relaxation will depend on the initial angular momentum distribution, as 
the resonant relaxation diffusion coefficient is a strong function of the angular momentum. Also,
complete thermalization of the S-star orbits may not be required. There are only a few dozen S-stars, so distinguishing 
a thermal distribution from one that is somewhat superthermal is challenging.

We determine the timescale for scalar resonant relaxation to approximately reproduce the observed S-star eccentricities by solving the angular momentum diffusion 
equation, viz.

\begin{equation}
	\frac{\partial P(j, t)}{\partial t}= \frac{1}{2} \frac{\partial}{\partial j} 
	\left[ j (D_{jj, RR}+ D_{jj, NRR})\frac{\partial}{\partial j}\left[\frac{P(j,t)}{j}\right]\right],
	\label{eq:evolve}
\end{equation} 
where $P(j,t)$ is the angular momentum distribution function and $j$ is the specific angular momentum, which we generally normalize by the circular angular momentum $j_c$. $D_{jj, NRR}$ is the nonresonant diffusion coefficient (which can be calculated using Appendix C of \citealt{bar-or&alexander2016}).\footnote{This equation is exact for 
  massless test particles and for particles of any mass interacting with a thermal eccentricity bath. In other cases, there are additional friction/polarization terms.} For the initial conditions, we use the angular momentum distributions from our binary disruption simulations. In particular, the initial distribution function is a Gaussian centered at $j/j_c=0.26$ with a standard deviation of $2.4\times 10^{-2}$.
We consider a few different diffusion coefficients corresponding to different models for the density of stars and remnants in the Galactic center. First, we consider the Fiducial$\times 10$ model from \citet{generozov+2018}. For technical reasons, we use a power-law approximation\footnote{Some of the functions in \texttt{SCRRPY} assume power-law density profiles.} for the density profile of each component in this model. In this approximation, the mass in stars ($m_\star$) and black holes ($m_\bullet$) within semimajor axis $a$ is

\begin{align}
&m_\star(<a)\approx 7.9\times 10^3 M_{\odot} \left(\frac{a}{0.1 \,\rm pc}\right)^{1.5}\nonumber\\
&m_\bullet(<a)\approx 3.8\times 10^4 M_{\odot} \left(\frac{a}{0.1 \,\rm pc}\right)^{1.2}.
\label{eq:fid10}
\end{align}
These profiles are accurate inside of $\sim$0.1 pc where the S-stars
are located, but overestimate the number of black holes at larger scales. However, 
this would not affect the results, as the torque on a particular star is dominated by orbits within a factor of two of its semimajor axis \citep{gurkan&hopman2007}.

We also consider the stellar density profiles from
\citet{antonini&merritt2013} to more directly compare to this work. They have $1 M_{\odot}$ stars and
$10 M_{\odot}$ black holes with the following profiles\footnote{Note that \citet{antonini&merritt2013} consider black holes and stars separately. Also, the profiles in \citet{antonini&merritt2013} are functions of radius. For these isotropic profiles, the mass enclosed within a given semimajor axis differs by $\sim 10\%$ from the mass enclosed within an equal radius (we neglect this correction here).}:
\begin{align}
&m_\star(<a)\approx 6.7\times 10^4 M_{\odot} \left(\frac{a}{0.1 \,\rm pc}\right)^{1.25}\nonumber\\
&m_\bullet(<a)\approx 2.4\times 10^4 M_{\odot} \left(\frac{a}{0.1 \,\rm pc}\right).
\label{eq:am13}
\end{align}
Finally, we consider a simple one-component model with only $1 M_{\odot}$ stars from \citet{bar-or&fouvry2018}
\begin{align}
    m_\star(<a)\approx 9.4\times 10^4  M_{\odot} \left(\frac{a}{0.1 \,\rm pc}\right)^{1.25}.
    \label{eq:bf18}
\end{align}
The left panels of Fig.~\ref{fig:rr_diff} show the second-order diffusion coefficient in these models for a couple of different semimajor axes. In all cases, the angular momentum diffusion coefficient falls off inside of some critical angular momentum $j_{\rm crit}$ (this is the ``Schwarzschild barrier'' from \citealt{merritt+2011}). Importantly, the location of the Schwarzschild barrier is a function of semimajor axis and the Galactic center density profile, so that binary disruption can deposit stars either inside or outside of it, with a drastic effect on the efficiency of angular momentum diffusion. The variation in $j_{\rm crit}$ is due to differences in the orbital precession rates. As shown by \citet{bar-or&fouvry2018}, $j_{\rm crit}$ is where the precession rate of orbits due to GR is equal to the coherence time for the system (that is a representative timescale over which orbits are fixed). Orbits that precess faster than the coherence time will experience no secular torque. The Schwarzschild barrier, $j_{\rm crit}$, can be estimated from the following equation

\begin{align}
    &\nu_{\rm gr}(a, j_{\rm crit}) = \frac{0.45}{T_c(a)}\nonumber\\
    & T_c\equiv \sqrt{\frac{\pi}{2}} \nu_p^{-1} (2 a),
\end{align}
where $a$ is the semimajor axis, $\nu_{\rm gr} (a, j_{\rm crit})$ is the orbital precession frequency due to GR, and $T_c$ is the coherence time for resonant relaxation. For a spherical, isotropic population the coherence time is the total orbital precession frequency $\nu_p$ evaluated at a semimajor axis of $2 a$, and an eccentricity of $1/\sqrt{2}$. This precession rate depends on the distributed mass profile within the Galactic center. Figure~\ref{fig:precRate} shows the precession rate as a function of semimajor axis for the models considered here.

Inside of the Schwarzschild barrier, nonresonant relaxation can be more efficient than resonant relaxation, as shown by the dashed red lines in the left panels of Fig.~\ref{fig:rr_diff}.

\begin{figure}
\includegraphics[width=8.5cm]{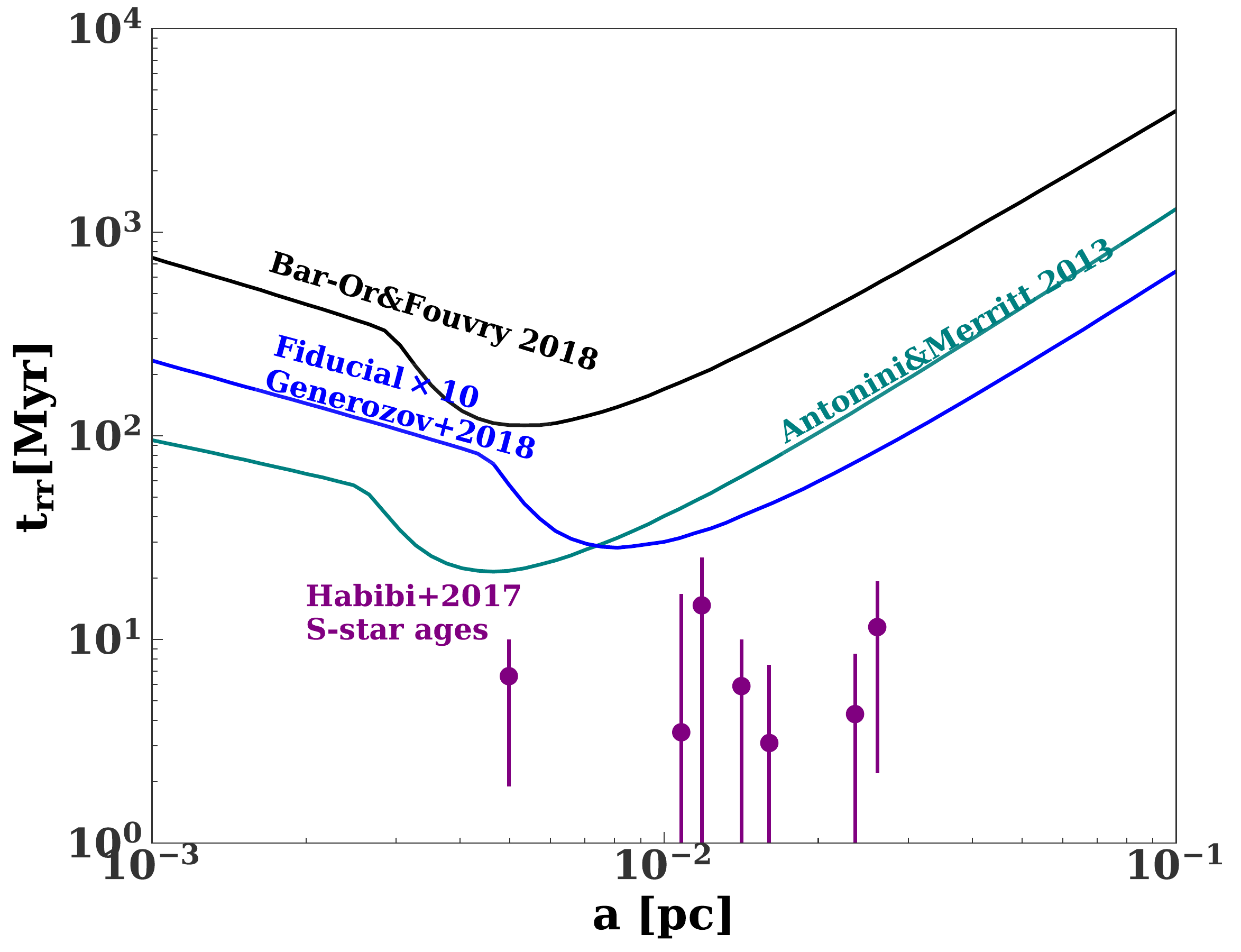}
\caption{\label{fig:trr}Isotropically averaged scalar resonant relaxation time (Equation~\ref{eq:trra}) as a function of semimajor axis. Different colors correspond to different models for the stellar density profile in the Galactic center (see the text for details). The purple points are the measured ages of eight of the S-stars from \citet{habibi+2017}.}\end{figure}

The right panels of Fig.~\ref{fig:rr_diff} show the evolution of the ``Hills injected distribution'' at a semimajor axis of 0.01 pc for different Galactic center models (equations~\ref{eq:fid10},~\ref{eq:am13}, and~\ref{eq:bf18}). The Fiducial$\times$10 and \citet{antonini&merritt2013} models have similar isotropically average resonant relaxation times at 0.01 pc (see Fig.~\ref{fig:trr}). However, the evolution is faster for the \citet{antonini&merritt2013} model, as binary disruptions deposit stars closer to the Schwarzschild barrier.

\begin{figure}
    \includegraphics[width=8.5cm]{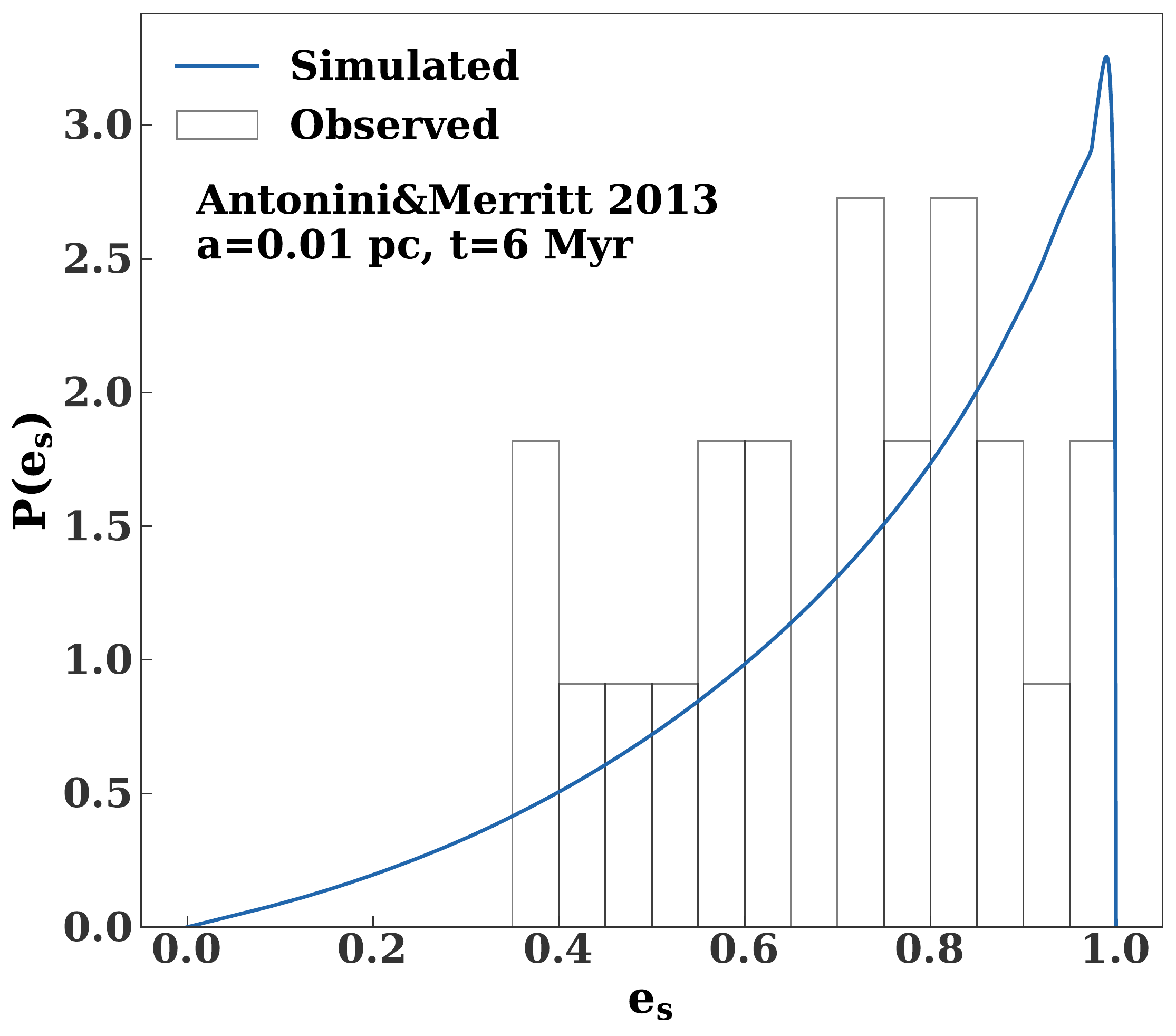}
    \caption{\label{fig:rr_snapshot} Eccentricity distribution at 0.01 pc in the \citet{antonini&merritt2013} model after 6 Myr of evolution (\emph{blue line}). This is the earliest snapshot for this model that is consistent with the observed eccentricity distribution of the S-stars (\emph{black, empty, histogram}).}
\end{figure}

\begin{deluxetable*}{lccccc}
    \tabletypesize{\small}
    \tablecaption{\label{tab:tss} Approximate Timescale Required to Reproduce the Observed S-star Eccentricities
      for Different Models of the Galactic Center Compared to the Isotropically Averaged Resonant Relaxation Time (Equation~\ref{eq:trra}).} 
    \tablecolumns{6}
    \tablehead{ & \multicolumn2c{a=0.005 pc} &  \colhead{\phantom{string} } &    \multicolumn2c{a=0.01 pc} \\
    \cline{2-3} \cline{5-6}
     & \colhead{Timescale to reproduce S-stars}   & $t_{\rm rr}$   & \colhead{\phantom{string}} &  \colhead{Timescale to reproduce S-stars}  & \colhead{$t_{\rm rr}$}\\
     \multicolumn1l{Model}  & (Myr)  & (Myr) & &  (Myr)  & (Myr)    }    
    \startdata
      \\
    Fiducial$\times$10      & 13  & 57  &  & 8 & 30  \\
    Antonini\&Merritt 2013  & 7 & 22    &  &  6 & 40  \\
    Bar-Or\&Fouvry 2018     & 41  & 110 &  & 24 & 170 \\
    \enddata
\end{deluxetable*}

We compare the evolving distribution functions in Fig.~\ref{fig:rr_diff} to the eccentricities of the 22 S-stars with semimajor axes less than or equal to 0.03 pc from Table 3 of \citet{gillessen+2017}. Specifically, we identify the earliest time where (i) the Kolmogorov--Smirnov (KS) test probability that the observed S-star eccentricities are consistent with the simulated distribution is at least 10\% and (ii) there is at least a five percent probability that there are two or fewer stars with eccentricities greater than equal to 0.95 as in the observed S-stars. Although simulated distributions correspond to one fixed semimajor axis, we always compare to the entire S-star sample within 0.03 pc. Table~\ref{tab:tss} shows the timescales required to satisfy these criteria for different Galactic center density profiles. Fig.~\ref{fig:rr_snapshot} shows a comparison between the observed S-star eccentricity distribution and a snapshot from the solutions in Fig.~\ref{fig:rr_diff}. This snapshot is the earliest one in the \citet{antonini&merritt2013} model where the above statistical criteria are satisfied. Overall, resonant relaxation can reproduce the observed S-star eccentricity distribution within $\sim10$ Myr assuming a realistic population of stellar mass black holes.

\subsection{Disk torques}
In this section, we quantify secular torques from the clockwise disk to see if
they could thermalize the S-star eccentricity distribution. First, we quantify the expected torque analytically. The torque on a particle well inside of the
disk may be estimated via spherical harmonic expansion, as shown below.

First, we replace the disk with a single particle of mass
$M_d$ on an elliptical orbit around the central SMBH. The spherical harmonic
expansion of this orbit's gravitational potential at a point $\mathbf{r}$ with
spherical polar coordinates $<r,
\theta, \phi>$ is

\begin{equation}
U_{\rm \ell m}=-G M_d\sum_{\ell=0}^{\infty} \sum_{m=-\ell}^{\ell} W_{\ell m} r^\ell Y_{\ell}^m(\theta, \phi) \frac{\exp[-i m \Phi] }{D^{\ell+1}},
\end{equation}
where $\Phi$ is the true anomaly of the particle, $D$ is its distance from
the center, and $Y_{\ell}^m$ is the spherical harmonic of degree $\ell$, $m$
(see, e.g., equation 3 of \citealt{fuller&lai2011}). Here $W_{\ell m}$ is a constant and
is nonzero only if $\ell+m$ is even. The physical potential is the real part of the 
above expression.

The orbit-averaged $\ell, m$ component of the potential is
\begin{equation}
\left<U_{\ell m}\right>= \frac{\int_{0}^{2 \pi} U_{\ell m}(D, \Phi) \frac{d \Phi}{\dot{\Phi}}}{P_{\rm orb}},
\end{equation}
where $P_{\rm orb}$ is the orbital period and 

\begin{align}
&D=\frac{a_d (1-e_d^2)}{1+e_d \cos (\Phi)}\\
&\dot{\Phi} = \frac{j_d}{D^2},
\end{align}
where $a_d$, $e_d$, $j_d$ are the semimajor axis, eccentricity, and the specific angular momentum of the disk orbit respectively ($j_d = {\rm Constant} \times \sqrt{a_d (1-e_d^2)}$ ).

The $\ell=0$ term does not depend on $\mathbf{r}$, and therefore does not contribute any force. The orbit--average of the $\ell=1$ and the $\ell=2, m=\pm 2$ terms is 0.

The orbit average of the $\ell=2, m=0$ component is nonzero, but the corresponding force is
radial in the plane of the orbit and would exert no torque. The leading-order
contribution to the torque comes from the $\ell,m=3,\pm1$ terms of the potential expansion.
In the plane of the orbit (where $\theta=\pi/2$), the $\phi$ component of the specific force from these terms is

\begin{align}
<f_{3,1}>= &\frac{3}{8} G M_d \sin(\phi) \frac{e_d r^2}{a_d^4 (1-e_d^2)^{5/2}} 
\end{align}
The corresponding specific torque is 
\begin{align} <\tau_{3,1}>=&\frac{3}{8} \underbrace{\frac{G M_d}{a_d}}_{\tau_d}
\sin(\phi) \frac{e_d}{(1-e_d^2)^{5/2}} \frac{r^3}{a_d^3}
\end{align}
Thus, the torques fall off as $r^3$ within the inner edge of the disk. The torque on a test orbit at an angle $\overline{\omega}$ with respect to the disk is

\begin{equation}
 <<\tau_{3,1}>> =f(e') \tau_d \frac{e_d}{(1-e_d^2)^{5/2}} \left(\frac{a'}{a_d}\right)^3 \sin (\overline{\omega}),
\end{equation} 
where the $a'$ and $e'$ are the semimajor axis and eccentricity of the test
orbit. The double brackets on the left side of the above
equation denote averages over the disk and test orbits.
If the test star was injected via the Hills mechanism, $e' \sim 0.97$ and $f(e')=1.6$.

For the S-stars $a'/a_d \approx 0.2$, and $<<\tau_{3,1}>> \approx
4.7\times 10^{-2} \tau_d$ for $e_d=0.7$. Here $\tau_d$ is the characteristic
torque at the inner edge of the disk. If the S-star and disk orbits remained
static, even this attenuated torque could perturb the S-stars' eccentricities
significantly over their lifetime. For a star with a semimajor axis of 0.01
pc and an eccentricity of 0.97, the characteristic time--scale to perturb its
angular momentum is $j/<<\tau_{3,1}>> \approx 8\times 10^4$ years.

\begin{figure}
\includegraphics[width=8.5cm]{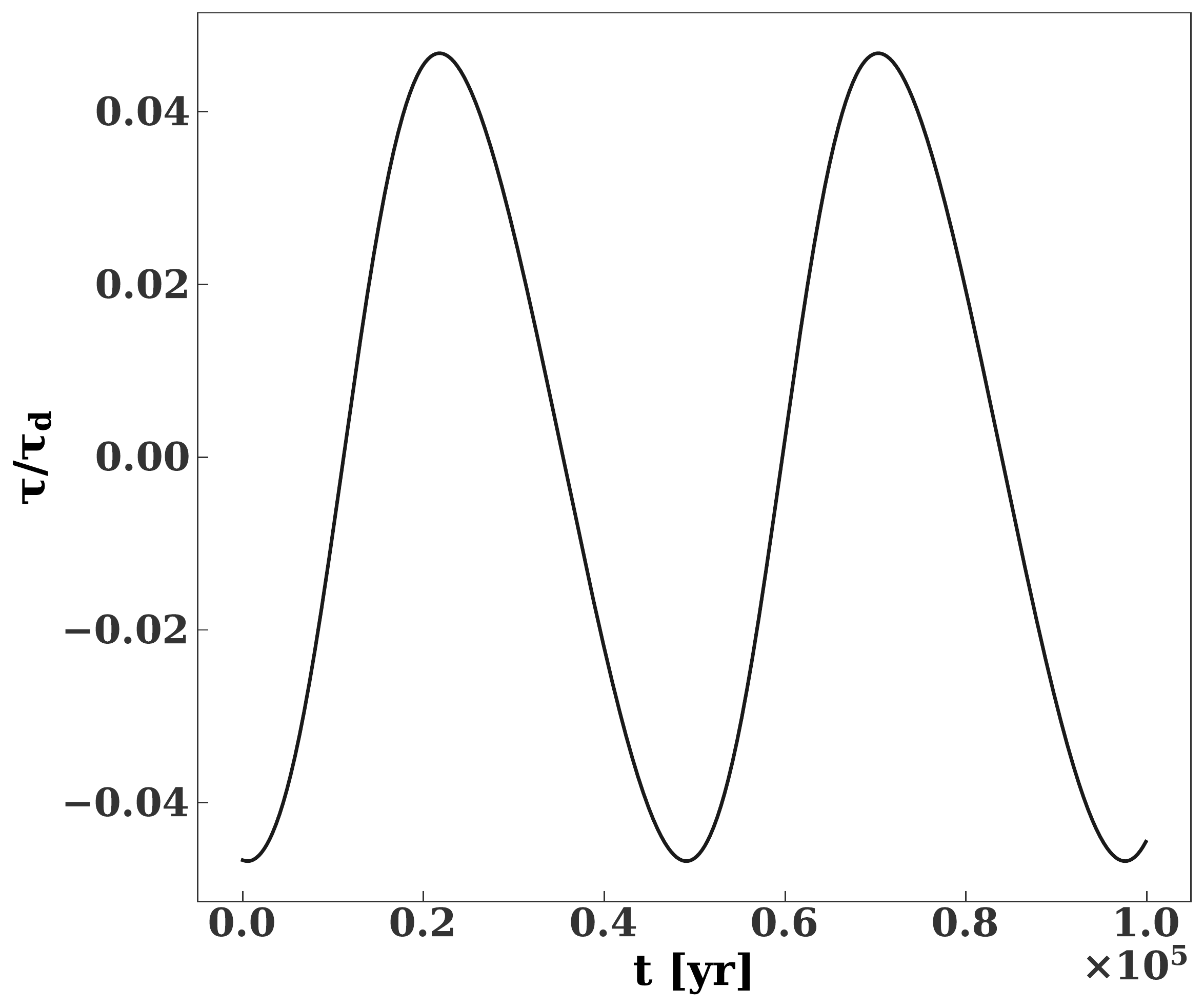}
\includegraphics[width=8.5cm]{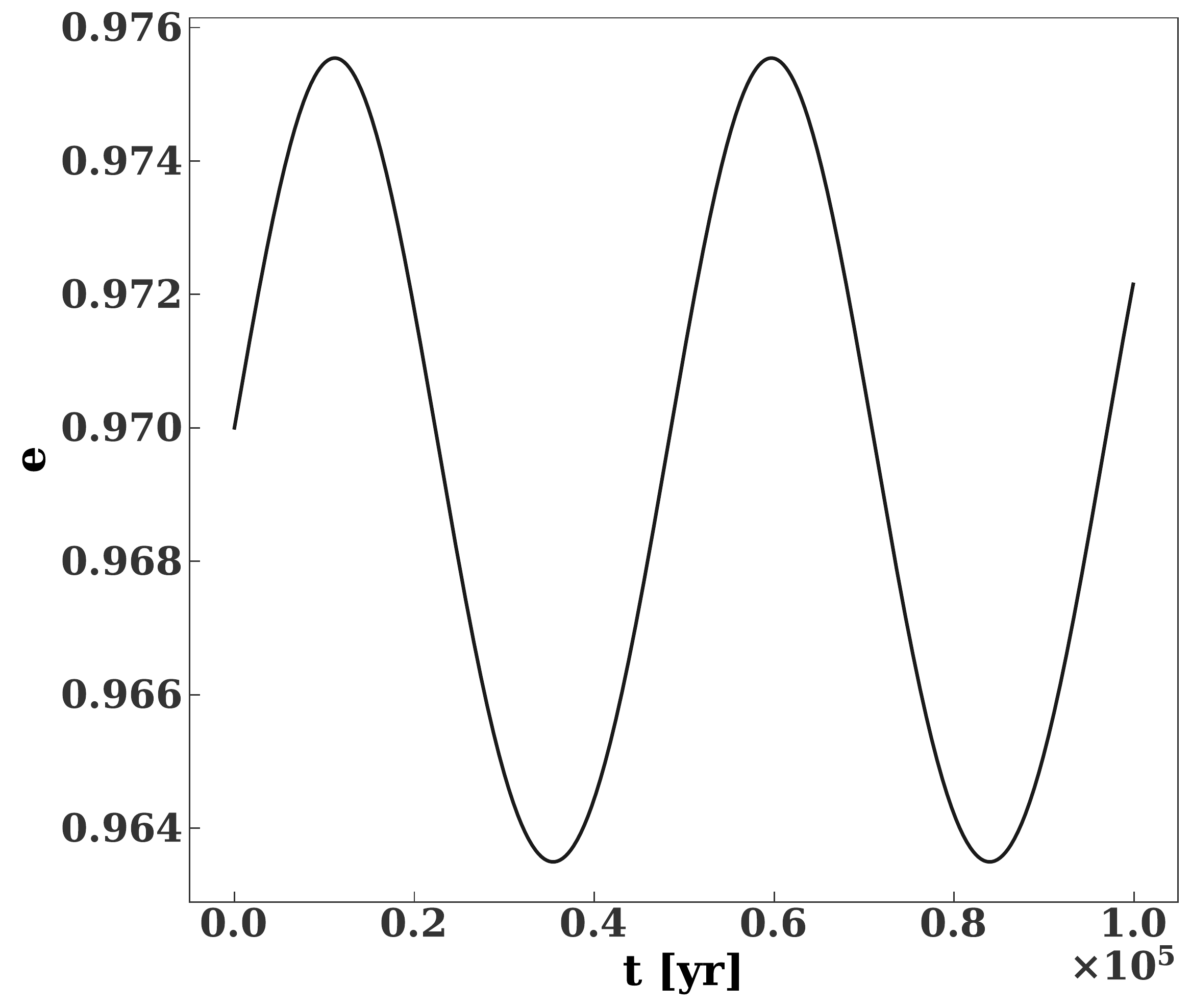}
\caption{\label{fig:torque} Top panel: torque from an idealized eccentric disk on a test orbit with a
  semimajor axis of 0.01 pc and an initial eccentricity 0.97 as a function of
time (see text for details). The torque is normalized by $\tau_d=\frac{G
m_{\rm d}}{a_{\rm d}}$, where $a_{\rm d}$ and $m_{\rm d}$ are the
characteristic semimajor axis and the mass of the disk, respectively. The
semimajor axis of this test orbit is five times smaller than the inner edge
of the disk. The test orbit experiences general relativistic precession, so
that the torque rapidly changes sign, and the eccentricity oscillates as shown
in the bottom panel.}
\end{figure}

However, the torques from the disk would be further attenuated by precession
of the disk and S-star orbits. The top panel of Fig.~\ref{fig:torque} shows
the specific torque on an $e'=0.97$, $a'=0.01$ pc test orbit inside of an
idealized, non--precessing eccentric disk with $e_d=0.7$, $a_d=0.05$ pc, and a
total mass of $10^4 M_{\odot}$. Here GR causes the highly eccentric test orbit to
precess $2 \pi$ radians over $4.7\times 10^4$ years. In principle, the test
orbit's eccentricity (and precession rate) are allowed to change. In practice,
the torque frequently reverses in sign, so the test orbit's eccentricity is
nearly constant, as shown in the bottom panel of Fig.~\ref{fig:torque}.

The eccentric disk will also evolve over time. We include this effect by
directly injecting remnant stars after binary disruptions in N-body
simulations of an eccentric disk (see $\S$~\ref{sec:instability} for a
description of the basic simulation setup). Specifically, whenever a particle
crosses the tidal radius\footnote{For all particles in
the simulation, $3\times 10^{-4}$ pc (see Equation~\ref{eq:rt}).} in a simulation it is replaced
with a child particle with a semimajor axis of $a'=0.01$ pc, and an
eccentricity of $e'=1-r_p/a'$. All of the other orbital elements are inherited
from the parent. For computational convenience, the child particles are taken
to be massless.

\begin{figure}
\centering
\includegraphics[width=8.5cm]{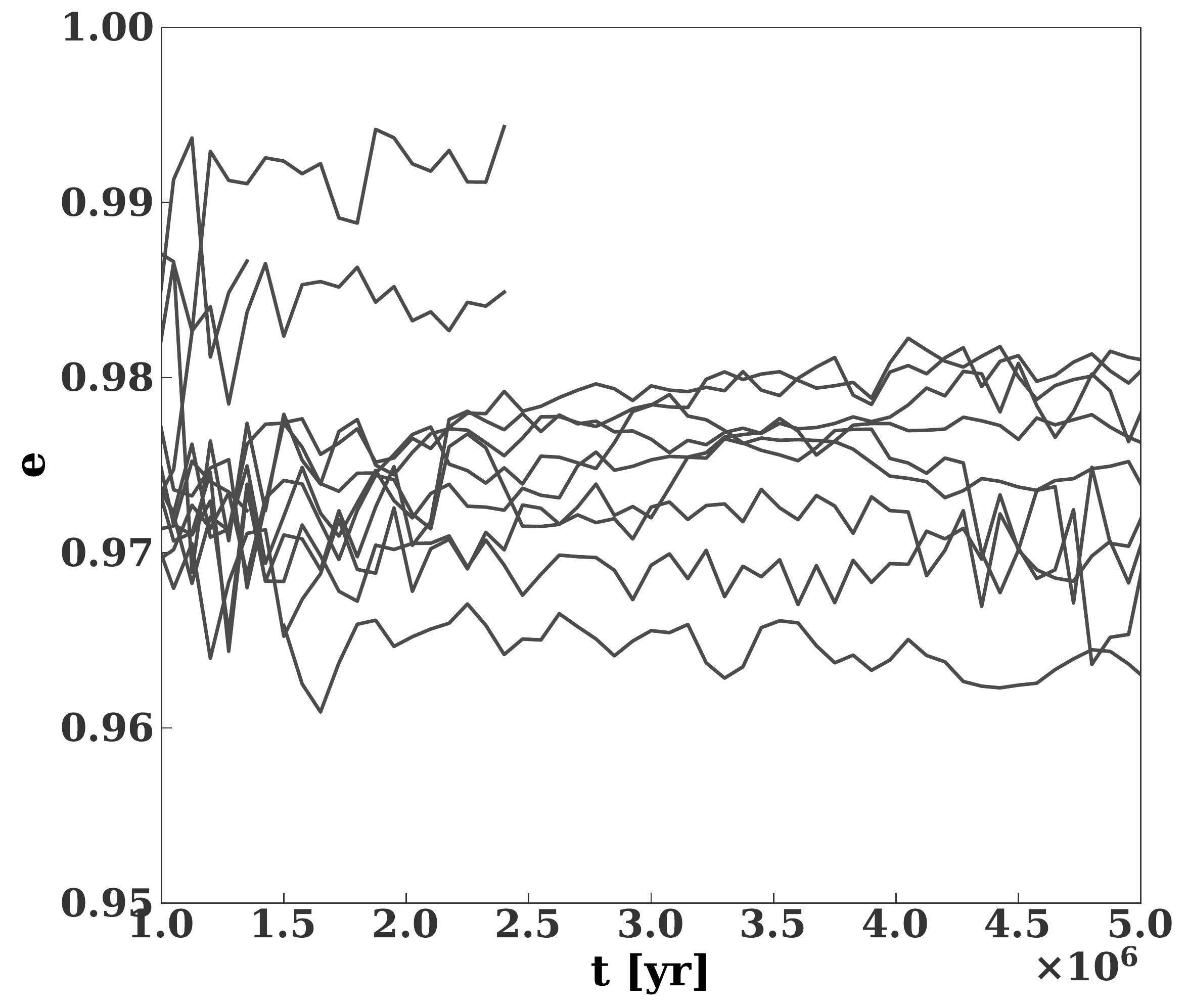}
\caption{\label{fig:bin_sep} Eccentricity evolution 
of a set of test particles injected into N-body simulations of an eccentric
disk (see Fig.~\ref{fig:nbodySim}). These particles are injected after binary
disruptions in the simulations and start with an eccentricity of $\sim 0.97$ and a
semimajor axis of 0.01 pc, which is five times smaller than the inner edge
of the disk. Note that these data are compiled from multiple simulations.} 
\end{figure}

The curves in Fig.~\ref{fig:bin_sep} show the postdisruption evolution of the
injected particles' eccentricities compiled from several simulations. The injected
stars retain eccentricities greater than 0.96 over 5 Myr.

\subsection{Effect of IMBHs}
\label{sec:imbh}
In this section, we demonstrate that an $\sim10^3 M_{\odot}$ IMBH with a
semimajor axis of $\sim$0.01 pc could thermalize the eccentricity
distribution of stars deposited via the Hills mechanism. Such IMBHs may originate from globular clusters. In particular, IMBHs 
may form from runaway black hole mergers in very dense globular clusters
\citep{antonini+2019} or other types of collisional runaways \citep{ebisuzaki+2001, portegies_zwart+2002, gurkan+2004, freitag+2006a, goswami+2012}. \citet{fragione+2018a,fragione+2018b}
studied the evolution of globular clusters with a primordial IMBH. In the Milky Way, most such 
IMBHs would be orphaned from their parent cluster (either by gravitational wave recoil kicks or tidal dissolution of the surrounding cluster), leaving a population 
of $\sim 1000$ bare IMBHs within the central kpc of the Galaxy. These IMBHs
may then inspiral toward the Galactic center via dynamical friction.
Past work has shown that IMBHs would stall at $\sim 0.01
(q/10^{-3})$ pc from the Galactic center (where $q$ is the mass ratio between
the IMBH and the central SMBH; see \citealt{merritt+2009} and the references
therein). As the IMBH is closer to the S-stars than the clockwise disk, it is
more effective at changing their eccentricity distribution.

We now review existing observational and theoretical constraints on the
presence of an IMBH in the Galactic center. Proper-motion studies of Sgr A*
can rule out an IMBH with mass $\gsim 10^{4} M_{\odot}$ 0.01 pc from the Galactic center \citep{reid&brunthaler2020}. Additionally, an IMBH of
$10^3 M_{\odot}$ inside of $1.3\times 10^{-4}$ pc would inspiral into the
central SMBH in less than 10 Myr due to gravitational wave emission (e.g.
\citealt{gualandris&merritt2009}).

Recently, \citet{naoz+2019} used observations of S2 to place
stringent constraints on the presence of an IMBH in the Galactic
Center. They found that a $10^5 M_{\odot}$ IMBH cannot have a semimajor axis
greater than 170 au ($8.2\times 10^{-4}$ pc). For a $10^3 M_{\odot}$ they found
the maximum semimajor axis from observational constraints is $\sim 1000$ au
($5\times 10^{-3}$ pc).

However, an IMBH at 0.01 pc is no longer inside of the orbit of S2, as assumed
in \citet{naoz+2019}. Also, in this case the IMBH-SMBH-S2 system is not
longer strongly hierarchical. Therefore, we perform direct three-body
integrations with \texttt{REBOUND} (with the ``GR'' effect included) to see
what effect an $10^3 M_{\odot}$ IMBH at 0.01 pc would have on the orientation
of S2's orbit. We find that that orbital precession rate is $\approx 0.01$
degrees per year in the case, which is consistent with existing observational
limits (note this is also the precession rate expected from GR effects alone). We
also find a $10^5 M_{\odot}$ IMBH at 170 au would cause S2 to precess by a few
tenths of a degree per year and can be ruled out from observations (as
expected from \citealt{naoz+2019}).

As we were completing revisions to this paper, the GRAVITY collaboration
reported a detection of apsidal precession in S2's orbit. The detected precession is consistent
with that expected from GR alone \citep{gravity+2020}. Combining the the latest constraints on 
S2's orbit with N-body simulations of an IMBH interacting with the S-star cluster \citep{gualandris+2010},
they find IMBH masses $\gsim 10^3 M_{\odot}$ are ruled out at 0.01 pc.

We perform N-body simulations with \texttt{AR--Chain} of 20 ten solar mass stars
interacting with an IMBH. \citet{merritt+2009} and \citet{gualandris&merritt2009} performed 
similar calculations. However, in our simulations, the stars are initially on highly eccentric
orbits with $e\approx 0.97$, as expected from the Hills
mechanism. More precisely, we draw the stars' semimajor axis ($a$) and
eccentricity from a distribution similar to that in Fig.~\ref{fig:dist},
with the restriction $0.005\,{\rm pc} \leq a
\leq 0.03\,{\rm pc}$. The stars are all initially in a plane with a
randomized orientation and mean anomaly. We consider four different IMBH
masses ($M_{\rm IMBH}$) between $3\times 10^2 M_{\odot}$ and $10^4 M_{\odot}$,
five different inclinations ($i_{\rm imbh}$) evenly spaced between
$5.7^{\circ}$ and $174.3^{\circ}$, and two eccentricities (0, 0.7). We perform
four different $\sim 10$ Myr simulations for each set of IMBH parameters.
Post-Newtonian effects are included for the black holes in these simulations
up to a PN order of 2.5. The IMBH parameters in our simulations are summarized in
Table~\ref{tab:imbh0}. 

\begin{deluxetable}{p{0.49\columnwidth}c}
\tablecaption{\label{tab:imbh0} IMBH Parameters for the Simulations in This Section.}
\tablewidth{0pt}
\tablehead{\multicolumn1l{$M_{\rm IMBH}$} & \colhead{$3\times 10^2$, $10^3$, $3\times 10^3$, $10^4 M_{\odot}$}}
\startdata
$i_{\rm IMBH}$ & $5.7$, $48$, $90$, $132$, $174^{\circ}$\\
$e_{\rm IMBH}$ & 0, 0.7\\
\enddata
\tablecomments{The semimajor axis is fixed to 0.01 pc for simplicity.}
\end{deluxetable}

To identify IMBH parameters that would reproduce the observations, we find snapshots where the following criteria are met.

\begin{enumerate}
	\item At least half of the simulated S-stars remain within
	0.03 pc.
	\item The KS and Anderson--Darling test probabilities that 
	the simulated and observed S-star eccentricity distributions
	are consistent is at least 10\%.
	\item There are no more than two stars with eccentricity greater than 0.95, 
	as in the observed population. For this and the preceding 
	condition, we only consider stars with semimajor axes within 0.03 pc
	\item The preceding criteria are met at least 5\% of the time after 4 Myr.
\end{enumerate}
The last condition is motivated by the best-fit age of the clockwise disk
from \citet{lu+2013}, and the (4.8 Myr) flight time of the
\citet{koposov+2019} hypervelocity star. (In the eccentric disk scenario, most
binary disruptions occur within $\sim 1$ Myr of disk formation; see
Fig.~\ref{fig:disTime}). The most massive IMBHs in our grid ($3\times 10^3
M_{\odot}$ and $10^4 M_{\odot}$) violate the first condition after 4 Myr, but
$3\times 10^3 M_{\odot}$ IMBHs can be effective on shorter time--scales.

Table~\ref{tab:imbh} summarizes the IMBH parameters for which at least
one of the four simulations we ran satisfies the above conditions. For circular
IMBHs, only simulations with $M_{\rm IMBH}=10^3 M_{\odot}$ and $i_{\rm
imbh}=5.7^{\circ}$ or $i_{\rm imbh}=48^{\circ}$ satisfy all of the above
criteria. A $10^3 M_{\odot}$ IMBH with an eccentricity of 0.7 satisfies the
criteria for all inclinations in our grid except for 48$^{\circ}$. Finally,
one simulation with an eccentric $300 M_{\odot}$ IMBH at an inclination of
$5.7^{\circ}$ satisfies these conditions. As shown in the last two columns,
these IMBHs can generally thermalize the S-stars' eccentricity distribution
over a few Myr, as required by observations.

\begin{deluxetable}{lccccc} 
\tablecaption{\label{tab:imbh} IMBH Parameters that Can Reproduce the Observed S-star Eccentricity Distribution, According to the Criteria Above. }
\tablewidth{0pt}
\tabletypesize{\small}
\tablehead{\colhead{$M_{\rm IMBH}$} & \colhead{$i_{\rm IMBH}$} & \colhead{$e_{\rm IMBH}$} & \colhead{$t_{\rm min}$} & \colhead{$t_{\rm max}$} & \colhead{Mergers}\\
\colhead{$(M_{\odot})$} & \colhead{(deg)} & & \colhead{(Myr)} & \colhead{(Myr)} &} 
\startdata
$3\times 10^2$ & 5.7 & 0.7 & 6.1 & 10 & 0\\
$10^3$ & 5.7 & 0 & 3.5 & 8.5 & 2 \\
$10^3$ & 48 & 0 & 7.3 & 9.4 & 1\\
$10^3$ & 5.7 & 0.7 & 4.3 & 9.9 & 0 \\
$10^3$ & 90 & 0.7 & 2.4 & 9.8 & 2\\ 
$10^3$ & 132 & 0.7 & 2.3 & 10 & 0\\
$10^3$ & 174 & 0.7 & 1.8 & 9.9 & 1--4\\
\enddata
\tablecomments{The fourth and fifth columns are the minimum and maximum times for which the first three criteria are satisfied. The last column shows the number of mergers (with respect to the SMBH) recorded in the simulation.}
\end{deluxetable}

\begin{figure*}
\includegraphics[width=8.5cm]{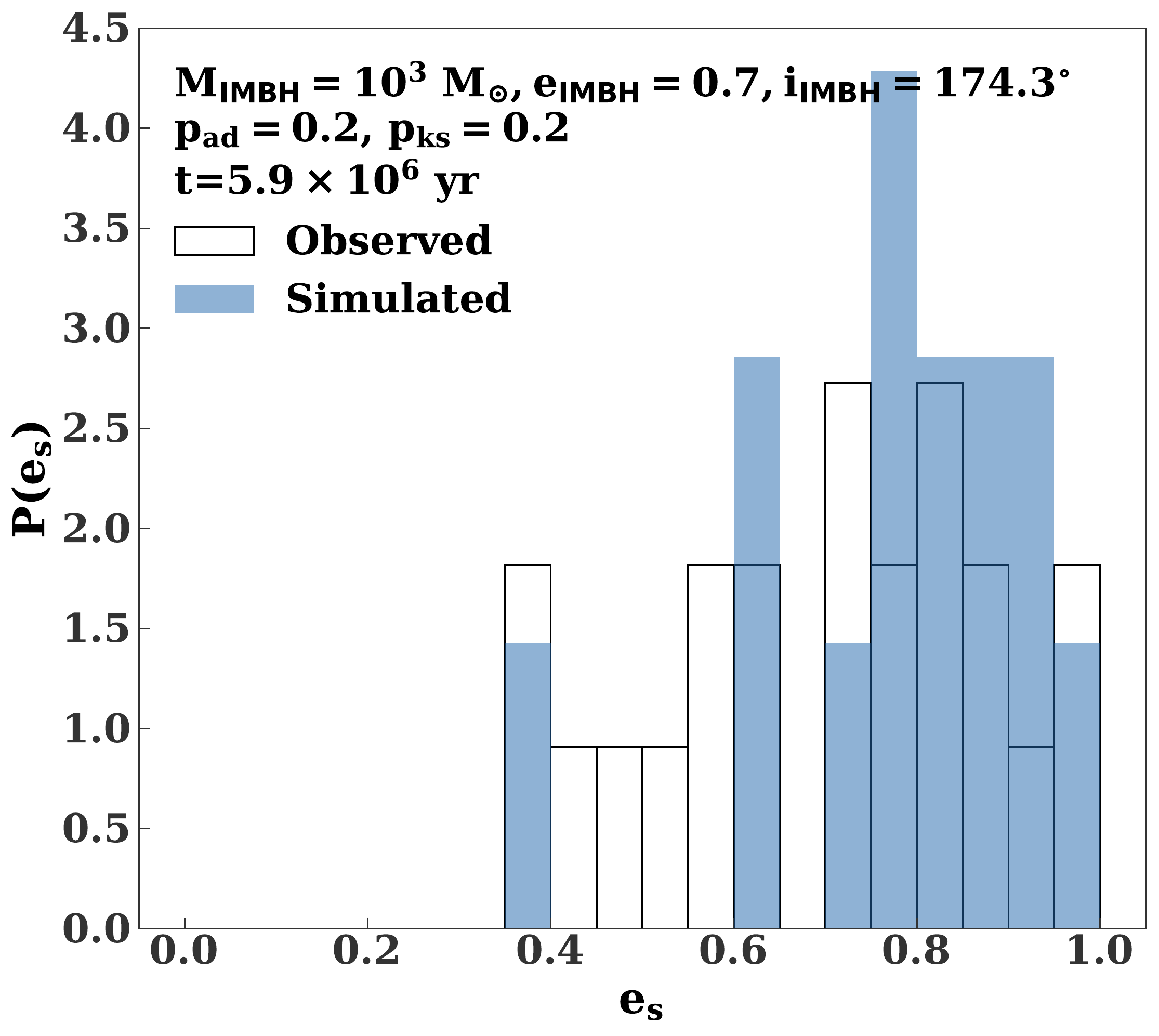}
\includegraphics[width=8.5cm]{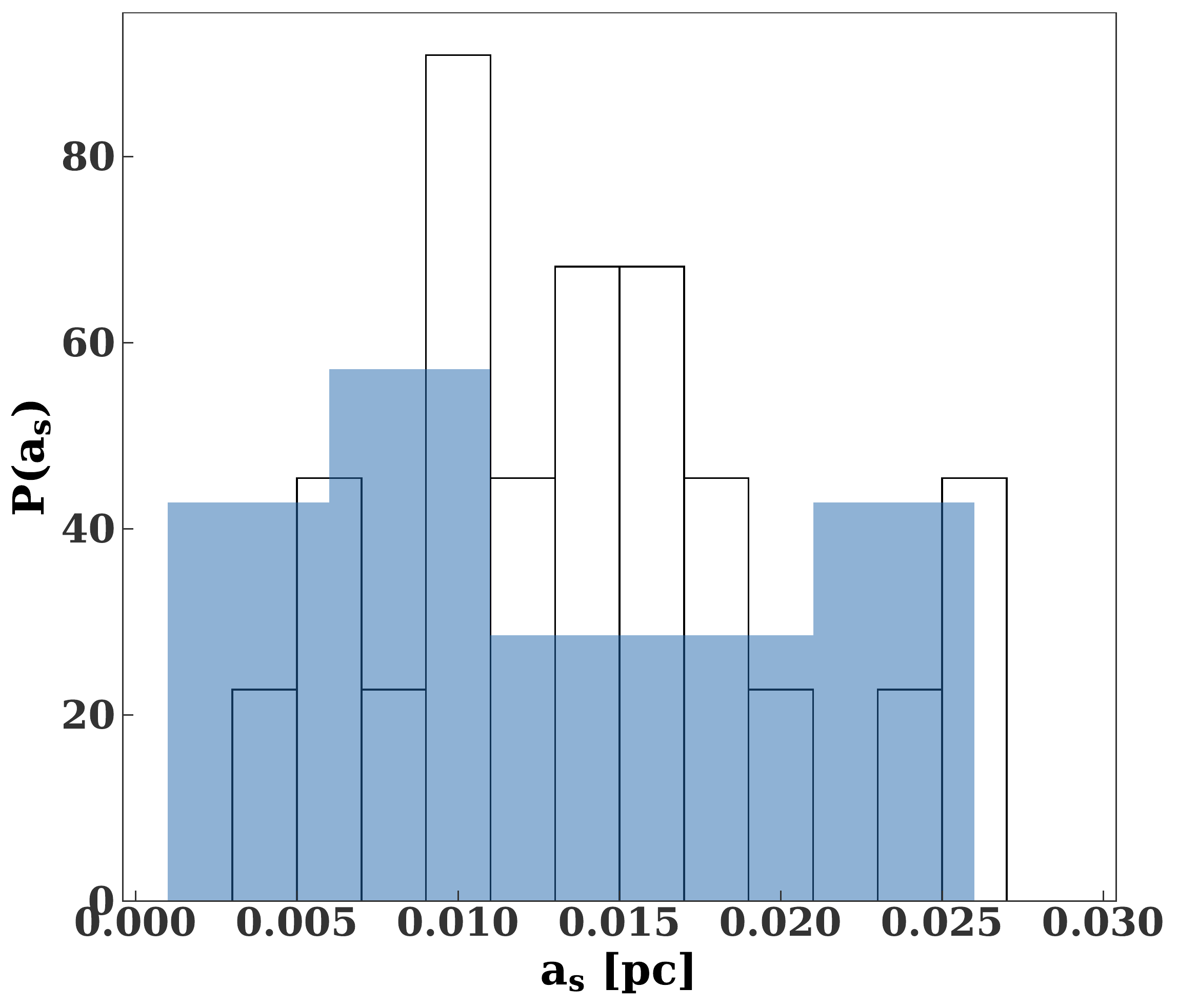}
\includegraphics[width=8.5cm]{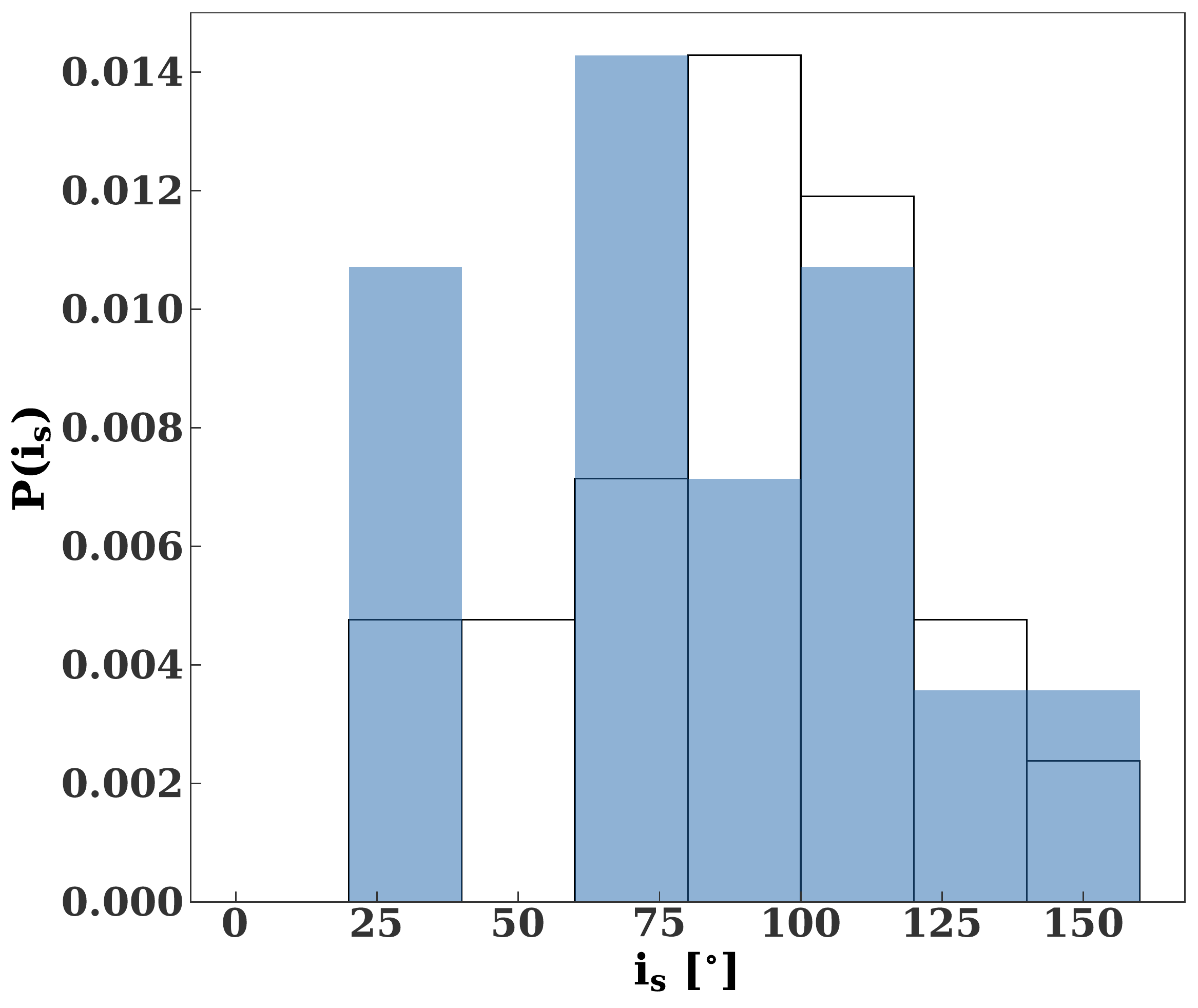}
\caption{\label{fig:imbh} Snapshot of orbital element distributions from an N-body simulation of 10 $M_{\odot}$ stars interacting with
an IMBH with the parameters in the top panel (blue filled histogram).
The stars all have an initial eccentricity $\gsim$0.96. The observed
eccentricity distribution of the S-stars from \citet{gillessen+2017} 
is shown by the black open histogram. The panels
show eccentricity (top left), semimajor axis (top right), and inclination (bottom). $p_{\rm KS}$ and $p_{\rm
AD}$ are the two-sample KS and and Anderson--Darling test p-values that the simulated and observed eccentricity distributions are consistent. Note that
for this plot, we exclude stars with semimajor axes greater than 0.03 pc.} 
\end{figure*}

\begin{figure}
\includegraphics[width=8.5cm]{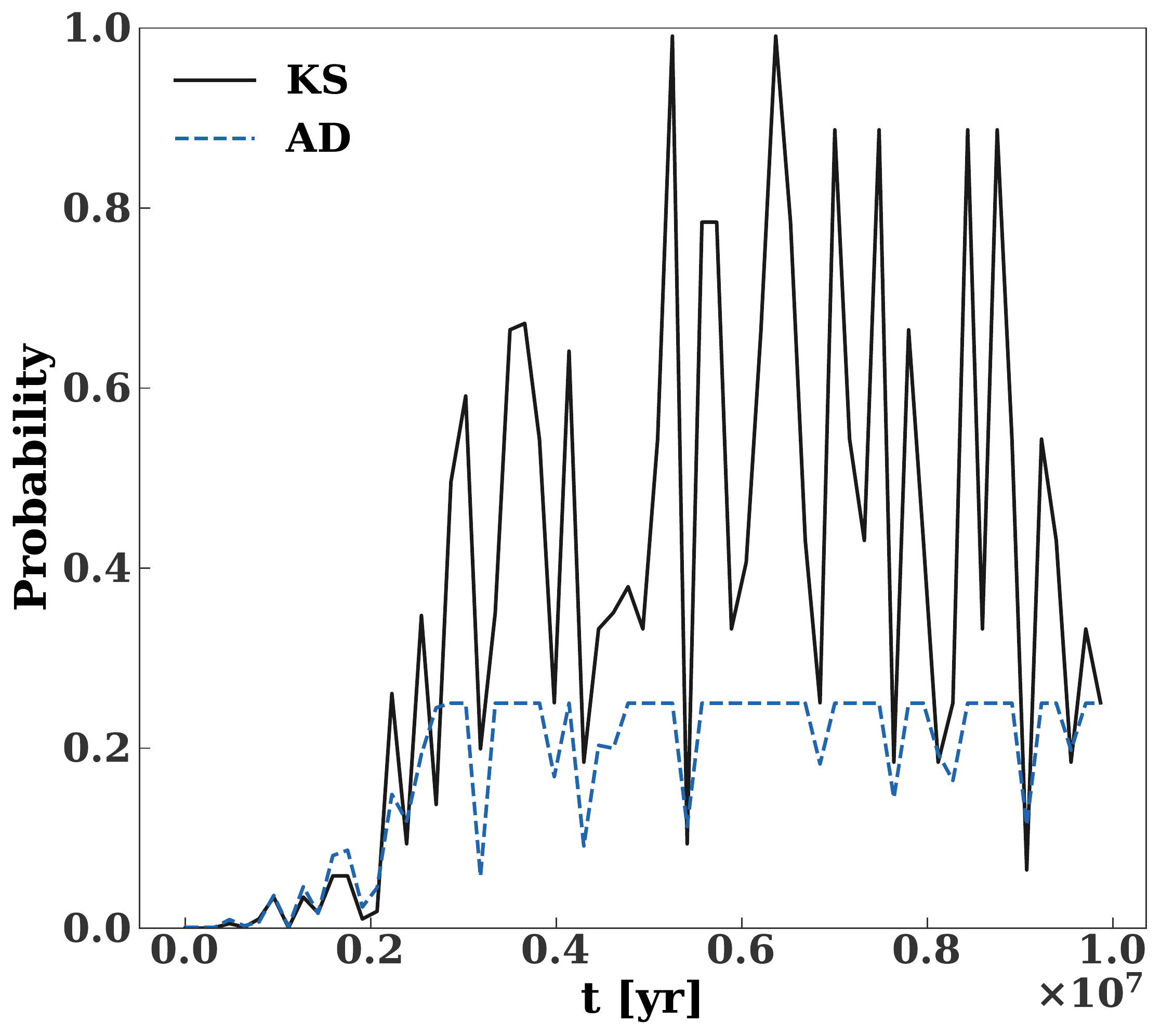}
\caption{\label{fig:ks} KS (black solid) and Anderson--Darling (AD)
(blue dashed) probabilities that the eccentricities from the
  simulation in Fig.~\ref{fig:imbh} are consistent with the
observed S-star eccentricities as a function of time. Note that implementation 
we use for the latter (``anderson\_ksamp'' in scipy.stats) caps the probability at 0.25. }
\end{figure}

The blue filled histograms in Fig.~\ref{fig:imbh}, show the eccentricity,
semimajor axis, and inclination distributions from an example simulation
snapshot where the above conditions are satisfied. The black open
histograms in this figure show the distribution of the S-stars with
semimajor axes inside of 0.03 pc. As shown in Fig.~\ref{fig:ks},
the eccentricity distribution from this simulation is statistically consistent
with observations for a range of times. Note that the distributions in 
Fig.~\ref{fig:imbh} only include stars with semimajor axes inside of 0.03 pc. Some stars in 
our simulations are kicked to larger semimajor axes. However, interactions with the 
clockwise disk (which are not included in these simulations), would be important for these 
stars.

In this case the IMBH also isotropizes the S-star orbits, so that their
inclination distribution is consistent with observations. In general,
the IMBHs in Table~\ref{tab:imbh} can isotropize the S-star orbits, except for the those with 
the lowest inclinations ($i=5.7^{\circ}$). However, as discussed in \S~\ref{sec:relax}, vector resonant
relaxation can also isotropize the S-star orbits. 

Finally, zero to two stars are unbound from the SMBH in these simulations. Also, a few ``mergers''
were recorded between stars and the central SMBH as summarized in the last column 
of Table~\ref{tab:imbh}.\footnote{In these simulations, mergers 
occur when stars pass within four Schwarzschild radii of the central black hole; that is, the 
default merger criterion in  \texttt{AR--Chain}. In fact, the cross section 
for stellar disruption is approximately an order of magnitude larger.}

\section{Discussion}
\label{sec:disc}
\subsection{Stellar collisions}
\label{sec:coll}
Approximately 20\% percent of our simulated binary--SMBH encounters result in
collisions between stars rather than tidal separation of the binary.
Table~\ref{tab:coll} shows the fraction of encounters that result in
collisions for different binary eccentricity distributions. We consider stars
to have collided if they enter into Roche lobe contact during the encounter
with the SMBH (see Equation~\ref{eq:aroche}).

The above statistics are based on one simulated encounter with the 
central SMBH per binary.
However, if binaries are in the empty loss cone, they would approach the tidal radius
gradually over several orbits (see \S~\ref{sec:lc}). In this process, the
binary could be excited to high eccentricities so that the stars collide
before the binary is disrupted \citep{antonini+2010,bradnick+2017}. This would
increase the ratio of collisions to disruptions. However,
\citet{antonini+2010} find that most additional collisions would come from
highly inclined binaries due to Kozai--Lidov resonances. Therefore, this
effect would not be present in our assumed coplanar configuration.

The ultimate fate of binaries that do collide depends on the collision velocity.
If this is smaller than the stellar escape speed, a merger will result. For faster 
collisions, tidal separation of the binary can still occur \citep{ginsburg&loeb2007}.
Binaries that do merge may account for G2-like objects \citep{witzel+2014}.

\begin{deluxetable*}{lcccc} 
\tablecaption{\label{tab:coll} Fraction of Monte Carlo Binary--SMBH Simulations Resulting in
Binary Disruptions (Column 1), Collisions (Column 2), and the Ratio of
Collisions to Disruptions (Column 3).}
\tablewidth{0pt}
\tabletypesize{\small}
\tablehead{\colhead{Eccentricity} & \colhead{Percent of Trials Resulting in Disruptions} & \colhead{Percent of Trials Resulting in Collisions} & \colhead{Collisions/Disruptions}}
\startdata
Circular & 50 & 20 & 0.40 \\
\hline
Thermal  & 49 & 16 & 0.33 \\
\enddata
\end{deluxetable*}

\subsection{Hypervelocity stars}
\begin{figure}
\includegraphics[width=8.5cm]{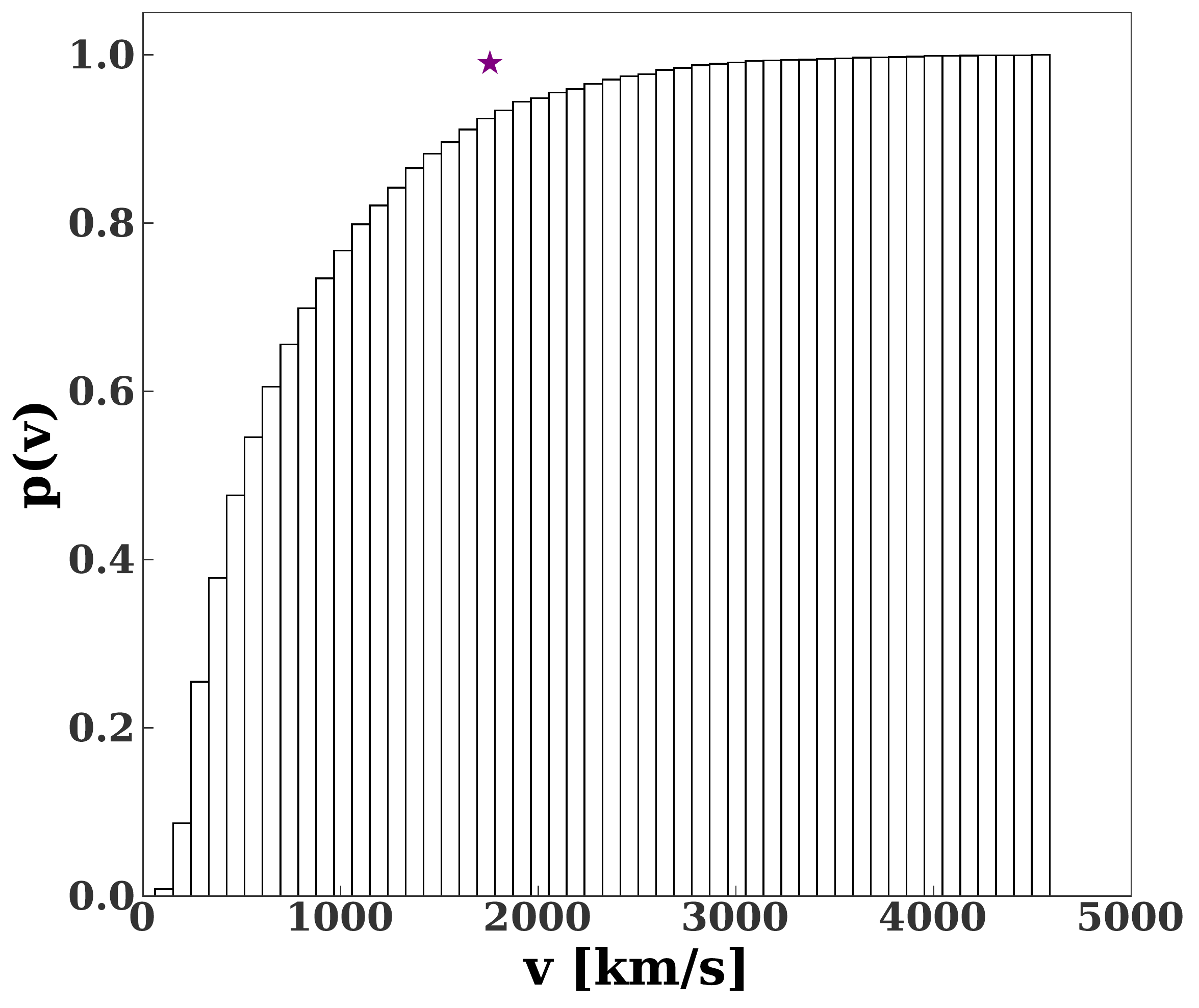}
\caption{\label{fig:hvs} Cumulative velocity distribution for ejected stars in
our Monte Carlo ensemble of binary--SMBH encounters (for binaries drawn from a
thermal eccentricity distribution). The purple star shows the ejection velocity
for the recently discovered hypervelocity star originating from the Galactic
Center \citep{koposov+2019}.}
\end{figure}

Many binary disruptions would result in a hypervelocity star ejected
from the Galactic center. The distribution of semimajor axes in Fig.~\ref{fig:dist} 
can be directly translated into a distribution of ejection velocities for such stars.
For a given S-star semimajor axis $a_s$, the ejection velocity of the corresponding hypervelocity star is
\begin{equation}
v_{ej}=\sqrt{\frac{G M m_{\rm s}}{a_s m_{\rm ej}}},
\end{equation}
where $M$ is the SMBH mass, and we have assumed the initial center-of-mass orbit of the binary was parabolic. Fig.~\ref{fig:hvs} shows the velocity
distribution of ejected stars from binaries with a thermal eccentricity
distribution (that is the model that best reproduces the observed S-star
semimajor axis distribution). For comparison, we also show the velocity of
the hypervelocity star detected by \citet{koposov+2019}. This star is in the
92nd percentile of the velocity distribution. If we instead use the distribution in Fig.~\ref{fig:paraDist}, where the binaries' center of mass has a more realistic eccentricity, this star moves to the \textbf{86th} percentile of the velocity distribution. In this case the ejected velocity is
\begin{equation}
v_{\rm ej}=\sqrt{\frac{-G M m_{\rm bin}}{m_{\rm ej} a_d}+ \frac{G M m_{\rm s}}{a_s m_{\rm ej}}},
\end{equation}
$a_d\approx 0.05$ pc is the initial semimajor axis of the binary's center-of-mass orbit.

Finally, for the ensemble of circular binary disruptions (see the top panel of Fig.~\ref{fig:dist}), the Koposov star would be in the 67th percentile of the velocity distribution.

\subsection{Masses of the S-stars}
\label{sec:smass}
We have proposed here that the S-stars originally come from the clockwise disk.
However, the clockwise disk has O-stars, while the S-stars
are all B stars with masses $\lsim 15 M_{\odot}$ \citep{habibi+2017}.

The simplest explanation would be a sampling effect: there are relatively few S-stars, and by chance no O-star binaries were disrupted. The chances that a random disk star has a mass below $\sim 15 M_\odot$ is

\begin{equation}
	p=\frac{\int_{1 M_{\odot}}^{15 M_{\odot}} m^{-1.7} dm}
	{\int_{1 M_{\odot}}^{60 M_{\odot}} m^{-1.7} dm},
\end{equation}
where we assume that the stars are drawn from the observed mass function in
\citet{lu+2013}, and that this mass function extends from 1 $M_{\odot}$ to the
main-sequence turnoff at the age of the disk ($\sim 4$ Myr). There are 22 S-stars inside of 0.03 pc. The chances that they would all be below 15 $M_{\odot}$ is
$p^{22}\approx 10\%$, which suggests sampling effects could be a plausible
explanation for the mass discrepancy between the S-stars and the clockwise
disk. However, the observed S-stars do not include any stars with masses of 
$\sim1 M_{\odot}$, which are too dim to detect. The dimmest S-star in Table 3 of \citet{gillessen+2017}
has a K-band magnitude of 17.8, which suggests a stellar mass of $\sim 3 M_{\odot}$ according to Table 1 of \citet{cai+2018}. Assuming only stars with masses $\gsim 3 M_{\odot}$ could be observed, a total of $\sim$50 disruptions are required to produce the observed S-stars, and the probability that
none of these disruptions include O stars is $\sim 0.5\%$.

However, the Hills mechanism has a built-in bias, and deposits massive
stars at larger semimajor axes. From Equation~\eqref{eq:as}, after the disruption, the bound star has a semimajor axis of

\begin{align}
	&a_s=\frac{\chi^2}{k'} \frac{m_s}{m_{\rm bin}} \frac{(1+q)^2}{2 q} \left(\frac{M}{m_{\rm bin}}\right)^{2/3} a_{\rm bin}=\nonumber\\
	&\begin{cases}
		\frac{1}{k} \frac{1+q}{2 q} \left(\frac{M}{m_{\rm bin}}\right)^{2/3} a_{\rm bin} &  {\rm if \,\, primary}\\
		\frac{1}{k} \frac{1+q}{2} \left(\frac{M}{m_{\rm bin}}\right)^{2/3} a_{\rm bin} & {\rm if \,\, secondary}. \\
	\end{cases}
	\label{eq:as2}
\end{align}
The semimajor axis of the bound star is a factor of $1/q$ larger if it is
the primary. (Note that in the limit of parabolic disruptions the primary and
secondary have an equal probability to be left bound to the SMBH; see \citealt{sari+2010, kobayashi+2012}.)
This effect could explain the dearth of O-stars among the S-stars. We leave a detailed exploration to future work.

\section{Conclusion}
Recent observations suggest a common origin for the S-star cluster and the
clockwise disk in the Galactic center. In particular, recent spectral
observations suggest that these two populations have consistent ages
\citep{habibi+2017}. Additionally, \citet{koposov+2019} discovered
a hypervelocity star that was ejected from the Galactic center 4.8 Myr ago (approximately 
when the clockwise disk was forming). This strongly suggests that stellar binaries 
from the clockwise disk were tidally disrupted by the central SMBH early in its history, which
would leave a compact cluster of stars inside of the disk (like the S-stars). 
Disk binaries can be pushed to tidal disruption via a
gravitational instability in the disk \citep{madigan+2009}. We perform a detailed study
of this scenario. Our results are summarized as follows.

\begin{enumerate}

	\item We quantify the plausible range of semimajor axes of binary stars
	in the clockwise disk (see Fig.~\ref{fig:sma}). We find that the
	evaporation of such binaries is dominated by other disk stars, rather than
	the isotropic star cluster in the Galactic center. 

	\item We simulate a large number of encounters between disk binaries and 
	the central SMBH. Tidal disruption of binary stars can reproduce the present-day
	semimajor axis distribution of the S-stars.

	\item The eccentricity distribution of the S-stars is more difficult to reproduce. If these 
	stars are injected via tidal disruption of binaries, they would start with
	a large eccentricity ($e\sim 0.97$). In contrast, the S-stars
	have a roughly thermal eccentricity distribution.

	\item  We have considered three possible mechanisms to thermalize the
	S-star eccentricity: (a) scalar resonant relaxation (b) torques from the
	parent clockwise disk, and (c) an IMBH. The flight time of the
	hypervelocity star discovered by \citet{koposov+2019} and the observed
	ages of the S-stars require that the S-stars reach
	their current distribution in $\lsim 10^7$ years. This is 
	a challenge, as highly eccentric orbits experience rapid general relativistic 
	precession, which suppresses secular torques. We find that torques 
	from the clockwise disk are unlikely to be effective on this timescale. However,
	scalar resonant relaxation by the isotropic cluster can reproduce the observed eccentricity distribution 
	of S-stars within $10^7$ years. Also, a
	$\sim 10^3 M_{\odot}$ IMBH at $\sim 0.01$ pc, could reproduce the S-star
	eccentricity distribution within a few million years. However, such IMBHs are, at best, marginally consistent with the latest observational limits \citep{gravity+2020}.

	\item After a binary is tidally disrupted, the primary and secondary have equal chances to remain bound to the SMBH (for parabolic disruptions). However, the primary is deposited at larger semimajor axes, which could explain the dearth of O stars among the S-stars.
\end{enumerate}	

\acknowledgements
\section*{ACKNOWLEDGEMENTS}
{
We thank the anonymous referee and Jean-Baptiste Fouvry for detailed comments that substantially improved the paper.

We thank Jason Dexter for suggesting IMBHs as an eccentricity relaxation mechanism and for other helpful comments. We thank Ben Bar-Or for looking over our multi-mass generalization of \texttt{SCRRPY}.

We thank Michi Baub\"{o}ck for clarifications on the latest GRAVITY results.

A.M. gratefully acknowledges support from the NASA Astrophysics Theory Program (ATP) under grant NNX17AK44G and the David and Lucile Packard Foundation.

This work utilized resources from the University of Colorado Boulder Research Computing Group, which is supported by the National Science Foundation (awards ACI-1532235 and ACI-1532236), the University of Colorado Boulder, and Colorado State University. 
}

\software{\texttt{AR--Chain} \citep{mikkola&merritt2008}, \texttt{REBOUND} \citep{rein.liu2012}, 
\texttt{REBOUNDX} \citep{tamayo+2019}, \texttt{SCRRPY} \citep{bar-or&fouvry2018}, 
\texttt{AstroPy} \citep{astropy+2018}, \texttt{Matplotlib} \citep{hunter+2007}, \texttt{NumPy}, \texttt{SciPy} \citep{2020SciPy-NMeth}, IPython \citep{perez+2007}}

\appendix
\section{Hills Approximation}
\label{app:hills}
In the Hills approximation the time derivative of the binary separation is \citep{sari+2010}
\begin{align}
\dot{\mathbf{r}}=\left(\frac{r_p}{r_m}\right)^3 \left[-\mathbf{r}+3 (\mathbf{r}\cdot \hat{\mathbf{r}}_m) \hat{\mathbf{r}}_m\right]-\frac{\mathbf{r}}{r^3},
\label{eq:hills}
\end{align}
where $\mathbf{r}_{m}$ is the unperturbed center-of-mass orbit and $r_p$ is
the pericenter distance. All distances and times are in units of
$\left(\frac{m_{\rm bin}}{M}\right)^{1/3} r_p$ and $\sqrt{r_p^3/G M}$, respectively.
Solving the above equation gives the separation of the binary stars as a
function of time. We can then use this separation to solve the positions of
the individual stars. We call this second step the ``Iterated Hills
Approximation.''

\begin{figure}
\includegraphics[width=8.5cm]{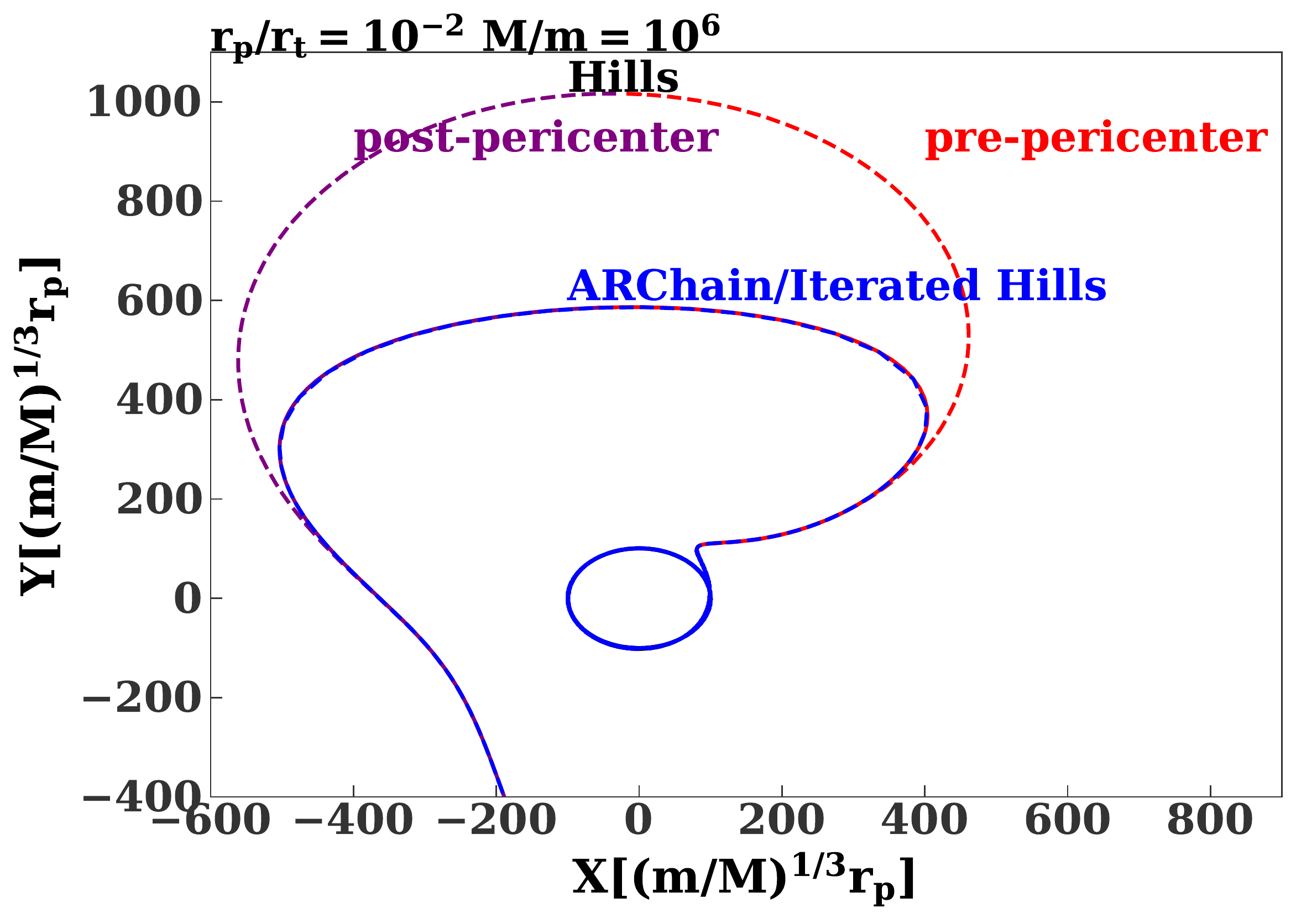}
\caption{\label{fig:archain} Binary separation for an example disruption. The top line is the solution from the Hills approximation, while
 the bottom line shows the solutions from \texttt{AR--Chain} and the
 iterated Hills approximation (see text for details).}
\end{figure}

The postdisruption orbital elements from the Hills approximation are in
good agreement with the results from three-body integrations. For deeply
penetrating disruptions ($r_p \ll r_t$) the separations from these two methods
diverge when the binary is near pericenter as shown in Fig.~\ref{fig:archain}.
However, the two solutions converge again at late times. The positions of the
individual stars from the ``Iterated Hills Approximation'' give a revised
estimate for the binary separation. This revised estimate always always falls
within $\lesssim$1\% of the \texttt{AR--Chain} solution in Fig.~\ref{fig:archain}.

\footnotesize{
\bibliographystyle{aastex}
\bibliography{master}

\begin{thebibliography}{}
\expandafter\ifx\csname natexlab\endcsname\relax\def\natexlab#1{#1}\fi
\providecommand{\url}[1]{\href{#1}{#1}}
\providecommand{\dodoi}[1]{doi:~\href{http://doi.org/#1}{\nolinkurl{#1}}}
\providecommand{\doeprint}[1]{\href{http://ascl.net/#1}{\nolinkurl{http://ascl.net/#1}}}
\providecommand{\doarXiv}[1]{\href{https://arxiv.org/abs/#1}{\nolinkurl{https://arxiv.org/abs/#1}}}

\bibitem[{{Alexander} {et~al.}(2017){Alexander}, {Wieringa}, {Berger},
  {Saxton}, \& {Komossa}}]{alexander+2017}
{Alexander}, K.~D., {Wieringa}, M.~H., {Berger}, E., {Saxton}, R.~D., \&
  {Komossa}, S. 2017, \apj, 837, 153, \dodoi{10.3847/1538-4357/aa6192}

\bibitem[{{Alexander}(2017)}]{alexander2017}
{Alexander}, T. 2017, \araa, 55, 17,
  \dodoi{10.1146/annurev-astro-091916-055306}

\bibitem[{{Alexander} \& {Pfuhl}(2014)}]{alexander&pfuhl2014}
{Alexander}, T., \& {Pfuhl}, O. 2014, \apj, 780, 148,
  \dodoi{10.1088/0004-637X/780/2/148}

\bibitem[{{Antonini} {et~al.}(2010){Antonini}, {Faber}, {Gualandris}, \&
  {Merritt}}]{antonini+2010}
{Antonini}, F., {Faber}, J., {Gualandris}, A., \& {Merritt}, D. 2010, \apj,
  713, 90, \dodoi{10.1088/0004-637X/713/1/90}

\bibitem[{{Antonini} {et~al.}(2019){Antonini}, {Gieles}, \&
  {Gualandris}}]{antonini+2019}
{Antonini}, F., {Gieles}, M., \& {Gualandris}, A. 2019, \mnras, 486, 5008,
  \dodoi{10.1093/mnras/stz1149}

\bibitem[{{Antonini} \& {Merritt}(2013)}]{antonini&merritt2013}
{Antonini}, F., \& {Merritt}, D. 2013, \apjl, 763, L10,
  \dodoi{10.1088/2041-8205/763/1/L10}

\bibitem[{{Antonini} \& {Perets}(2012)}]{antonini&perets2012}
{Antonini}, F., \& {Perets}, H.~B. 2012, \apj, 757, 27,
  \dodoi{10.1088/0004-637X/757/1/27}

\bibitem[{{Bar-Or} \& {Alexander}(2016)}]{bar-or&alexander2016}
{Bar-Or}, B., \& {Alexander}, T. 2016, \apj, 820, 129,
  \dodoi{10.3847/0004-637X/820/2/129}

\bibitem[{Bar-Or \& Fouvry(2018)}]{bar-or&fouvry2018}
Bar-Or, B., \& Fouvry, J.-B. 2018, \apjl, 860, L23,
  \dodoi{10.3847/2041-8213/aac88e}

\bibitem[{{Bartko} {et~al.}(2009){Bartko}, {Martins}, {Fritz}, {Genzel},
  {Levin}, {Perets}, {Paumard}, {Nayakshin}, {Gerhard}, {Alexander},
  {Dodds-Eden}, {Eisenhauer}, {Gillessen}, {Mascetti}, {Ott}, {Perrin},
  {Pfuhl}, {Reid}, {Rouan}, {Sternberg}, \& {Trippe}}]{bartko+2009}
{Bartko}, H., {Martins}, F., {Fritz}, T.~K., {et~al.} 2009, \apj, 697, 1741,
  \dodoi{10.1088/0004-637X/697/2/1741}

\bibitem[{{Bonnell} \& {Rice}(2008)}]{Bonnell2008}
{Bonnell}, I.~A., \& {Rice}, W.~K.~M. 2008, Science, 321, 1060,
  \dodoi{10.1126/science.1160653}

\bibitem[{{Bradnick} {et~al.}(2017){Bradnick}, {Mandel}, \&
  {Levin}}]{bradnick+2017}
{Bradnick}, B., {Mandel}, I., \& {Levin}, Y. 2017, \mnras, 469, 2042,
  \dodoi{10.1093/mnras/stx1007}

\bibitem[{{Cai} {et~al.}(2018){Cai}, {Liu}, \& {Wang}}]{cai+2018}
{Cai}, R.-G., {Liu}, T.-B., \& {Wang}, S.-J. 2018, arXiv e-prints,
  arXiv:1808.03164.
\newblock \doarXiv{1808.03164}

\bibitem[{Do {et~al.}(2013)Do, Lu, Ghez, Morris, Yelda, Martinez, Wright, \&
  Matthews}]{do+2013}
Do, T., Lu, J.~R., Ghez, A.~M., {et~al.} 2013, \apj, 764, 154,
  \dodoi{10.1088/0004-637X/764/2/154}

\bibitem[{{Drake}(1965)}]{drake1965}
{Drake}, F.~D. 1965, {The Radio Search for Intelligent Extraterrestrial Life},
  323--345

\bibitem[{{Dremova} {et~al.}(2019){Dremova}, {Dremov}, \&
  {Tutukov}}]{dremova+2019}
{Dremova}, G.~N., {Dremov}, V.~V., \& {Tutukov}, A.~V. 2019, Astronomy Reports,
  63, 862, \dodoi{10.1134/S1063772919100032}

\bibitem[{{Ebisuzaki} {et~al.}(2001){Ebisuzaki}, {Makino}, {Tsuru}, {Funato},
  {Portegies Zwart}, {Hut}, {McMillan}, {Matsushita}, {Matsumoto}, \&
  {Kawabe}}]{ebisuzaki+2001}
{Ebisuzaki}, T., {Makino}, J., {Tsuru}, T.~G., {et~al.} 2001, \apjl, 562, L19,
  \dodoi{10.1086/338118}

\bibitem[{Eggleton(1983)}]{eggleton1983}
Eggleton, P.~P. 1983, \apj, 268, 368, \dodoi{10.1086/160960}

\bibitem[{{Foote} {et~al.}(2020){Foote}, {Generozov}, \&
  {Madigan}}]{foote+2020}
{Foote}, H.~R., {Generozov}, A., \& {Madigan}, A.-M. 2020, \apj, 890, 175,
  \dodoi{10.3847/1538-4357/ab6c66}

\bibitem[{{Fragione} \& {Antonini}(2019)}]{fragione&antonini2019}
{Fragione}, G., \& {Antonini}, F. 2019, \mnras, 488, 728,
  \dodoi{10.1093/mnras/stz1723}

\bibitem[{{Fragione} {et~al.}(2017){Fragione}, {Capuzzo-Dolcetta}, \&
  {Kroupa}}]{fragione+2017}
{Fragione}, G., {Capuzzo-Dolcetta}, R., \& {Kroupa}, P. 2017, \mnras, 467, 451,
  \dodoi{10.1093/mnras/stx106}

\bibitem[{{Fragione} {et~al.}(2018{\natexlab{a}}){Fragione}, {Ginsburg}, \&
  {Kocsis}}]{fragione+2018a}
{Fragione}, G., {Ginsburg}, I., \& {Kocsis}, B. 2018{\natexlab{a}}, \apj, 856,
  92, \dodoi{10.3847/1538-4357/aab368}

\bibitem[{{Fragione} {et~al.}(2018{\natexlab{b}}){Fragione}, {Leigh},
  {Ginsburg}, \& {Kocsis}}]{fragione+2018b}
{Fragione}, G., {Leigh}, N. W.~C., {Ginsburg}, I., \& {Kocsis}, B.
  2018{\natexlab{b}}, \apj, 867, 119, \dodoi{10.3847/1538-4357/aae486}

\bibitem[{{Fragione} \& {Sari}(2018)}]{fragione&sari2018}
{Fragione}, G., \& {Sari}, R. 2018, \apj, 852, 51,
  \dodoi{10.3847/1538-4357/aaa0d7}

\bibitem[{{Freitag} {et~al.}(2006){Freitag}, {G{\"u}rkan}, \&
  {Rasio}}]{freitag+2006a}
{Freitag}, M., {G{\"u}rkan}, M.~A., \& {Rasio}, F.~A. 2006, \mnras, 368, 141,
  \dodoi{10.1111/j.1365-2966.2006.10096.x}

\bibitem[{{Fuller} \& {Lai}(2011)}]{fuller&lai2011}
{Fuller}, J., \& {Lai}, D. 2011, \mnras, 412, 1331,
  \dodoi{10.1111/j.1365-2966.2010.18017.x}

\bibitem[{{Generozov} {et~al.}(2018){Generozov}, {Stone}, {Metzger}, \&
  {Ostriker}}]{generozov+2018}
{Generozov}, A., {Stone}, N.~C., {Metzger}, B.~D., \& {Ostriker}, J.~P. 2018,
  \mnras, 478, 4030, \dodoi{10.1093/mnras/sty1262}

\bibitem[{{Ghez} {et~al.}(2003){Ghez}, {Duch{\^e}ne}, {Matthews}, {Hornstein},
  {Tanner}, {Larkin}, {Morris}, {Becklin}, {Salim}, {Kremenek}, {Thompson},
  {Soifer}, {Neugebauer}, \& {McLean}}]{ghez+2003}
{Ghez}, A.~M., {Duch{\^e}ne}, G., {Matthews}, K., {et~al.} 2003, \apjl, 586,
  L127, \dodoi{10.1086/374804}

\bibitem[{{Gillessen} {et~al.}(2017){Gillessen}, {Plewa}, {Eisenhauer}, {Sari},
  {Waisberg}, {Habibi}, {Pfuhl}, {George}, {Dexter}, \& {von
  Fellenberg}}]{gillessen+2017}
{Gillessen}, S., {Plewa}, P.~M., {Eisenhauer}, F., {et~al.} 2017, \apj, 837,
  30, \dodoi{10.3847/1538-4357/aa5c41}

\bibitem[{{Ginsburg} \& {Loeb}(2006)}]{ginsburg&loeb2006}
{Ginsburg}, I., \& {Loeb}, A. 2006, \mnras, 368, 221,
  \dodoi{10.1111/j.1365-2966.2006.10091.x}

\bibitem[{{Ginsburg} \& {Loeb}(2007)}]{ginsburg&loeb2007}
---. 2007, \mnras, 376, 492, \dodoi{10.1111/j.1365-2966.2007.11461.x}

\bibitem[{{Goswami} {et~al.}(2012){Goswami}, {Umbreit}, {Bierbaum}, \&
  {Rasio}}]{goswami+2012}
{Goswami}, S., {Umbreit}, S., {Bierbaum}, M., \& {Rasio}, F.~A. 2012, \apj,
  752, 43, \dodoi{10.1088/0004-637X/752/1/43}

\bibitem[{{Gravity Collaboration} {et~al.}(2020){Gravity Collaboration},
  {Abuter}, {Amorim}, {Baub{\"o}ck}, {Berger}, {Bonnet}, {Brand ner},
  {Cardoso}, {Cl{\'e}net}, {de Zeeuw}, {Dexter}, {Eckart}, {Eisenhauer},
  {F{\"o}rster Schreiber}, {Garcia}, {Gao}, {Gendron}, {Genzel}, {Gillessen},
  {Habibi}, {Haubois}, {Henning}, {Hippler}, {Horrobin}, {Jim{\'e}nez-Rosales},
  {Jochum}, {Jocou}, {Kaufer}, {Kervella}, {Lacour}, {Lapeyr{\`e}re}, {Le
  Bouquin}, {L{\'e}na}, {Nowak}, {Ott}, {Paumard}, {Perraut}, {Perrin},
  {Pfuhl}, {Rodr{\'\i}guez-Coira}, {Shangguan}, {Scheithauer}, {Stadler},
  {Straub}, {Straubmeier}, {Sturm}, {Tacconi}, {Vincent}, {von Fellenberg},
  {Waisberg}, {Widmann}, {Wieprecht}, {Wiezorrek}, {Woillez}, {Yazici}, \&
  {Zins}}]{gravity+2020}
{Gravity Collaboration}, {Abuter}, R., {Amorim}, A., {et~al.} 2020, 636, L5,
  \dodoi{10.1051/0004-6361/202037813}

\bibitem[{{Gualandris} {et~al.}(2010){Gualandris}, {Gillessen}, \&
  {Merritt}}]{gualandris+2010}
{Gualandris}, A., {Gillessen}, S., \& {Merritt}, D. 2010, \mnras, 409, 1146,
  \dodoi{10.1111/j.1365-2966.2010.17373.x}

\bibitem[{{Gualandris} {et~al.}(2012){Gualandris}, {Mapelli}, \&
  {Perets}}]{gualandris+2012}
{Gualandris}, A., {Mapelli}, M., \& {Perets}, H.~B. 2012, \mnras, 427, 1793,
  \dodoi{10.1111/j.1365-2966.2012.22133.x}

\bibitem[{{Gualandris} \& {Merritt}(2009)}]{gualandris&merritt2009}
{Gualandris}, A., \& {Merritt}, D. 2009, \apj, 705, 361,
  \dodoi{10.1088/0004-637X/705/1/361}

\bibitem[{{G{\"u}rkan} {et~al.}(2004){G{\"u}rkan}, {Freitag}, \&
  {Rasio}}]{gurkan+2004}
{G{\"u}rkan}, M.~A., {Freitag}, M., \& {Rasio}, F.~A. 2004, \apj, 604, 632,
  \dodoi{10.1086/381968}

\bibitem[{G{\"u}rkan \& Hopman(2007)}]{gurkan&hopman2007}
G{\"u}rkan, M.~A., \& Hopman, C. 2007, \mnras, 379, 1083,
  \dodoi{10.1111/j.1365-2966.2007.11982.x}

\bibitem[{{Habibi} {et~al.}(2017){Habibi}, {Gillessen}, {Martins},
  {Eisenhauer}, {Plewa}, {Pfuhl}, {George}, {Dexter}, {Waisberg}, {Ott}, {von
  Fellenberg}, {Baub{\"o}ck}, {Jimenez-Rosales}, \& {Genzel}}]{habibi+2017}
{Habibi}, M., {Gillessen}, S., {Martins}, F., {et~al.} 2017, \apj, 847, 120,
  \dodoi{10.3847/1538-4357/aa876f}

\bibitem[{Hailey {et~al.}(2018)Hailey, Mori, Bauer, Berkowitz, Hong, \&
  Hord}]{hailey+2018}
Hailey, C.~J., Mori, K., Bauer, F.~E., {et~al.} 2018, Nature, 556, 70.
\newblock \url{http://dx.doi.org/10.1038/nature25029}

\bibitem[{{Hills}(1988)}]{hills1988}
{Hills}, J.~G. 1988, \nat, 331, 687, \dodoi{10.1038/331687a0}

\bibitem[{Hunter(2007)}]{hunter+2007}
Hunter, J.~D. 2007, Computing in Science \& Engineering, 9, 90,
  \dodoi{10.1109/MCSE.2007.55}

\bibitem[{{Kobayashi} {et~al.}(2012){Kobayashi}, {Hainick}, {Sari}, \&
  {Rossi}}]{kobayashi+2012}
{Kobayashi}, S., {Hainick}, Y., {Sari}, R., \& {Rossi}, E.~M. 2012, \apj, 748,
  105, \dodoi{10.1088/0004-637X/748/2/105}

\bibitem[{{Kocsis} \& {Tremaine}(2011)}]{kocsis&tremaine2011}
{Kocsis}, B., \& {Tremaine}, S. 2011, \mnras, 412, 187,
  \dodoi{10.1111/j.1365-2966.2010.17897.x}

\bibitem[{{Koposov} {et~al.}(2020){Koposov}, {Boubert}, {Li}, {Erkal}, {Da
  Costa}, {Zucker}, {Ji}, {Kuehn}, {Lewis}, {Mackey}, {Simpson}, {Shipp},
  {Wan}, {Belokurov}, {Bland-Hawthorn}, {Martell}, {Nordlander}, {Pace}, {De
  Silva}, {Wang}, \& {S5 collaboration}}]{koposov+2019}
{Koposov}, S.~E., {Boubert}, D., {Li}, T.~S., {et~al.} 2020, \mnras, 491, 2465,
  \dodoi{10.1093/mnras/stz3081}

\bibitem[{{Kozai}(1962)}]{kozai1962}
{Kozai}, Y. 1962, \aj, 67, 591, \dodoi{10.1086/108790}

\bibitem[{{Levin}(2007)}]{levin2007}
{Levin}, Y. 2007, \mnras, 374, 515, \dodoi{10.1111/j.1365-2966.2006.11155.x}

\bibitem[{{Levin} \& {Beloborodov}(2003)}]{levin&beloborodov03}
{Levin}, Y., \& {Beloborodov}, A.~M. 2003, \apjl, 590, L33,
  \dodoi{10.1086/376675}

\bibitem[{{Lidov}(1962)}]{lidov1962}
{Lidov}, M.~L. 1962, \planss, 9, 719, \dodoi{10.1016/0032-0633(62)90129-0}

\bibitem[{{L{\"o}ckmann} {et~al.}(2009){L{\"o}ckmann}, {Baumgardt}, \&
  {Kroupa}}]{lockmann+2009}
{L{\"o}ckmann}, U., {Baumgardt}, H., \& {Kroupa}, P. 2009, \mnras, 398, 429,
  \dodoi{10.1111/j.1365-2966.2009.15157.x}

\bibitem[{{Lu} {et~al.}(2013){Lu}, {Do}, {Ghez}, {Morris}, {Yelda}, \&
  {Matthews}}]{lu+2013}
{Lu}, J.~R., {Do}, T., {Ghez}, A.~M., {et~al.} 2013, \apj, 764, 155,
  \dodoi{10.1088/0004-637X/764/2/155}

\bibitem[{{Madigan} {et~al.}(2018){Madigan}, {Halle}, {Moody}, {McCourt},
  {Nixon}, \& {Wernke}}]{madigan+2018}
{Madigan}, A.-M., {Halle}, A., {Moody}, M., {et~al.} 2018, \apj, 853, 141,
  \dodoi{10.3847/1538-4357/aaa714}

\bibitem[{{Madigan} {et~al.}(2011){Madigan}, {Hopman}, \&
  {Levin}}]{madigan+2011}
{Madigan}, A.-M., {Hopman}, C., \& {Levin}, Y. 2011, \apj, 738, 99,
  \dodoi{10.1088/0004-637X/738/1/99}

\bibitem[{Madigan {et~al.}(2009)Madigan, Levin, \& Hopman}]{madigan+2009}
Madigan, A.-M., Levin, Y., \& Hopman, C. 2009, \apjl, 697, L44,
  \dodoi{10.1088/0004-637X/697/1/L44}

\bibitem[{{Mapelli} {et~al.}(2012){Mapelli}, {Hayfield}, {Mayer}, \&
  {Wadsley}}]{mapelli+2012}
{Mapelli}, M., {Hayfield}, T., {Mayer}, L., \& {Wadsley}, J. 2012, \apj, 749,
  168, \dodoi{10.1088/0004-637X/749/2/168}

\bibitem[{{Merritt}(2013)}]{merritt2013}
{Merritt}, D. 2013, {Dynamics and Evolution of Galactic Nuclei} (Princeton
  University Press)

\bibitem[{{Merritt} {et~al.}(2011){Merritt}, {Alexander}, {Mikkola}, \&
  {Will}}]{merritt+2011}
{Merritt}, D., {Alexander}, T., {Mikkola}, S., \& {Will}, C.~M. 2011, \prd, 84,
  044024, \dodoi{10.1103/PhysRevD.84.044024}

\bibitem[{{Merritt} {et~al.}(2009){Merritt}, {Gualandris}, \&
  {Mikkola}}]{merritt+2009}
{Merritt}, D., {Gualandris}, A., \& {Mikkola}, S. 2009, \apjl, 693, L35,
  \dodoi{10.1088/0004-637X/693/1/L35}

\bibitem[{Mikkola \& Merritt(2008)}]{mikkola&merritt2008}
Mikkola, S., \& Merritt, D. 2008, \aj, 135, 2398,
  \dodoi{10.1088/0004-6256/135/6/2398}

\bibitem[{{Mori} {et~al.}(2019){Mori}, {Hailey}, {Mandel}, {Schutt},
  {Bachetti}, {Coerver}, {Baganoff}, {Dykaar}, {Grindlay}, {Haggard}, {Heuer},
  {Hong}, {Hord}, {Jin}, {Nynka}, {Ponti}, \& {Tomsick}}]{mori+2019}
{Mori}, K., {Hailey}, C.~J., {Mandel}, S., {et~al.} 2019, \apj, 885, 142,
  \dodoi{10.3847/1538-4357/ab4b47}

\bibitem[{{Naoz}(2016)}]{naoz2016}
{Naoz}, S. 2016, \araa, 54, 441, \dodoi{10.1146/annurev-astro-081915-023315}

\bibitem[{{Naoz} {et~al.}(2018){Naoz}, {Ghez}, {Hees}, {Do}, {Witzel}, \&
  {Lu}}]{naoz+2018}
{Naoz}, S., {Ghez}, A.~M., {Hees}, A., {et~al.} 2018, \apjl, 853, L24,
  \dodoi{10.3847/2041-8213/aaa6bf}

\bibitem[{{Naoz} {et~al.}(2020){Naoz}, {Will}, {Ramirez-Ruiz}, {Hees}, {Ghez},
  \& {Do}}]{naoz+2019}
{Naoz}, S., {Will}, C.~M., {Ramirez-Ruiz}, E., {et~al.} 2020, \apjl, 888, L8,
  \dodoi{10.3847/2041-8213/ab5e3b}

\bibitem[{{Nayakshin} \& {Cuadra}(2005)}]{nayakshin&cuadra2005}
{Nayakshin}, S., \& {Cuadra}, J. 2005, \aap, 437, 437,
  \dodoi{10.1051/0004-6361:20042052}

\bibitem[{{Nayakshin} \& {Zubovas}(2018)}]{nayakshin&zubovas2018}
{Nayakshin}, S., \& {Zubovas}, K. 2018, \mnras, 478, L127,
  \dodoi{10.1093/mnrasl/sly082}

\bibitem[{{Paumard} {et~al.}(2006){Paumard}, {Genzel}, {Martins}, {Nayakshin},
  {Beloborodov}, {Levin}, {Trippe}, {Eisenhauer}, {Ott}, {Gillessen}, {Abuter},
  {Cuadra}, {Alexander}, \& {Sternberg}}]{paumard+2006}
{Paumard}, T., {Genzel}, R., {Martins}, F., {et~al.} 2006, \apj, 643, 1011,
  \dodoi{10.1086/503273}

\bibitem[{Perets {et~al.}(2007)Perets, Hopman, \& Alexander}]{perets+2007}
Perets, H.~B., Hopman, C., \& Alexander, T. 2007, \apj, 656, 709,
  \dodoi{10.1086/510377}

\bibitem[{P\'erez \& Granger(2007)}]{perez+2007}
P\'erez, F., \& Granger, B.~E. 2007, Computing in Science and Engineering, 9,
  21, \dodoi{10.1109/MCSE.2007.53}

\bibitem[{Peters(1964)}]{peters1964}
Peters, P.~C. 1964, Physical Review, 136, 1224,
  \dodoi{10.1103/PhysRev.136.B1224}

\bibitem[{{Portegies Zwart} \& {McMillan}(2002)}]{portegies_zwart+2002}
{Portegies Zwart}, S.~F., \& {McMillan}, S. L.~W. 2002, \apj, 576, 899,
  \dodoi{10.1086/341798}

\bibitem[{{Prodan} {et~al.}(2015){Prodan}, {Antonini}, \&
  {Perets}}]{prodan+2015}
{Prodan}, S., {Antonini}, F., \& {Perets}, H.~B. 2015, \apj, 799, 118,
  \dodoi{10.1088/0004-637X/799/2/118}

\bibitem[{{Rauch} \& {Tremaine}(1996)}]{rauch&tremaine1996}
{Rauch}, K.~P., \& {Tremaine}, S. 1996, \na, 1, 149,
  \dodoi{10.1016/S1384-1076(96)00012-7}

\bibitem[{{Reid} \& {Brunthaler}(2020)}]{reid&brunthaler2020}
{Reid}, M.~J., \& {Brunthaler}, A. 2020, \apj, 892, 39,
  \dodoi{10.3847/1538-4357/ab76cd}

\bibitem[{{Rein} \& {Liu}(2012)}]{rein.liu2012}
{Rein}, H., \& {Liu}, S.~F. 2012, \aap, 537, A128,
  \dodoi{10.1051/0004-6361/201118085}

\bibitem[{{Rein} \& {Spiegel}(2015)}]{rein.spiegel2015}
{Rein}, H., \& {Spiegel}, D.~S. 2015, \mnras, 446, 1424,
  \dodoi{10.1093/mnras/stu2164}

\bibitem[{{Sari} \& {Fragione}(2019)}]{sari&fragione2019}
{Sari}, R., \& {Fragione}, G. 2019, \apj, 885, 24,
  \dodoi{10.3847/1538-4357/ab43df}

\bibitem[{Sari {et~al.}(2010)Sari, Kobayashi, \& Rossi}]{sari+2010}
Sari, R., Kobayashi, S., \& Rossi, E.~M. 2010, \apj, 708, 605,
  \dodoi{10.1088/0004-637X/708/1/605}

\bibitem[{{Sch{\"o}del} {et~al.}(2018){Sch{\"o}del}, {Gallego-Cano}, {Dong},
  {Nogueras-Lara}, {Gallego-Calvente}, {Amaro-Seoane}, \&
  {Baumgardt}}]{schodel+2017}
{Sch{\"o}del}, R., {Gallego-Cano}, E., {Dong}, H., {et~al.} 2018, \aap, 609,
  A27, \dodoi{10.1051/0004-6361/201730452}

\bibitem[{{Stephan} {et~al.}(2016){Stephan}, {Naoz}, {Ghez}, {Witzel},
  {Sitarski}, {Do}, \& {Kocsis}}]{stephan+2016}
{Stephan}, A.~P., {Naoz}, S., {Ghez}, A.~M., {et~al.} 2016, \mnras, 460, 3494,
  \dodoi{10.1093/mnras/stw1220}

\bibitem[{{\v{S}ubr} \& {Haas}(2016)}]{subr&haas2016}
{\v{S}ubr}, L., \& {Haas}, J. 2016, \apj, 828, 1,
  \dodoi{10.3847/0004-637X/828/1/1}

\bibitem[{{Tamayo} {et~al.}(2020){Tamayo}, {Rein}, {Shi}, \& {Hernand
  ez}}]{tamayo+2019}
{Tamayo}, D., {Rein}, H., {Shi}, P., \& {Hernand ez}, D.~M. 2020, \mnras, 491,
  2885, \dodoi{10.1093/mnras/stz2870}

\bibitem[{{The Astropy Collaboration} {et~al.}(2018){The Astropy
  Collaboration}, {Price-Whelan}, {Sip{\H o}cz}, {G{\"u}nther}, {Lim},
  {Crawford}, \& {Contributors}}]{astropy+2018}
{The Astropy Collaboration}, {Price-Whelan}, A.~M., {Sip{\H o}cz}, B.~M.,
  {et~al.} 2018, \aj, 156, 123, \dodoi{10.3847/1538-3881/aabc4f}

\bibitem[{{Trani} {et~al.}(2019){Trani}, {Fujii}, \& {Spera}}]{trani+2019}
{Trani}, A.~A., {Fujii}, M.~S., \& {Spera}, M. 2019, \apj, 875, 42,
  \dodoi{10.3847/1538-4357/ab0e70}

\bibitem[{Virtanen {et~al.}(2020)Virtanen, Gommers, Oliphant, Haberland, Reddy,
  Cournapeau, Burovski, Peterson, Weckesser, Bright, Walt, Brett, Wilson, K,
  Mayorov, Nelson, Jones, Kern, Larson, C, Polat, Feng, Moore, VanderPlas,
  Laxalde, Perktold, Cimrman, Henriksen, E, Harris, Archibald, Ribeiro,
  Pedregosa, Mulbregt, Vijaykumar, Bardelli, Rothberg, Hilboll, Kloeckner,
  Scopatz, Lee, Rokem, C, Fulton, Masson, Häggström, Fitzgerald, Nicholson,
  Hagen, Pasechnik, Olivetti, Martin, Wieser, Silva, Lenders, Wilhelm, G,
  Price, Ingold, Allen, Lee, Audren, Probst, Dietrich, Silterra, Webber,
  Slavič, Nothman, Buchner, Kulick, Schönberger, Cardoso, Reimer, Harrington,
  Rodríguez, Nunez-Iglesias, Kuczynski, Tritz, Thoma, Newville, Kümmerer,
  Bolingbroke, Tartre, Pak, Smith, Nowaczyk, Shebanov, Pavlyk, Brodtkorb, Lee,
  McGibbon, Feldbauer, Lewis, Tygier, Sievert, Vigna, Peterson, More, Pudlik,
  Oshima, Pingel, Robitaille, Spura, Jones, Cera, Leslie, Zito, Krauss,
  Upadhyay, Halchenko, \& Vázquez-Baeza}]{2020SciPy-NMeth}
Virtanen, P., Gommers, R., Oliphant, T.~E., {et~al.} 2020, Nature Methods, 17,
  261, \dodoi{10.1038/s41592-019-0686-2}

\bibitem[{{Wernke} \& {Madigan}(2019)}]{wernke&madigan2019}
{Wernke}, H.~N., \& {Madigan}, A.-M. 2019, \apj, 880, 42,
  \dodoi{10.3847/1538-4357/ab2711}

\bibitem[{{Witzel} {et~al.}(2014){Witzel}, {Ghez}, {Morris}, {Sitarski},
  {Boehle}, {Naoz}, {Campbell}, {Becklin}, {Canalizo}, {Chappell}, {Do}, {Lu},
  {Matthews}, {Meyer}, {Stockton}, {Wizinowich}, \& {Yelda}}]{witzel+2014}
{Witzel}, G., {Ghez}, A.~M., {Morris}, M.~R., {et~al.} 2014, \apjl, 796, L8,
  \dodoi{10.1088/2041-8205/796/1/L8}

\bibitem[{{Yelda} {et~al.}(2014){Yelda}, {Ghez}, {Lu}, {Do}, {Meyer}, {Morris},
  \& {Matthews}}]{yelda+2014}
{Yelda}, S., {Ghez}, A.~M., {Lu}, J.~R., {et~al.} 2014, \apj, 783, 131,
  \dodoi{10.1088/0004-637X/783/2/131}

\bibitem[{{Yu} \& {Tremaine}(2003)}]{yu&tremaine2003}
{Yu}, Q., \& {Tremaine}, S. 2003, \apj, 599, 1129, \dodoi{10.1086/379546}

\end{thebibliography}
}

\end{document}